\documentclass[nonblindrev]{informs3}

\DoubleSpacedXI 

\usepackage[plainpages=false,hyperfootnotes=false]{hyperref}
\hypersetup{
  colorlinks   = true, 
  urlcolor     = blue, 
  linkcolor    = red, 
  citecolor   = blue 
}

\DeclareMathOperator*{\corr}{\hbox{corr}}
\DeclareMathOperator*{\cov}{\hbox{cov}}
\DeclareMathOperator*{\var}{\hbox{var}}
\DeclareMathOperator*{\std}{\text{std}}
\DeclareMathOperator*{\E}{\mathbb{E}}
\DeclareMathOperator*{\N}{\mathbb{N}}

\DeclareMathOperator*{\M}{{\mathcal{M}}}

\DeclareMathOperator*{\arginf}{\text{arg}\min}

\newcommand{\K}{{\mathcal{K}}}
\newcommand{\Rr}{{\boldsymbol{R}}}
\newcommand{\x}{{\boldsymbol{x}}}

\newcommand{\Id}{{\mathds{1}}}

\usepackage{todonotes}
\usepackage{dsfont}
\usepackage{caption}
\usepackage{subcaption}
\usepackage{dcolumn}
\usepackage{tabularx}
\usepackage{booktabs}
\usepackage{multirow, stackrel}

\usepackage{natbib}
 \bibpunct[, ]{(}{)}{,}{a}{}{,}%
 \def\bibfont{\small}%
\TheoremsNumberedThrough     
\ECRepeatTheorems

\EquationsNumberedThrough    

\begin{document}

\setcounter{page}{1}
\VOLUME{}%
\NO{}%
\MONTH{}
\YEAR{}
\FIRSTPAGE{}%
\LASTPAGE{}%
\SHORTYEAR{}
\ISSUE{} %
\LONGFIRSTPAGE{} %
\DOI{}%

\RUNAUTHOR{Bernard, Pesenti and Vanduffel}

\RUNTITLE{Robust Distortion Risk Measures}

\TITLE{Robust Distortion Risk Measures}

\ARTICLEAUTHORS{%
\AUTHOR{Carole Bernard}
\AFF{Department of Accounting, Law and Finance, Grenoble Ecole de Management, France, \\
Department of Economics and Political
Science, Vrije Universiteit Brussel, Belgium} 
\AUTHOR{Silvana M. Pesenti}
\AFF{Department of Statistical Sciences, University of Toronto, Canada, \EMAIL{silvana.pesenti@utoronto.ca}}
\AUTHOR{Steven Vanduffel}
\AFF{Department of Economics and Political
Science, Vrije Universiteit Brussel, Belgium}
03. February 2023
\footnote{First version 18. August 2020; revised version 18. May 2022} 
} 

\ABSTRACT{%
The robustness of risk measures to changes in underlying loss distributions (distributional uncertainty) is of crucial importance in making well-informed decisions. In this paper, we quantify, for the class of  distortion risk measures with an absolutely continuous distortion function, its robustness to distributional uncertainty by deriving its largest (smallest) value when the underlying loss distribution has a known mean and variance and, furthermore, lies within a ball - specified through the Wasserstein distance - around a reference distribution. We employ the technique of isotonic projections to provide for these distortion risk measures a complete characterisation of sharp bounds on their value, and we obtain quasi-explicit bounds in the case of Value-at-Risk and Range-Value-at-Risk. We extend our results to account for uncertainty in the first two moments and provide applications to portfolio optimisation and to model risk assessment. 
}%


\HISTORY{Earlier versions have been presented at the 
\textit{Statistical Society of Canada Annual Meeting} (virtual), the \textit{Conference on Quantitative Risk Management and Financial Technology} (Waterloo, Canada), the  \textit{INFORMS Annual Meeting} (Seattle, USA), the \textit{Quantact lab} (Montr\'eal, Canada), the \textit{Workshop on Insurance Mathematics} (London, Canada), the \textit{Bahnhofs Colloquium} of the Swiss Actuarial Society (Zurich, Switzerland), the \textit{University of Waterloo} (Canada), the \textit{Quantact} (Montr\'eal, Canada), and the \textit{University of Siegen} (Germany).}

\maketitle

%


\section{Introduction}
Many decisions in financial risk management are based on evaluations of so-called risk or performance measures, e.g., Value-at-Risk (VaR) and Tail Value-at-Risk (TVaR) \citep{SolvencyII, BCBS13}, which are limit cases of  Range Value-at-Risk (RVaR) \citep{cont2010robustness}. Risk managers rely on such measures, as they provide a classification of risk severities by associating to every loss distribution a real number \citep{Artzner1999MF}. Many risk measures are, however, highly sensitive to changes in the underlying loss distribution, and this vulnerability can contribute  to adverse decisions; see \cite{cont2010robustness}, \cite{Kou2013MOR}, \cite{Embrechts2015FS}, and \cite{Pesenti2016DM}. Here, we quantify distributional uncertainty via bounds on the values of risk measures. Specifically, we study \textit{worst-case} and \textit{best-case values} of risk measures -- that is, the largest (resp.\,smallest) value a risk measure can attain when the loss distribution belongs to a set of plausible alternative distributions. 

The literature on risk bounds presents two main approaches. The first  studies risk bounds for the aggregate portfolio loss $S=X_1 +\ldots + X_n$ given that the components have known marginal distributions but unknown interdependence; see,  among others, \cite{denuit1999stochastic}, \cite{embrechts2006bounds}, \cite{wang2011complete}, and \cite{embrechts2013model}. Although explicit results can be derived in certain cases of interest, the bounds obtained are typically too wide to be practically useful. Specifically, the upper bound on a distortion risk measure is always at least as large as its value under the assumption of a comonotonic dependence, a situation that is arguably very extreme. Papers that study bounds with additional dependence constraints or by imposing modelling structures include \cite{bernard2017risk},
\cite{lux2019value}, and \cite{Wang2019dual}. Another stream of the literature is concerned with deriving risk bounds of the aggregate loss $S$ under partial knowledge of its moments. Since only information on the moments of $S$ is required, these risk bounds apply to non-linear loss models, such as re-insurance portfolios, and to loss models for which the explicit aggregation is not known analytically, e.g., when $S$ is obtained via simulation models. As (higher) moments of the aggregate loss $S$ depend both on the marginal distributions and their interdependence, one thus retains some aspects of the marginal distributions and dependence information. Relevant papers include \cite{hurlimann2002analytical} for the case of VaR and \cite{cornilly2018IME} and \cite{Zhu2018SSRN} for the case of (concave) distortion risk measures. Risk bounds that use additional refined information, such as uni-modality or symmetry, can be found in, e.g., \cite{li2018SIAM}.

This paper is situated in the second stream, in that we consider risk bounds of an aggregate loss $S$ under knowledge of its first two moments. However, we additionally impose a probability distance constraint on $S$, specified via the Wasserstein distance.  Specifically, all alternative aggregate loss distributions, over which the worst- and best- cases are sought -- the \textit{uncertainty set} -- lie within a tolerance distance from a reference loss distribution. Thus, the size of the tolerance distance determines the uncertainty around the reference distribution, which may also entail uncertainties on its components. Indeed, if the tolerance is small enough, the uncertainty set contains only the reference distribution, whereas if the tolerance distance increases to infinity, we recover bounds with moments constraints only. Therefore, the tolerance distance allows for a refined notion of model risk, which results in a continuous spectrum of risk bounds interpolating from the risk of the reference distribution to the bounds on the aggregate loss under the knowledge of the first two moments only.

Worst- (best-) case values under probability distance constraints have been considered by \cite{Glasserman2014QF}
and \cite{Lam2016MOR}, who use the Kullback-Leibler divergence, as well as by \cite{Blanchet2019MOR}, who utilise distances stemming from mass transportation. These papers, however, consider the expected value of a function of the loss random variable, whereas here we study the class of distortion risk measures. Moreover, we use a distance that allows for comparison of distributions on differing support, which is important in our context as  it is not clear a priori whether the worst-case distribution shares the same support as the reference distribution. For example, a discrete reference distribution, together with a particular choice of distortion risk measure (e.g., inverse-S shaped), will result in a continuous worst-case distribution.
In this work, we quantify the distance between the alternative distribution function and the reference distribution via the Wasserstein distance of order 2. The Wasserstein distance has been widely applied to model distributional uncertainty in financial contexts (see e.g., \cite{Pflug2007QF}, \cite{bartl2020computational} and \cite{chen2021sharing}), and as a distance from reference model (see e.g., \cite{Pesenti2021SSRN} and \cite{pesenti2021Wasser}).

A related stream of literature studies distributionally robust optimisation, leading to an  optimum in the worst case. Relevant literature, in which the worst-case value of a distortion risk measure is considered includes \cite{Cai2020SSRN} and \cite{Pesenti2020optimizing}. \cite{Cai2020SSRN} consider distortion risk measures under the assumption that the uncertainty set of the components $X_1, \ldots, X_n$ is characterised by the knowledge of their moments, the components having compact support and additional convex constraints. \cite{Pesenti2020optimizing} consider signed Choquet Integrals (see Section~\ref{sec: Choquet} for a definition) and uncertainty sets characterised by so-called \textit{closedness-under-concentration}. Neither of these works considers the Wasserstein distance, and the \textit{closedness-under-concentration} condition is not compatible with uncertainty sets that are $\sqrt{\varepsilon}$-Wasserstein balls around an (arbitrary) reference distribution.

In this paper, we derive for any given \textit{distortion risk measure} (with absolutely continuous distortion function) its worst- and best-case values  when the loss distribution has known first two moments and lies within a tolerance distance from a reference distribution, specified through the Wasserstein distance. To derive these bounds, we use the technique of isotonic projections (\cite{Nemeth2003isothonic}). As a result, we obtain  quasi analytical best- and worst-case bounds that significantly improve existing moment bounds. We find that for small Wasserstein tolerance distances, the worst-case distribution function (the distribution function attaining the worst-case value) is close (in a probabilistic sense) to the reference distribution, whereas for large tolerance distances the worst-case distribution function is no longer affected by the reference distribution. Thus, when the Wasserstein tolerance distance goes to infinity, the reference distribution becomes irrelevant, and we recover, as corollaries to our results, known moment bounds on distortion risk measures; see \cite{Li2018OR} and  \cite{Zhu2018SSRN}. Indeed, \cite{Li2018OR} considers worst-case concave distortion risk measures with given mean and standard deviation, and \cite{Zhu2018SSRN} consider the best-case and worst-case of the entire class of distortion risk measure under the knowledge of the first two moments. Neither considers a Wasserstein constraint.

We further apply our results to obtain quasi-explicit bounds on RVaR and (via a direct proof) VaR. We find that for small Wasserstein tolerance distances, the worst-case distribution functions for RVaR and VaR are no longer two-point distributions, thus making the bounds attractive for financial risk management applications. 

This paper is structured as follows: Section~\ref{sec:prel} introduces the necessary notation and formulates the problem. The main results -- the quasi-analytical best- and worst-case values of distortion risk measures under a Wasserstein distance constraint -- are presented in Section~\ref{sec: bounds-RM}. Section \ref{sec: extension} contains two extensions. In Section~\ref{sec: VaR}, we calculate bounds of VaR and RVaR. In Section~\ref{sec:Wasserstein-only}, we extend our results  to complete moment uncertainty -- that is, when the uncertainty set is a $\sqrt{\varepsilon}$-Wasserstein ball around a reference distribution without any moment constraints. We provide applications of the risk bounds to portfolio optimisation and to model risk assessment of an insurance portfolio, including rationales for choosing the Wasserstein distance, in Section~\ref{sec: application}. All proofs are relegated to the appendix.

\section{Problem Formulation \label{sec:prel}}
We consider an atomless probability space $(\Omega, \mathcal{A}, \mathbb{P})$ and let $L^2 = L^2(\Omega, \mathcal{A}, \mathbb{P})$ be the set of all square integrable random variables on that space. We denote by $\mathcal{M}^2 = \{\, G(x) = \mathbb{P}(X \leq x) ~|~ X \in L^2\, \}$ the corresponding space of distribution functions with finite second moment. The (left-continuous) inverse, also called the quantile function, of a distribution function $G \in \mathcal{M}^2$ is $G^{-1}(u) = \inf\{\, y \in \mathbb{R}~|~ G(y) \geq u\,\}$, and we denote its right-continuous inverse by $G^{-1,+}(u) = \inf\{ \,y \in \mathbb{R}~|~ G(y) > u\,\}, ~ 0 < u< 1$. Throughout the exposition, we write $U \sim \mathcal{U}(0,1)$ for a standard uniform random variable on $(0,1)$.

\subsection{Distortion Risk Measures}\label{sec: risk measures}
In this section, we recall the class of \emph{distortion risk measures} that contains risk measures commonly encountered in financial applications, such as VaR and TVaR. A distortion risk measure, evaluated at a distribution function $G \in \mathcal{M}^2$, is defined via the Choquet integral
\begin{align*}
H_g(G) 
	&= - \int_{-\infty}^0 1 - g(1 -  G(x)) \,\mathrm{d}x + \int_0^{+ \infty}g(1 - G(x))\,\mathrm{d}x\,,
\end{align*}
whenever at least one of the two integrals is finite. The function $g \colon [0,1] \to [0,1]$ refers to a \emph{distortion function} -- that is, a non-decreasing function satisfying $g(0) = 0$ and $g(1) = 1$. If $g$ is absolutely continuous, then the distortion risk measure $H_g$ has the following representation \citep{Dhaene2012EAJ}
\begin{align}
H_g(G) &= \int_0^1 \gamma(u)G^{-1}(u) \,\mathrm{d}u\,, \label{eq: distortion Choquet integral}
\end{align}
with \textit{weight function} $\gamma(u) = \partial_- g(x)|_{x = 1 - u}, ~ 0 < u < 1$, which satisfies $\int_0^1 \gamma(u)\mathrm{d}u = 1$ and where $\partial_-$ denotes the derivative from the left. In what follows, we may sometimes write $H_g(G^{-1})$ instead of $H_g(G)$. 
\begin{assumption}\label{asm: Choquet integral}
We assume that representation \eqref{eq: distortion Choquet integral} holds and that $\int_0^1 | \gamma (u)|^2 \mathrm{d}u  < + \infty$. 
\end{assumption}

Throughout the paper, we use the following three concave distortion risk measures to illustrate our results: the dual power distortion with parameter $\beta>0$, 
\begin{equation}\label{eqa}
	g(x)=1-(1-x)^\beta\,, \quad \text{and} 
	\quad \gamma(u)=\beta u^{\beta-1}\,;
\end{equation}
the Wang transform \citep{Wang1996ASTIN} with parameter $0<q_{0}<1$,
\begin{equation}\label{eqb}
g(x)=\Phi\left(\Phi^{-1}(x)+\Phi^{-1}(q_{0})\right)\,, \quad \text{and}
\quad  \gamma(u)=\frac{\phi\left(\Phi^{-1}(1-u)+\Phi^{-1}(q_{0})\right)}{\phi\left(\Phi^{-1}(1-u)\right)}\,,
\end{equation}
where $\Phi$ and $\phi$ denote the standard normal distribution and its density, respectively; and the Tail Value-at-Risk ($\text{TVaR}_\alpha$), also called Expected Shortfall, at level $0<\alpha<1$  \citep{Acerbi2002JBF} with
\begin{equation}\label{eqc}
g(x)=\min\big\{\tfrac{x}{1-\alpha}, ~1\big\} \quad \text{and}
	\quad \gamma(u)=\tfrac{1}{1-\alpha} \Id_{(\alpha , 1)}(u)\,. 
\end{equation}
In Section~\ref{sec: VaR}, we also consider the case of Value-at-Risk ($\text{VaR}_\alpha$), $0<\alpha<1$, a distortion risk measure with $g(x) =  \Id_{(1-\alpha , 1]}(x)$ that does not satisfy Assumption \ref{asm: Choquet integral}, i.e. it does not admit a representation as in \eqref{eq: distortion Choquet integral} as a Lebesgue Integral.

\subsection{Modelling Distributional Uncertainty} \label{sec: worst case rm}

In financial risk management, distributional uncertainty is prevalent and bounds on the value of a risk measure, so-called \emph{worst- and best-case values} or \emph{upper and lower bounds}, are valuable tools for decision making. A worst- (best-) case value of a risk measure is defined as the largest (smallest) value a risk measure can attain when the underlying distribution belongs to a set of alternative distributions. Here, we consider subsets of $\mathcal{M}^2$ that are characterised via a tolerance distance to a \emph{reference distribution}, specified through the Wasserstein distance of order 2.  Distribution functions belonging to such an uncertainty set may be viewed as alternatives to the reference distribution, representing model misspecification or alternative model assumptions. To formalise this, we recall the definition of the Wasserstein distance. The Wasserstein distance exhibits desirable properties, such as no restriction on the support of distribution functions, which is in contrast to the Kullback-Leibler divergence; see \cite{Villani2008book} for a discussion of the Wasserstein distance. 

\begin{definition}[Wasserstein distance of order 2] The Wasserstein distance of order 2 between $G_1, G_2 \in \mathcal{M}^2$ is \citep{Villani2008book}
	\begin{align*}
	d_W(G_1,G_2) &= \inf \left.\left\{ \Big[\,\mathbb{E}\big(\,(X_1 - X_2)^2\,\big)\, \Big]^\frac12  \,\right| \,X_1 \sim G_1, ~X_2 \sim G_2 \right\}\, ,
	\end{align*}
where the infimum is taken over all bivariate distributions with marginals $G_1$ and $G_2$. 
\end{definition}
For distributions on the real line, the Wasserstein distance admits the following well-known representation \citep{dall1956SNS}
\begin{equation*}
    d_W(G_1,G_2) = \left(\int_0^1 \left(G_1^{-1}(u) - G_2^{-1}(u) \right)^2 \,\mathrm{d}u\right)^\frac12\, .
\end{equation*}
Hence, the Wasserstein distance between $G_1$ and $G_2$ is uniquely determined by their corresponding quantile functions, and we may write $d_W(G_1^{-1}, G_2^{-1})$ instead of $d_W(G_1, G_2)$. 

We denote by $F \in \mathcal{M}^2$ the reference distribution and its first two moments by $\int x \,\mathrm{d} F(x)= \mu_F \in \mathbb{R}$ and $\int x^2 \,\mathrm{d}F(x) = \mu_F^2 +  \sigma_F^2, ~ \sigma_F>0$, respectively. Throughout, we fix the reference distribution $F$ and consider distributional uncertainty sets of the type
\begin{equation*}
\mathcal{M}_\varepsilon(\mu, \sigma) 
= \left\{ G \in \mathcal{M}^2 ~\left|~ \int x\, \mathrm{d}G(x) = \mu, \,\int x^2\, \mathrm{d}G(x) = \mu^2 + \sigma^2,  ~ d_W(F,G) \leq \sqrt{\,\varepsilon\,} \right\}\right.,
\end{equation*}
where $\mu\in\mathbb{R}$, $\sigma>0$, and $0 \le \varepsilon \le + \infty$. The set  $\mathcal{M}_\varepsilon(\mu, \sigma)$ thus contains all distribution functions whose first two moments are $\mu$ and $\mu^2 + \sigma^2$, respectively, and that lie within a $\sqrt{\,\varepsilon\,}$-Wasserstein ball around the reference distribution $F$. The $\varepsilon$-constraint in the definition of $\mathcal{M}_\varepsilon(\mu, \sigma)$ is redundant when $\varepsilon = +\infty$, and in this special case we denote the uncertainty set $\mathcal{M}_\infty(\mu, \sigma)$ by $\mathcal{M}(\mu, \sigma)$. Note that in applications, one may often encounter that $(\mu,\sigma)=(\mu_F,\sigma_F)$, but our set-up allows them to be different, which could for example be useful when their values are estimated using different data sources.

The best- and worst-case values of a distortion risk measure $H_g$ over the distributional uncertainty set $\mathcal{M}_\varepsilon(\mu, \sigma)$ are respectively defined by

\begin{subequations}\label{eq: prob opt}
\begin{minipage}{0.45\textwidth}
\begin{equation}\label{eq:problem inf}
\inf_{G \in \mathcal{M}_\varepsilon(\mu, \sigma)} H_g(G)
\end{equation} 
\end{minipage}
\begin{minipage}{0.45\textwidth}
\begin{equation}\label{eq:problem sup}
\sup_{G \in \mathcal{M}_\varepsilon(\mu, \sigma)} H_g(G)\,.
\end{equation}
\end{minipage}%
\end{subequations}\\%

\noindent
In addition to the  best- and worst-case values, we also study \emph{best-case} and \emph{worst-case distribution functions} if they exist -- that is, the distribution functions attaining \eqref{eq:problem inf} and \eqref{eq:problem sup}, respectively
\begin{subequations}\label{eq: prob arg}
\begin{minipage}{0.45\textwidth}
\begin{equation}\label{eq:problem arginf}
\argmin_{G \in \mathcal{M}_\varepsilon(\mu, \sigma)} H_g(G)
\end{equation} 
\end{minipage}
\begin{minipage}{0.45\textwidth}
\begin{equation}\label{eq:problem argsup}
\argmax_{G \in \mathcal{M}_\varepsilon(\mu, \sigma)} H_g(G)\,.
\end{equation}
\end{minipage}%
\end{subequations}\\%

\noindent
Note that the square-integrability of $\gamma$ in Assumption \ref{asm: Choquet integral} is not restrictive. Indeed if $\int_0^1 | \gamma (u)|^2 \mathrm{d}u  = + \infty$, then problems \eqref{eq:problem inf} and \eqref{eq:problem sup} are both equal to $+\infty$. We provide a proof in Appendix \ref{sec: proofs}, Lemma \ref{lemma-asm-1}.
We impose two additional conditions that are necessary in order to ensure that the considered optimisation problems are relevant and well-posed. 
\begin{assumption}\label{asm: gamma not 1}
We assume that $\gamma(\cdot) \neq 1$; otherwise $H_g(G) = \mu$ for all $G \in \mathcal{M}_\varepsilon(\mu, \sigma)$ and problems \eqref{eq: prob opt} and \eqref{eq: prob arg} are obsolete. 
\end{assumption}
\begin{assumption}\label{asm: epsilon big enough}
We assume that $\mathcal{M}_\varepsilon(\mu, \sigma) $ contains at least two elements, and hence infinitely many; otherwise problems \eqref{eq: prob opt} and \eqref{eq: prob arg} are either ill-posed or trivial.
\end{assumption}

In this regard, Assumption \ref{asm: epsilon big enough} is equivalent to assuming that $\varepsilon > (\mu_F - \mu)^2 + (\sigma_F - \sigma)^2$. To see this, note that for any $G \in \mathcal{M}_\varepsilon(\mu, \sigma) $ 
\begin{align*}
d_W(F, G)^2
&= \int_0^1 \left(F^{-1}(u) - G^{-1}(u) \right)^2 \,\mathrm{d}u\\
& = \mu_F^2 + \sigma_F^2 + \mu^2 + \sigma^2
- 2 \, \cov\left(	F^{-1}(U), G^{-1}(U)\right) - 2 \mu \mu_F\\
& = (\mu_F - \mu)^2 + (\sigma_F - \sigma)^2 + 2\, \sigma \sigma_F(1 - \corr(F^{-1}(U), G^{-1} (U)))\\
& \geq (\mu_F - \mu)^2 + (\sigma_F - \sigma)^2.  	
\end{align*}
Hence, if $\varepsilon < (\mu_F - \mu)^2 + (\sigma_F - \sigma)^2$, then $\mathcal{M}_\varepsilon(\mu, \sigma)= \emptyset$, and if $\varepsilon = (\mu_F - \mu)^2 + (\sigma_F - \sigma)^2$, then $\mathcal{M}_\varepsilon(\mu, \sigma)$ is a singleton, containing only one distribution with quantile function 
\begin{equation*} 
G^{-1}(u)= \mu+\sigma \left(\frac{F^{-1}(u)-\mu_F}{\sigma_F}\right), \quad 0 < u < 1. \quad 
\end{equation*}
Moreover, in this special case, $G$ coincides with the reference distribution $F$ if and only if $\mu=\mu_F$ and $\sigma=\sigma_F$. Thus, from now on, we assume that $\varepsilon> (\mu_F - \mu)^2 + (\sigma_F - \sigma)^2$.

\section{Bounds on Distortion Risk Measures}\label{sec: bounds-RM}

For ease of exposition, we first study  worst-case values \eqref{eq:problem sup} and worst-case distribution functions \eqref{eq:problem argsup} for the  class of \emph{concave distortion risk measures} that are distortion risk measures with concave distortion function $g$. The corresponding results for the case of  general distortion risk measures are presented in Section~\ref{sec: distortion} and those for best-case values \eqref{eq:problem inf} and best-case distributions \eqref{eq:problem arginf} in Section~\ref{sec: best}.

\subsection{Upper Bounds on Concave Distortion Risk Measures}\label{sec: concave}

Concave distortion risk measures are a subset of the class of distortion risk measures that are coherent and include the widely used TVaR. Furthermore, concave distortion risk measures span the class of law-invariant, coherent, and comonotone additive risk measures \citep{Kusuoka2001AME}. For the special case of concave distortion risk measures, we obtain analytic solutions to problems \eqref{eq:problem sup} and \eqref{eq:problem argsup}, which are presented in the following theorem.
 
\begin{theorem}[Worst-Case Values and Quantiles of Concave Distortions]\label{thm: concave}
Let Assumptions \ref{asm: Choquet integral}, \ref{asm: gamma not 1}, and \ref{asm: epsilon big enough} be fulfilled.
Further, let $g$ be a concave distortion function and denote by $c_0 = \corr(F^{-1}(U), \gamma(U))$.\footnote{If $c_0 = 1$, then only case 2. applies.} Then, the following statements hold:  
\begin{enumerate}
	\item If $(\mu_F - \mu)^2 + (\sigma_F - \sigma)^2 < \varepsilon < (\mu_F - \mu)^2 + (\sigma_F - \sigma)^2 + 2\sigma \sigma_F(1 - c_0)$, then the solution to \eqref{eq:problem argsup} is unique and has a quantile function given by
	\begin{equation*}
	h_\lambda(u) = 
	\mu + \sigma \left(\frac{\gamma(u) + \lambda F^{-1}(u)-a_\lambda}{b_\lambda}\right), \quad  0 < u < 1\,,
	\end{equation*}
	where $\lambda>0$ denotes the unique positive solution to $ d_W(F^{-1}, h_\lambda)= \sqrt{\varepsilon} $, which is explicitly given by
\begin{equation}
    \lambda=\frac{K}{\sigma_F^2}\sqrt{\frac{C_{\gamma, F}^2-V\sigma_F^2}{K^2-\sigma^2\sigma_F^2}}-\frac{C_{\gamma, F}}{\sigma_F^2},
	\label{lamSOLth1}
\end{equation}
where $V=\var(\gamma(U))$, $C_{\gamma, F}=\cov\left(	F^{-1}(U),  \gamma(U)\right)$, and 
$K=\frac{\mu_F^2 +\sigma_F^2 + \mu^2 + \sigma^2- 2 \mu \mu_F-\varepsilon}{2}\ge0$ and where $a_\lambda = \E(\gamma(U) + \lambda F^{-1}(U)) $ and $b_\lambda ={\std(\gamma(U) + \lambda F^{-1}(U))}$. The corresponding worst-case value, i.e., the solution to \eqref{eq:problem sup}, is 
	\begin{align*}
	H_g(h_\lambda) 
	&= \mu  + \frac{\sigma}{b_\lambda}\left(V+\lambda C_{\gamma, F}\right) 
	= \mu  + \sigma \std\left(\gamma(U)\right) \corr\left(\gamma(U), \gamma(U) + \lambda F^{-1}(U)\right),
	\end{align*} 
	and  $H_g(h_\lambda)$  
	is continuous in $\varepsilon$.
		
	\item If $(\mu_F - \mu)^2 + (\sigma_F - \sigma)^2 + 2\sigma \sigma_F(1 - c_0) \leq \varepsilon$, then case $1.$ applies with $\lambda = 0$.
\end{enumerate}
\end{theorem}
The worst-case value provided in Theorem~\ref{thm: concave} is sharp for all $(\mu_F - \mu)^2 + (\sigma_F - \sigma)^2  < \varepsilon \le + \infty$ and attained by the worst-case quantile function $h_\lambda$. We observe that $h_\lambda$ is a weighted average of the reference quantile function $F^{-1}$ and the (non-decreasing) weight function $\gamma$ from the distortion risk measure. As the worst-case value $H_g$ is non-decreasing in the tolerance distance $\varepsilon$, we obtain that $\corr\left(\gamma(U), \gamma(U) + \lambda F^{-1}(U)\right)$ is non-decreasing in $\varepsilon$, which in turn implies that $\lambda$ is non-increasing in $\varepsilon$. Hence, we obtain that the influence of the reference distribution on the worst-case quantile function $h_\lambda$ diminishes with increasing tolerance distance $\varepsilon$. Furthermore, for $\varepsilon$ sufficiently large, i.e., under case 2. of Theorem~\ref{thm: concave}, $\lambda$ is zero and the worst-case quantile function is independent of the reference distribution. Note that when $\varepsilon=(\mu_F - \mu)^2 + (\sigma_F - \sigma)^2 + 2\sigma \sigma_F(1 - c_0)$,  the expression of $\lambda$ in \eqref{lamSOLth1} is still valid and simplifies to $\lambda=0$.

As a direct corollary we obtain the worst-case values of TVaR. The proof is omitted as it follows by direct calculations from Theorem~\ref{thm: concave}.

\begin{corollary}[Worst-case Value of TVaR]\label{cor:TVaR-WC-explicit}
Let Assumption \ref{asm: epsilon big enough} be fulfilled. Then, the worst-case value of the $\text{TVaR}_\alpha$ is given by
\begin{equation*}
    \sup_{G \in \mathcal{M}_\varepsilon(\mu, \sigma)}TVaR_{\alpha}(G)
    =
    \mu + \sigma \frac{\frac{\alpha}{1 - \alpha} +\lambda \left(TVaR_\alpha (F) - \mu_F\right)}{\sqrt{ \frac{\alpha}{ 1 - \alpha} +2\lambda \left(TVaR_\alpha (F) - \mu_F\right) + \lambda^2 \sigma_F^2}}\,,
\end{equation*}
Here, if $ \varepsilon < (\mu_F - \mu)^2  + (\sigma_F - \sigma)^2 + 2 \sigma \sigma_F (1 - c_0)$ then $\lambda$ is given by equation \eqref{lamSOLth1} in Theorem~\ref{thm: concave}, and if  $\varepsilon \geq (\mu_F - \mu)^2  + (\sigma_F - \sigma)^2 + 2 \sigma \sigma_F (1 - c_0)$, then $\lambda = 0.$ 
\end{corollary}

Theorem~\ref{thm: concave} generalises results that first appeared in \cite{Li2018OR}. This author derived sharp upper bounds on $H_g$ under the knowledge of the first two moments, i.e., without considering an $\sqrt{\varepsilon}$-Wasserstein distance constraint or a reference distribution. Applying Theorem~\ref{thm: concave} with $\varepsilon = + \infty$, we obtain his result, which is stated here for completeness; see also \cite{cornilly2018IME} and \cite{Zhu2018SSRN}.   
\begin{corollary}[Worst-Case Values for $\varepsilon = + \infty$; \cite{Li2018OR}]\label{cor: properties}
Let Assumptions \ref{asm: Choquet integral} and \ref{asm: gamma not 1} be fulfilled. Further, let $g$ be a concave distortion function; then
\begin{equation*}
\sup_{G \in \mathcal{M}(\mu, \sigma)} H_g(G) = \mu  + \sigma \std\left(\gamma(U)\right)=  
\mu  + \sigma \sqrt{\,\int_{0}^{1}(\gamma(p)-1)^2\,dp\,}\,.
\end{equation*}
The bound is sharp and obtained by a unique maximising distribution function with quantile function $h_0$ defined in Theorem~\ref{thm: concave}. 
\end{corollary}

\begin{example}
We illustrate the worst-case values of Theorem~\ref{thm: concave} for three concave distortion risk measures: the dual power distortion, the Wang transform, and TVaR defined in \eqref{eqa}, \eqref{eqb}, and \eqref{eqc}, respectively. The reference distribution $F$ is chosen to be standard normal, and we further set $\mu=\mu_{F} = 0$ and $\sigma=\sigma_{F} = 1$.
\begin{figure}[!htbp]
\begin{tabular}{ccc}
Dual Power Distortion & Wang Distortion & TVaR distortion\\
\includegraphics[width=5cm, height=5.5cm]{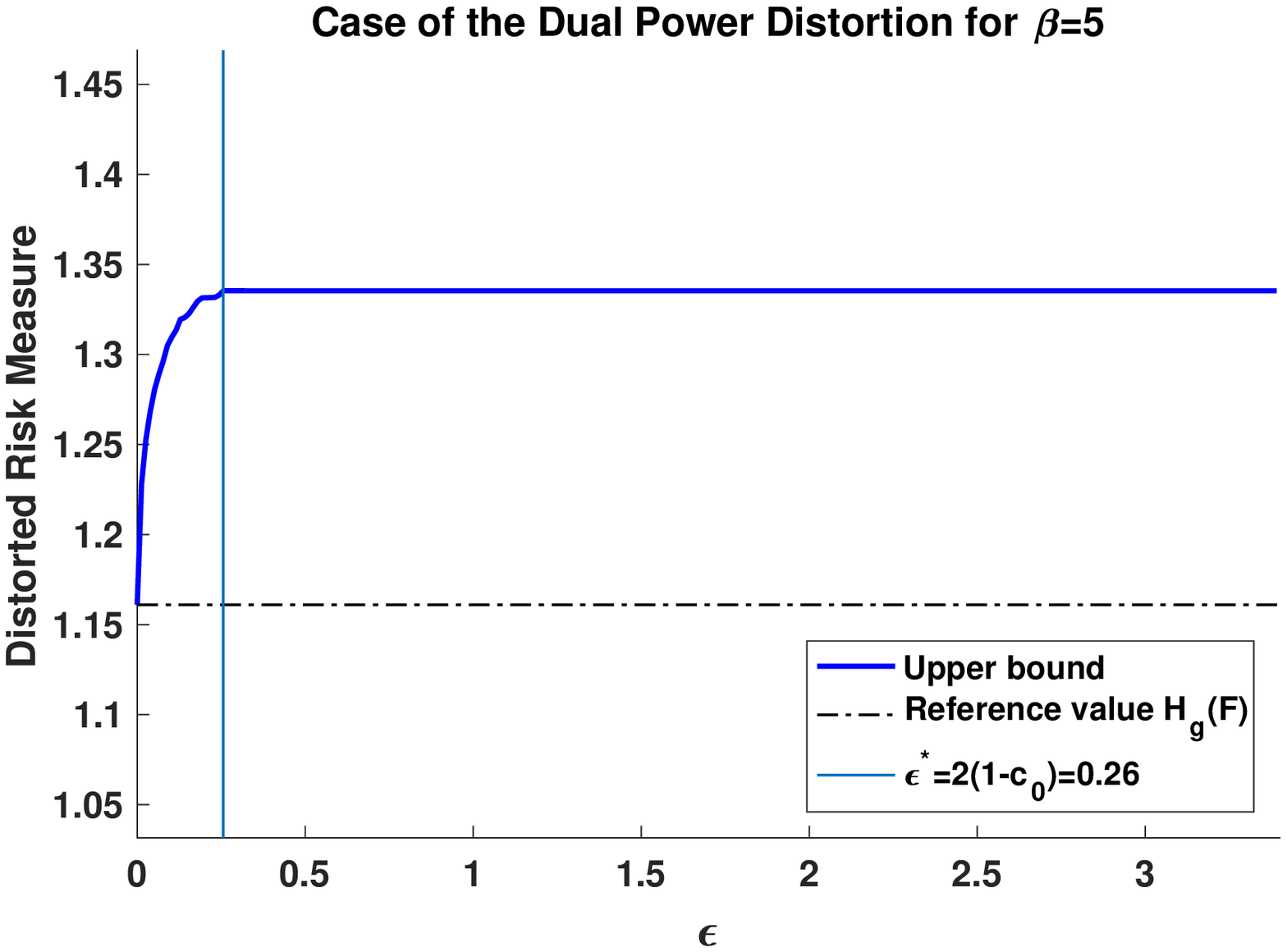}
 &\includegraphics[width=5cm, height=5.5cm]{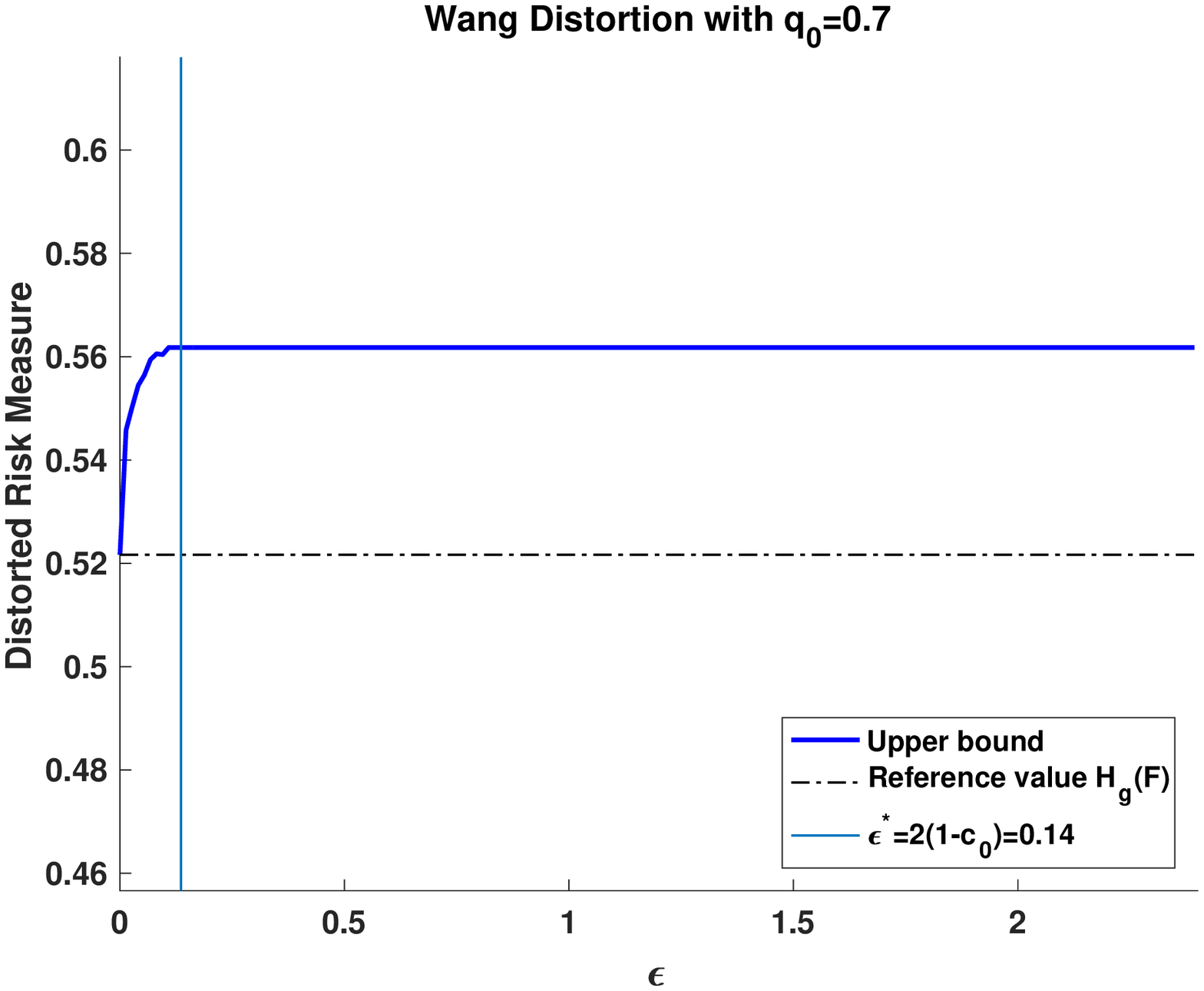}&
\includegraphics[width=5cm, height=5.5cm]{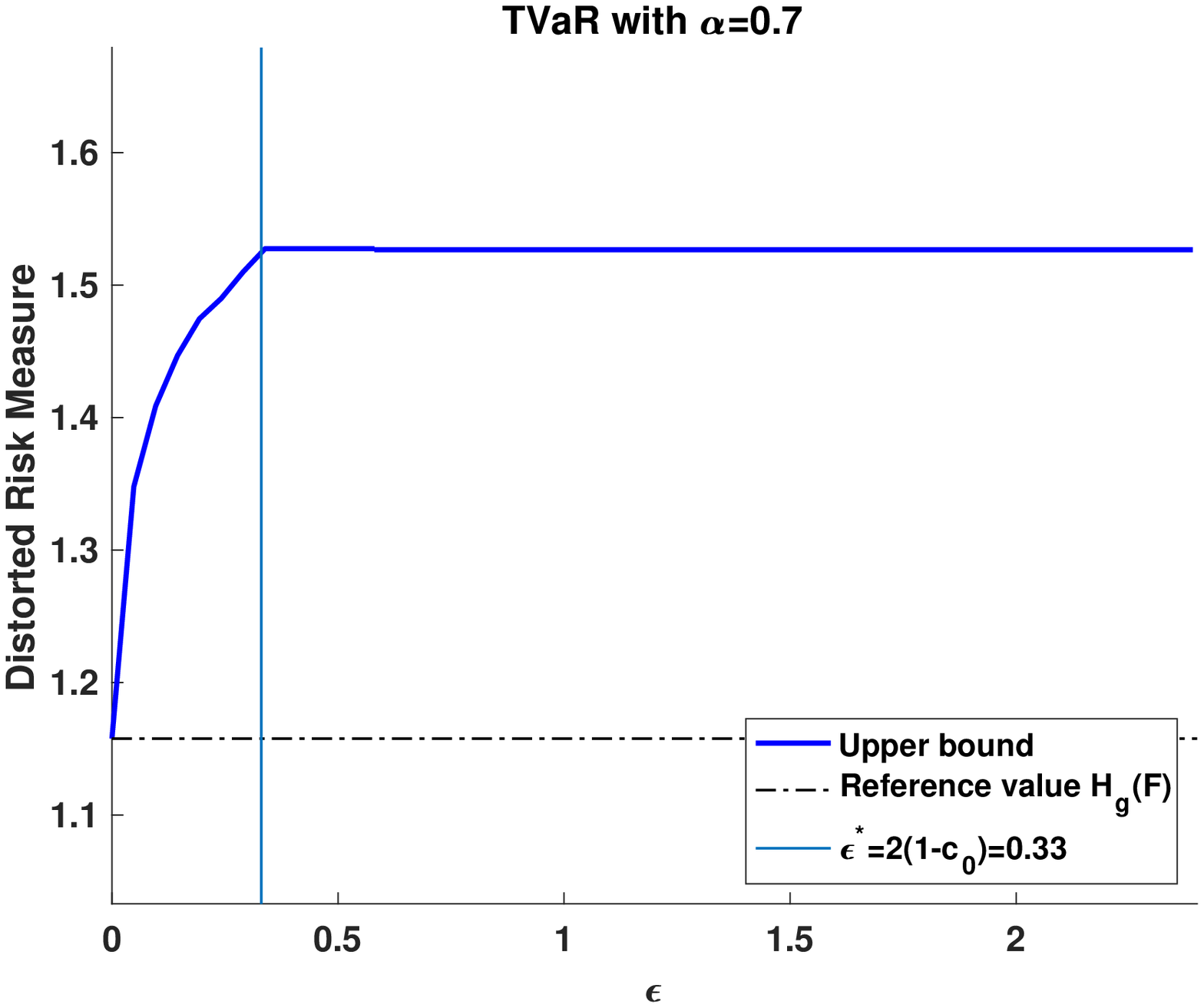}\\
 \includegraphics[width=5cm, height=5.5cm]{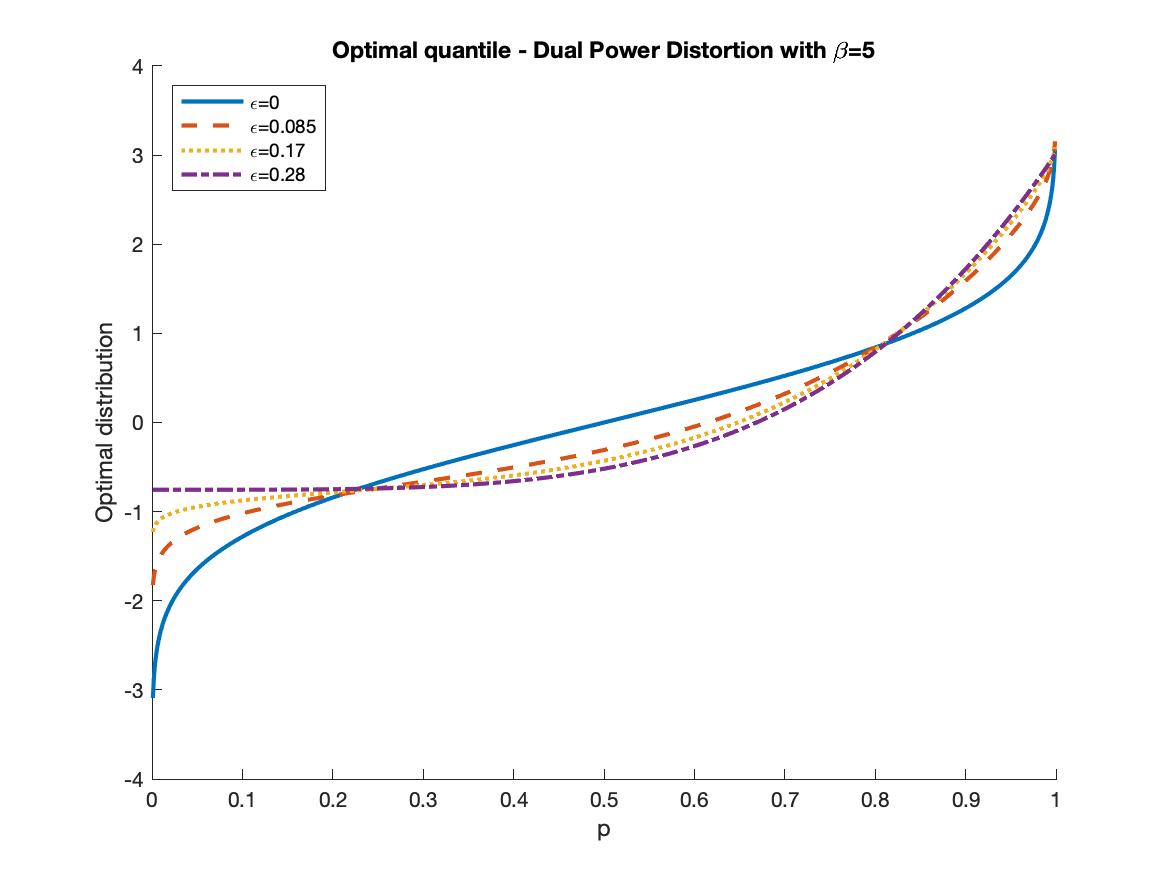} &\includegraphics[width=5cm, height=5.5cm]{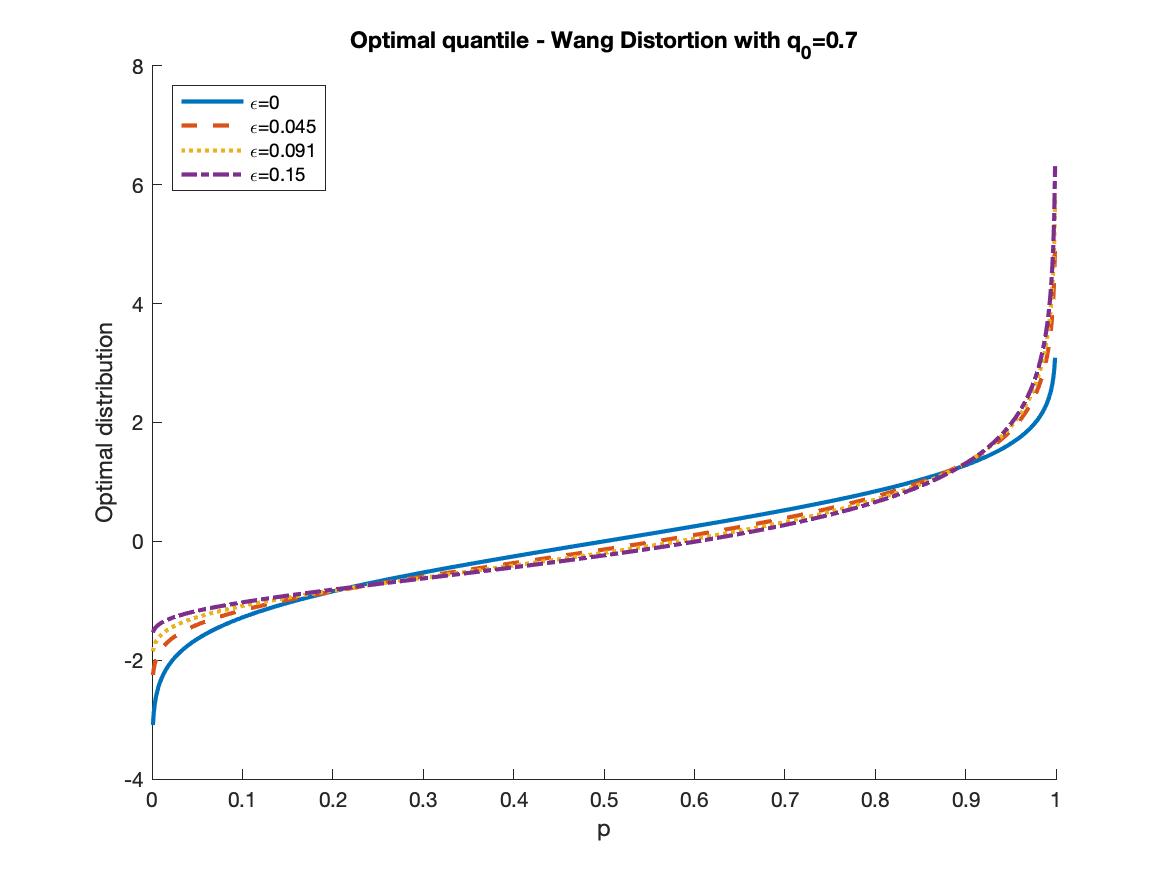} 
&\includegraphics[width=5cm, height=5.5cm]{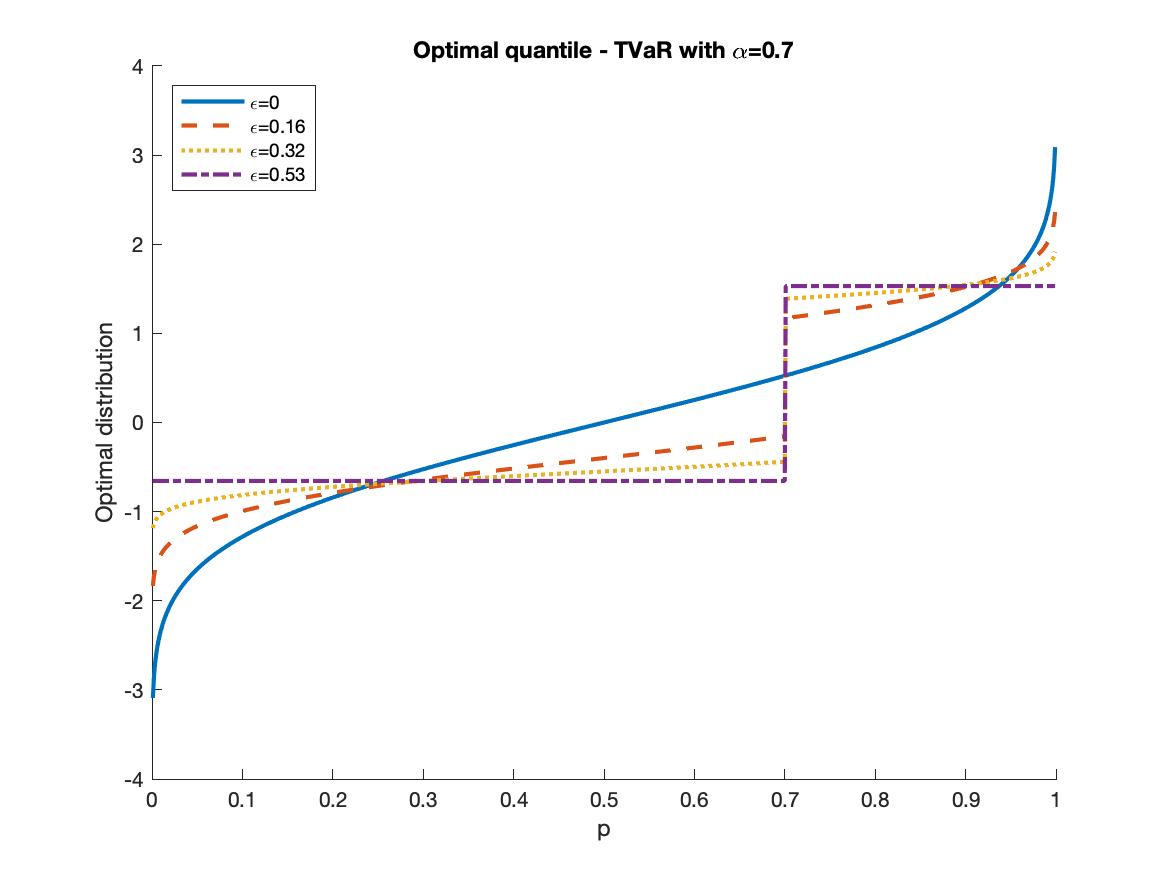} 
\vspace{-1em}
\end{tabular}
\caption{Worst-case values and quantile functions of three concave distortion risk measures; the dual power distortion (left), the Wang transform (middle), and the TVaR (right). The top panels display the worst-case values as a function of $\varepsilon$; the bottom panels display the worst-case quantile functions for selected values of $\varepsilon$.\label{fig: concave0}}
\end{figure}
In the top panels of Figure \ref{fig: concave0}, we observe that the upper bounds are indeed non-decreasing and continuous functions of $\varepsilon$. The turquoise coloured vertical lines display $\varepsilon^* = (\mu_F - \mu)^2 + (\sigma_F - \sigma)^2 + 2\sigma \sigma_F(1 - c_0) = 2\,(1 - c_0) $. The parameter value $\varepsilon^*$ indicates the transition from case 1. to case 2. in Theorem~\ref{thm: concave}; indeed, for $\varepsilon\ge \varepsilon^*$ case 2. of Theorem~\ref{thm: concave} applies and the worst-case value is independent of $\varepsilon$. For $\text{TVaR}_\alpha$ with $\alpha=0.7$ (top right panel), the worst-case value is equal to $\mu + \sigma\sqrt{\,\frac{\alpha}{1 - \alpha}\,}=\sqrt{\,\frac{\alpha}{1 - \alpha}\,} = 1.53$, and we recover the well-known Cantelli upper bound.

The bottom panels of Figure \ref{fig: concave0} display the worst-case quantile functions for selected values of $\varepsilon$. We observe that for $\varepsilon=0$, the worst-case quantile functions are equal to the reference quantile functions (blue solid line). When the Wasserstein distance $\varepsilon$ increases,
the influence of the reference distribution diminishes and, if $\varepsilon > 2\,(1 - c_0)$, the worst-case quantile function (violet dash-dotted line) is independent of the reference distribution. This can  be seen clearly for TVaR, where the violet dash-dotted quantile function corresponds to a two-point distribution. 
\end{example}

\subsection{Upper Bounds on General Distortion Risk Measures}\label{sec: distortion}

To characterise the worst-case values and quantile functions of general distortion risk measures we introduce the concept of isotonic projections; see e.g., \cite{Nemeth2003isothonic}. For this, we denote the space of square-integrable, non-decreasing, and left-continuous functions on $(0,1)$ by
\begin{equation}\label{defK}
    \K = \left\{\, k \colon (0,1) \to \mathbb{R} ~\left|~ \int_0^1 k(u)^2 \mathrm{d}u < + \infty, \; k\;\text{ non-decreasing \& left-continuous} \, \right\}\right.. 
\end{equation}  

\begin{definition}\label{def: isotonic}
For $\lambda \geq 0$, denote by $k_\lambda^\uparrow$ the \emph{isotonic projection} of $\gamma + \lambda F^{-1}$ onto the space of square-integrable non-decreasing and left-continuous functions on $(0,1)$ -- that is, the unique solution to 
\begin{equation}\label{eq: isotonic non-decreasing}
k^\uparrow_\lambda = \argmin_{k \in \K} \;|| \gamma + \lambda F^{-1} - k||^2\,,
\end{equation} 
where $|| \cdot ||$ denotes the $L^2$ norm. Whenever $\lambda = 0$, we write $\gamma^\uparrow$ instead of $k^{\uparrow}_0$, as $k^{\uparrow}_0$ is indeed the isotonic projection of $\gamma$ alone. 
\end{definition}
The isotonic projection differs from the isotonic regression \citep{barlow1972statistical}, which is a metric projection onto the set of finite dimensional and component-wise increasing vectors. We refer to Appendix \ref{sec: isotone projection} for further details and properties of isotonic projections.

We further define 
\begin{equation*}
\lambda^\uparrow = \inf\left\{ \,\lambda \geq 0 ~\left|~k_\lambda^\uparrow \text{ is not constant} \,\right\}\right.
\end{equation*} 
and set $\inf{\emptyset} = +\infty$. Finally, we define for $\lambda > \lambda^\uparrow$ the quantile function 
\begin{equation}\label{eq: hup}
h^{\uparrow}_\lambda(u) =
\mu + \sigma  \left(\frac{k_\lambda^\uparrow(u)-a_\lambda}{b_\lambda}\right) ,\quad  0  < u < 1\,,
\end{equation} 
with $a_\lambda = \E\left(k_\lambda^\uparrow(U)\right)$ and $b_\lambda =\std\left(k_\lambda^\uparrow(U)\right)$. Note that $h^\uparrow_\lambda$ is well-defined whenever $\lambda > \lambda^\uparrow$, as the function $k^\uparrow_\lambda$ is non-constant for all $\lambda > \lambda^\uparrow$; see also Proposition~\ref{prop:iso-property-appendix} in Appendix \ref{sec: isotone projection}.

Before stating the worst-case value of general distortion risk measures, we need a further assumption.
\begin{assumption} \label{asm: lambda finite}
We assume that $\lambda^\uparrow =0$. 
\end{assumption}
Note that Assumption \ref{asm: lambda finite} is a requirement on the weight function and the reference distribution. Assumption \ref{asm: lambda finite} is  satisfied, for instance, when $\gamma$ is bounded and the reference distribution has a quantile function that is unbounded to the right, i.e. $F^{-1}(u)\rightarrow+\infty$ for $u\rightarrow 1$.

\begin{theorem}[Worst-Case Values and Quantiles of General Distortions]\label{thm: general}
Let Assumptions \ref{asm: Choquet integral}, \ref{asm: gamma not 1}, \ref{asm: epsilon big enough}, and \ref{asm: lambda finite} be fulfilled.
Further, let $g$ be a distortion function and denote $c_0 = \lim_{\lambda \searrow 0}\;\corr\left(F^{-1}(U), \,k^{\uparrow}_{\lambda}(U)\right)$. Then, the following statements hold: 

\begin{enumerate}
	\item If $(\mu_F - \mu)^2 + (\sigma_F - \sigma)^2 < \varepsilon < (\mu_F - \mu)^2 + (\sigma_F - \sigma)^2 + 2\sigma \sigma_F(1 - c_{0})$, then a solution to \eqref{eq:problem argsup} is unique and is given by the quantile function $h^{\uparrow}_\lambda$ defined in \eqref{eq: hup}, where $\lambda > 0$ is the unique positive solution to $d_W(F^{-1}, \, h_\lambda^\uparrow)  = \sqrt{\,\varepsilon\,}$. 
	Furthermore, the worst-case value, i.e., the solution to \eqref{eq:problem sup}, is
	\begin{equation*}
	H_g(h^{\uparrow}_\lambda) 
	= \mu  + \sigma \std(\gamma(U)) \corr\left(\gamma(U), \,k^\uparrow_{\lambda}(U)\right)
	\end{equation*} 
	and $H_g(h^{\uparrow}_\lambda)$ is continuous in $\varepsilon$.

	\item Let $(\mu_F - \mu)^2 + (\sigma_F - \sigma)^2 + 2\sigma \sigma_F (1 - c_{0}) \leq \varepsilon $. If  $\gamma^\uparrow$ is not constant, then case $i)$ applies with $\lambda = 0$. If $\gamma^\uparrow$ is constant, then the solution to \eqref{eq:problem sup} is equal to $\mu$ but cannot be attained in general. 	 
\end{enumerate}
\end{theorem}
\begin{remark}
If $\gamma$ is non-increasing, equivalently the distortion function $g$ is convex, then its isotonic projection $\gamma^\uparrow$ is constant and thus equal to 1. Hence, for $\varepsilon$ sufficiently large, case 2. of Theorem~\ref{thm: general} implies that the upper bound is equal to $\mu$. However, the well-known Hoeffding bounds for the expectation of a product of two random variables with given marginal distributions imply that for any admissible quantile function $G^{-1}$, it holds that $\int_{0}^1{\gamma(u)G^{-1}(u)du} <  \int_{0}^1{G^{-1}(u)du}=\mu$ (note that $\int_{0}^1{\gamma(u)du}=1$ and $\gamma$ is assumed to be non-constant). Hence, the upper bound $\mu$ is not attained. However, if for example there exists $0<p<\frac{1}{2}$ such that $\int_0^p{\gamma(u)du}=\int_{1-p}^1{\gamma(u)du}$ and $\gamma^\uparrow=1$, then the upper bound $\mu$ is attained. To see this, consider the quantile function $G^{-1}(u)=\mu+\sigma F^{-1}(u)$ in which $F^{-1}(u)$ takes value $\frac{-1}{\sqrt{2p}}$ on $(0,p)$, value $\frac{1}{\sqrt{2p}}$ on $(1-p,p)$, and value zero otherwise. It verifies that the mean of $G^{-1}$ is  $\mu$ whereas its variance is equal to $\sigma^2$. Moreover, $\int_{0}^1{\gamma(u)G^{-1}(u)du}=\mu$, thus $G^{-1}$ attains the worst-case bound. An example of $\gamma$ satisfying these conditions is if $\gamma(u)=1-2u$ on $(0,1/2)$ and $\gamma(u)=2u-1$ on $[1/2,1)$.
\end{remark}

If $\gamma$ is non-decreasing, equivalently  $g$ is concave, then $\gamma + \lambda F^{-1}$ is non-decreasing and thus equal to its isotonic projection. Hence, Theorem~\ref{thm: general} reduces to Theorem~\ref{thm: concave} in the case of concave distortion risk measures. Theorem~\ref{thm: general} generalises results by \cite{Zhu2018SSRN}, who analyse problem \eqref{eq:problem sup} in the special case in which the Wasserstein constraint is redundant, which we state in the subsequent corollary.
\begin{corollary}[Worst-Case Values for $\varepsilon = + \infty$]\label{cor: momentbound-general}
Let Assumptions \ref{asm: Choquet integral}, \ref{asm: gamma not 1}, and \ref{asm: lambda finite} be fulfilled. Further, let $g$ be a distortion function; then, the following statements hold:
\begin{enumerate}
	\item If $\gamma^{\uparrow}$ is not constant, then 	
\begin{equation}\label{eq:problem sup bpv}
\sup_{G \in \mathcal{M}(\mu, \sigma)} H_g(G)=\mu  + \sigma \std\left(\gamma^\uparrow(U)\right) = \mu  + \sigma \sqrt{\,\int_{0}^{1}{(\gamma^{\uparrow}(p)-1)^2\,dp\,}}\,.
\end{equation}
The bound is sharp and attained by a distribution with quantile function $h^{\uparrow}_{0}$ defined in \eqref{eq: hup}. 
	\item If $\gamma^{\uparrow}$ is constant, then 
	\begin{equation*}
	\sup_{G \in \mathcal{M}(\mu, \sigma)} H_g(G) = \mu  
	\end{equation*}
	and the bound cannot be attained in general.
\end{enumerate}		
\end{corollary}

\subsection{Lower Bounds on General Distortion Risk Measures}\label{sec: best}
In a similar way as in Section~\ref{sec: distortion}, we first introduce the notation and assumptions needed to derive the best-case values and quantile functions of general distortion risk measures. 

\begin{definition}\label{def: isotonic2}
For $\lambda \geq 0$, denote by $k_\lambda^{\downarrow}$ the isotonic projection of $\gamma - \lambda F^{-1}$ onto the set of square-integrable non-increasing functions and left-continuous -- that is, the unique solution to 
\begin{equation}\label{eq: isotonic non-increasing}
k^\downarrow_\lambda = \argmin_{-k \in \K} || \gamma - \lambda F^{-1} - k||^2\,.
\end{equation}
Whenever $\lambda = 0$, we write $\gamma^\downarrow$ instead of $k^{\downarrow}_0$, as $k^{\downarrow}_0$ is indeed the isotonic projection of $\gamma$ onto the non-increasing functions. 
\end{definition}
Further, we define
\begin{equation*}
\lambda^\downarrow = \inf\left\{ \,\lambda \geq 0 ~\left|~k_\lambda^\downarrow \text{ is not constant} \,\right\}\right.
\end{equation*} 
and for all $\lambda> \lambda^{\downarrow}$ the quantile function 
\begin{equation}\label{eq: hdown}
h^{\downarrow}_\lambda(u) =
\mu + \sigma  \left(\frac{a_\lambda- k_\lambda^{\downarrow}(u)}{b_\lambda}\right) ,\quad 0  < u < 1,
\end{equation}
with $a_\lambda = \E(k_\lambda^{\downarrow}(U))$ and $b_\lambda =\std(k_\lambda^{\downarrow}(U))$. Note that $h^\downarrow_\lambda$ is well-defined whenever $\lambda > \lambda^\downarrow$, as the function $k^\downarrow_\lambda$ is non-constant for all $\lambda > \lambda^\downarrow$; see also Proposition~\ref{prop:iso-property-appendix} in Appendix \ref{sec: isotone projection}.

\begin{assumption} \label{asm2: lambda finite}
We assume that $\lambda^\downarrow=0$.
\end{assumption}
The next theorem states the best-case values and quantile functions of general distortion risk measures.

\begin{theorem}[Best-Case Values and Quantiles of General Distortions]\label{thm: lower bound}
Let Assumptions \ref{asm: Choquet integral}, \ref{asm: gamma not 1}, \ref{asm: epsilon big enough}, and \ref{asm2: lambda finite} be fulfilled.
Further, let $g$ be a distortion function and denote $c_{0} = \lim_{\lambda \searrow 0}\;\corr(F^{-1}(U),\,- k_\lambda^{\downarrow}(U))$. Then, the following statements hold: 
\begin{enumerate}
	\item If $(\mu_F - \mu)^2 + (\sigma_F - \sigma)^2 < \varepsilon  < (\mu_F - \mu)^2 + (\sigma_F - \sigma)^2 + 2\sigma \sigma_F(1 - c_{0})$, then a solution to \eqref{eq:problem arginf} is unique and is given by the quantile function $h^{\downarrow}_\lambda$ defined in \eqref{eq: hdown}, where $ \lambda > 0$ is the unique positive solution to $d_W(F^{-1},\, h_\lambda^\downarrow) = \sqrt{\,\varepsilon\,}$. The corresponding best-case value, i.e., the solution to \eqref{eq:problem inf}, is 
	\begin{equation*}
	H_g(h^{\downarrow}_\lambda) 
	= \mu - \sigma \std(\gamma(U)) \corr\left(\gamma(U),\, k^\downarrow_{\lambda}(U)\right),
	\end{equation*}
	and $H_g$ is continuous in $\varepsilon$ 
	\item Let $(\mu_F - \mu)^2 + (\sigma_F - \sigma)^2 + 2\sigma \sigma_F (1 - c_{0}) \leq \varepsilon $. If $\gamma^\downarrow$ is not constant, then case $i)$ applies with $\lambda = 0$. If $\gamma^\downarrow$ is constant, then the infimum is equal to $\mu$ but cannot be attained. 	 
\end{enumerate}
\end{theorem}
Theorem~\ref{thm: lower bound} implies that for $\varepsilon $ that is sufficiently large, i.e.,  in case 2., the lower bound of any concave distortion risk measure is equal to $\mu$ and not attained. The next corollary collects the results for $\varepsilon = + \infty$.
\begin{corollary}[Best-Case Values for $\varepsilon = + \infty$]
\label{cor: momentbound-generallower}
Let Assumptions \ref{asm: Choquet integral}, \ref{asm: gamma not 1}, and \ref{asm2: lambda finite} be fulfilled.
Further, let $g$ be a distortion function; then, the following statements hold:
\begin{enumerate}
	\item If $\gamma^{\downarrow}$ is not constant, then 	
\begin{equation*}
\inf_{G \in \mathcal{M}(\mu, \sigma)} H_g(G) = \mu  - \sigma \std(\gamma(U)) \corr\left(\gamma(U), \,\gamma^\downarrow(U)\right)\,.
\end{equation*}
The bound is sharp and is attained by a distribution with quantile function $h^{\downarrow}_{0}$ defined in \eqref{eq: hdown}. 
\item If $\gamma^{\downarrow}$ is constant, then 
	\begin{equation*}
	\sup_{G \in \mathcal{M}(\mu, \sigma)} H_g(G) = \mu  \end{equation*}
	and the bound cannot be attained.
\end{enumerate}		
\end{corollary}

\begin{remark}
\label{remark1}
Distortion risk measures have an alternative representation that makes it possible to write minimisation problems in terms of maximisation problems. Specifically,
\begin{equation}\label{eq: distortion dual}
\inf_{G_X \in \mathcal{M}(\mu, \sigma)} H_g(G_X)
= -\sup_{G_{-X} \in \mathcal{M} (-\mu, \sigma)} H_{\bar{g}}(G_{-X}) 
\end{equation}
where $\bar{g}$ is the dual distortion of $g$, given by $\bar{g}(x) = 1 - g(1 - x)$, and $G_{X}$, $G_{-X}$ denote the distribution functions of the random variables $X$ and $-X$, respectively. Equation \eqref{eq: distortion dual} follows from the fact that $H_g(G_X) = - H_{\bar{g}}(G_{-X}) $; see e.g., \cite{Dhaene2012EAJ}. This observation makes it possible to obtain Corollary~\ref{cor: momentbound-generallower}, the statements on lower bounds, from Corollary~\ref{cor: momentbound-general}, the corresponding statements on upper bounds, in a more direct manner. Such reasoning, however, cannot be extended when dealing with the Wasserstein distance constraints considered in this paper. 
\end{remark} 

\begin{remark}
\label{sec: Choquet}
Theorems \ref{thm: general} and \ref{thm: lower bound} can be generalised to \emph{signed Choquet integrals}. A signed Choquet integral is, for a distribution function $G \in \mathcal{M}^2$, defined by $H_g(G) = - \int_{-\infty}^0 g(1) - g(1 -  G(x)) \,\mathrm{d}x + \int_0^{+ \infty}g(1 - G(x))\,\mathrm{d}x$, where $g \colon [0,1] \to \mathbb{R}$ is of bounded variation with $g(0) = 0$. Signed Choquet integrals generalise the class of distortion risk measures in that the distortion function $g$ can be decreasing. For absolutely continuous $g$, the signed Choquet integral admits a representation as in \eqref{eq: distortion Choquet integral}, see \cite{Wang2019SSRN}, thus, Theorems \ref{thm: general} and \ref{thm: lower bound} can be extended to signed Choquet integrals with absolutely continuous distortion function. 
\end{remark}

\begin{example}[Continued]
Figure \ref{fig: lower-Concave} illustrates the best- and worst-case values (Theorems \ref{thm: general} and \ref{thm: lower bound}) of three concave distortion risk measures: the dual power distortion, the Wang transform, and TVaR, with a standard normal reference distribution, $\mu_F = \mu = 0$, and $\sigma_F = \sigma = 1$. In the upper panels of Figure \ref{fig: lower-Concave}, the solid blue line corresponds to the upper bound and the dashed blue line corresponds to the lower bound. The black line depicts the reference risk measure $H_g(F)$ and the vertical turquoise lines display the critical $\varepsilon$ value for the transition between case 1. and case 2. in Theorem~\ref{thm: general} (upper bound, solid line) and Theorem~\ref{thm: lower bound} (lower bound, dashed line).

As all three risk measures are concave, their lower bounds -- for $\varepsilon$ sufficiently large (case 2. in Theorem~\ref{thm: lower bound}) -- are all equal to $\mu = 0$ but not attained. That the lower bounds are not attained can be seen in the lower panels, where the corresponding best-case quantile functions are displayed. The best-case quantile functions become flatter for larger $\varepsilon$, and for $\varepsilon$ sufficiently large the quantile functions are no longer defined.
\begin{figure}[!htbp]
\centering
\begin{tabular}{ccc}
Dual Power Distortion & Wang Distortion & TVaR distortion\\
\includegraphics[width=4.5cm, height=5cm]{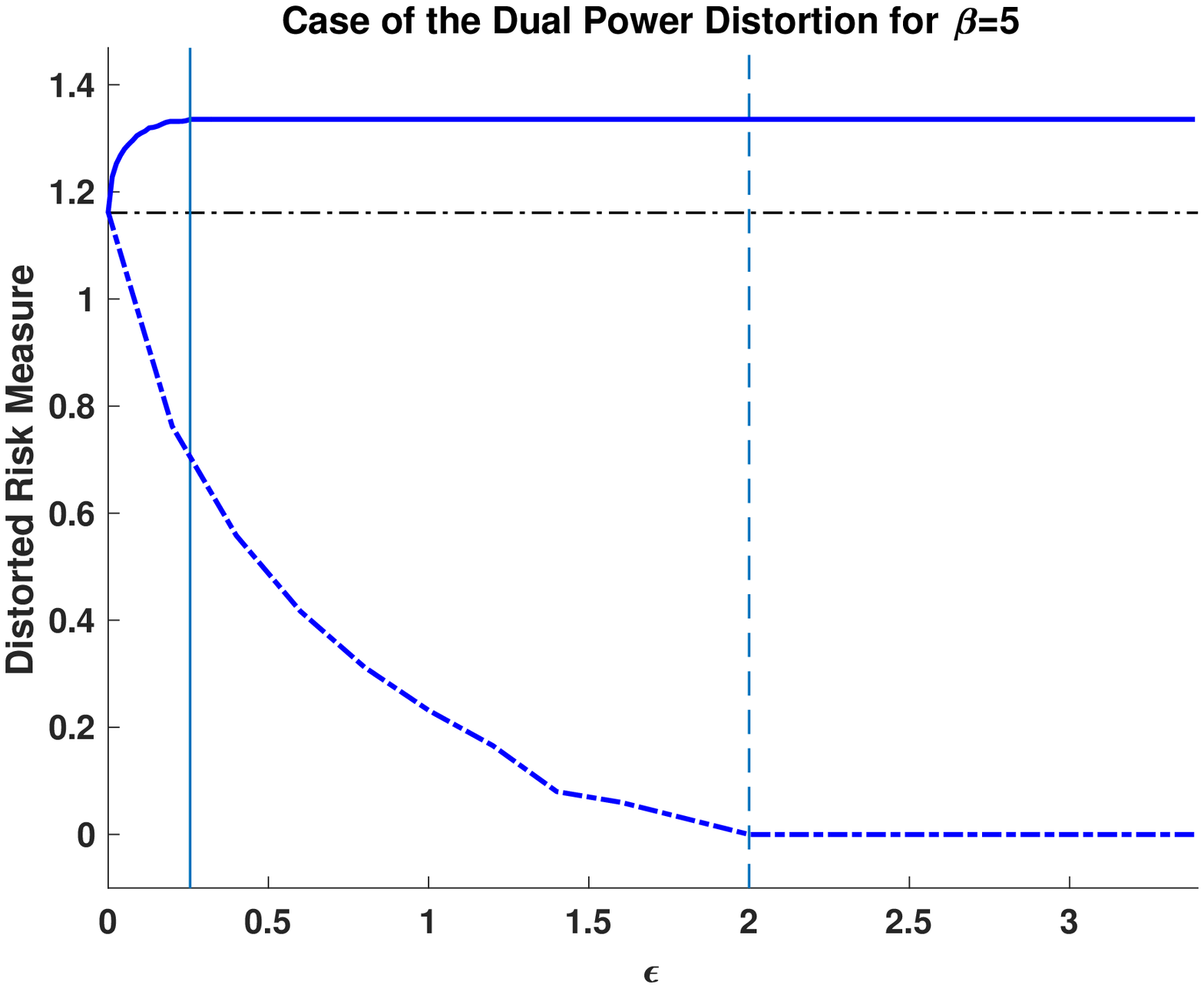} &
\includegraphics[width=4.5cm, height=5cm]{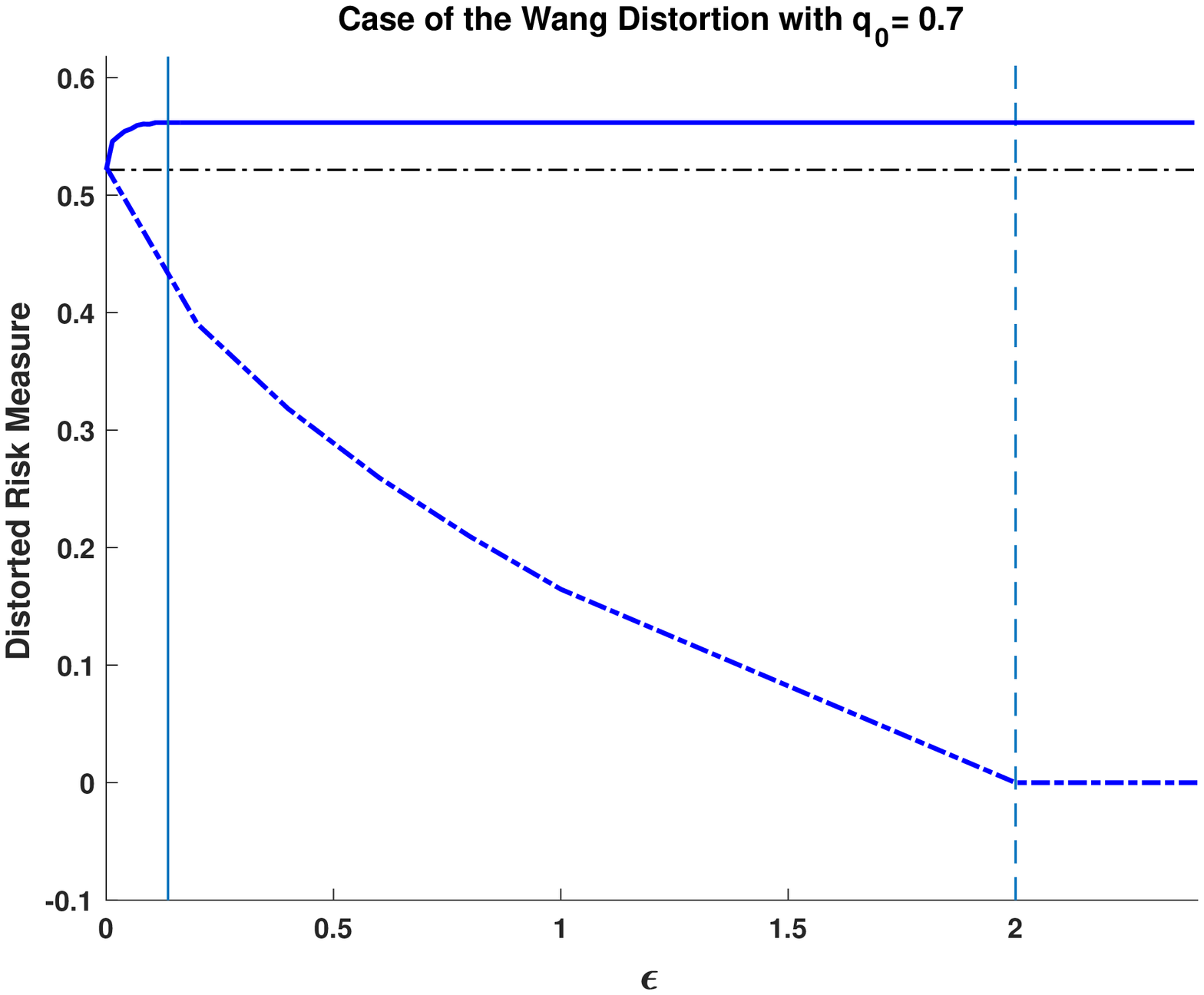}
&
\includegraphics[width=4.5cm, height=5cm]{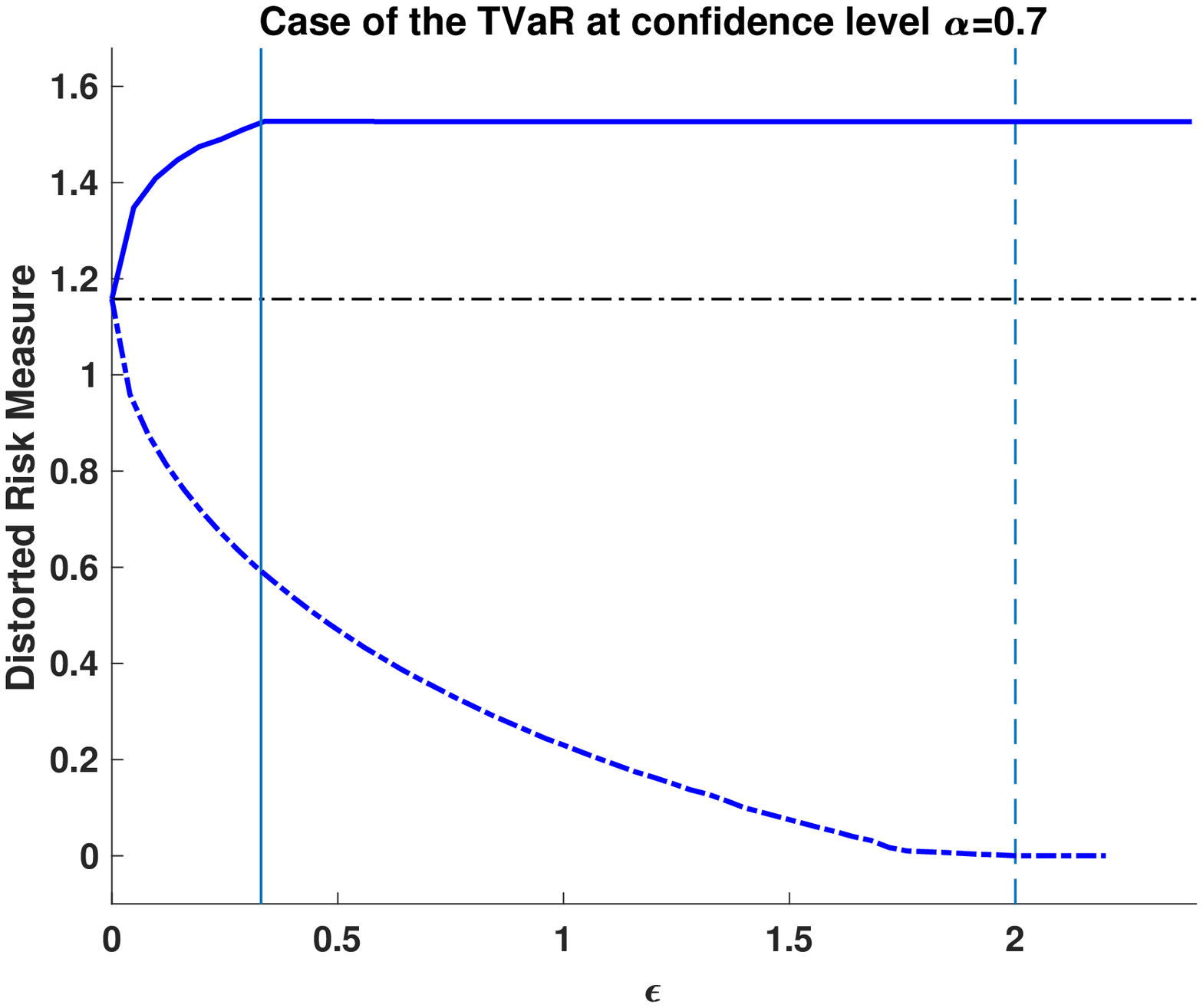}\\
 \includegraphics[width=4.5cm, height=5cm]{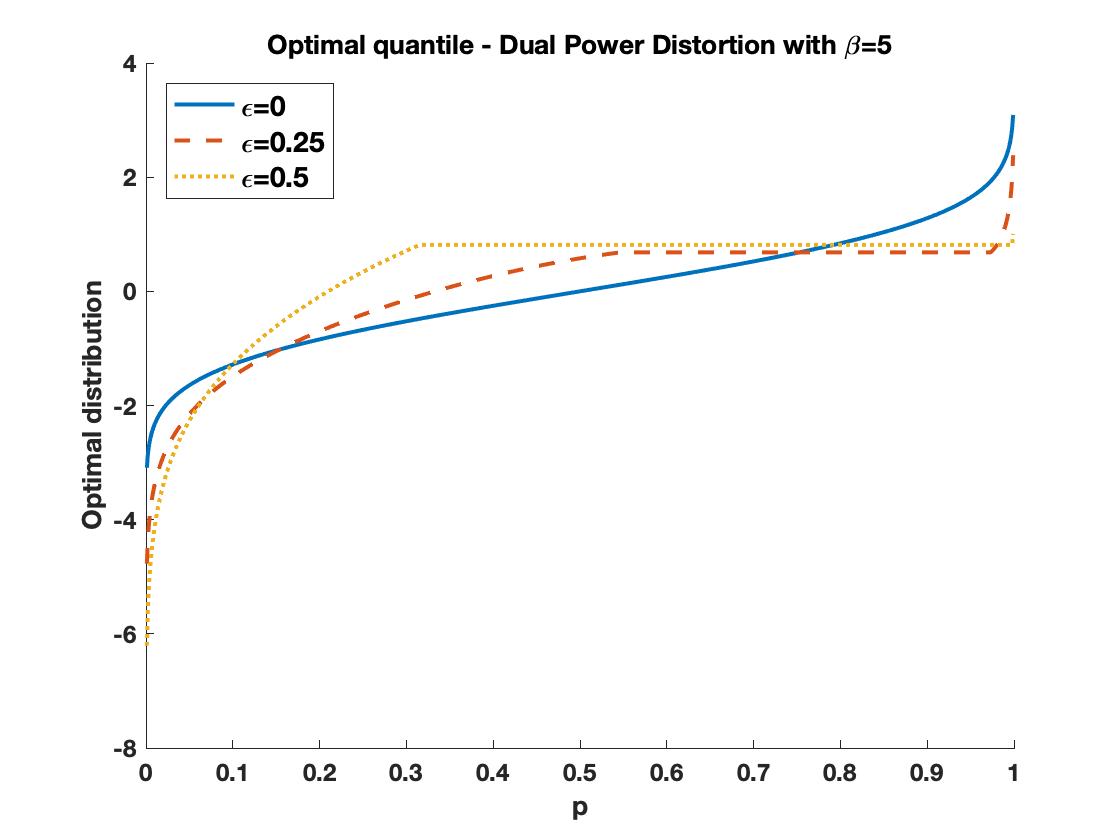} &
 \includegraphics[width=4.5cm, height=5cm]{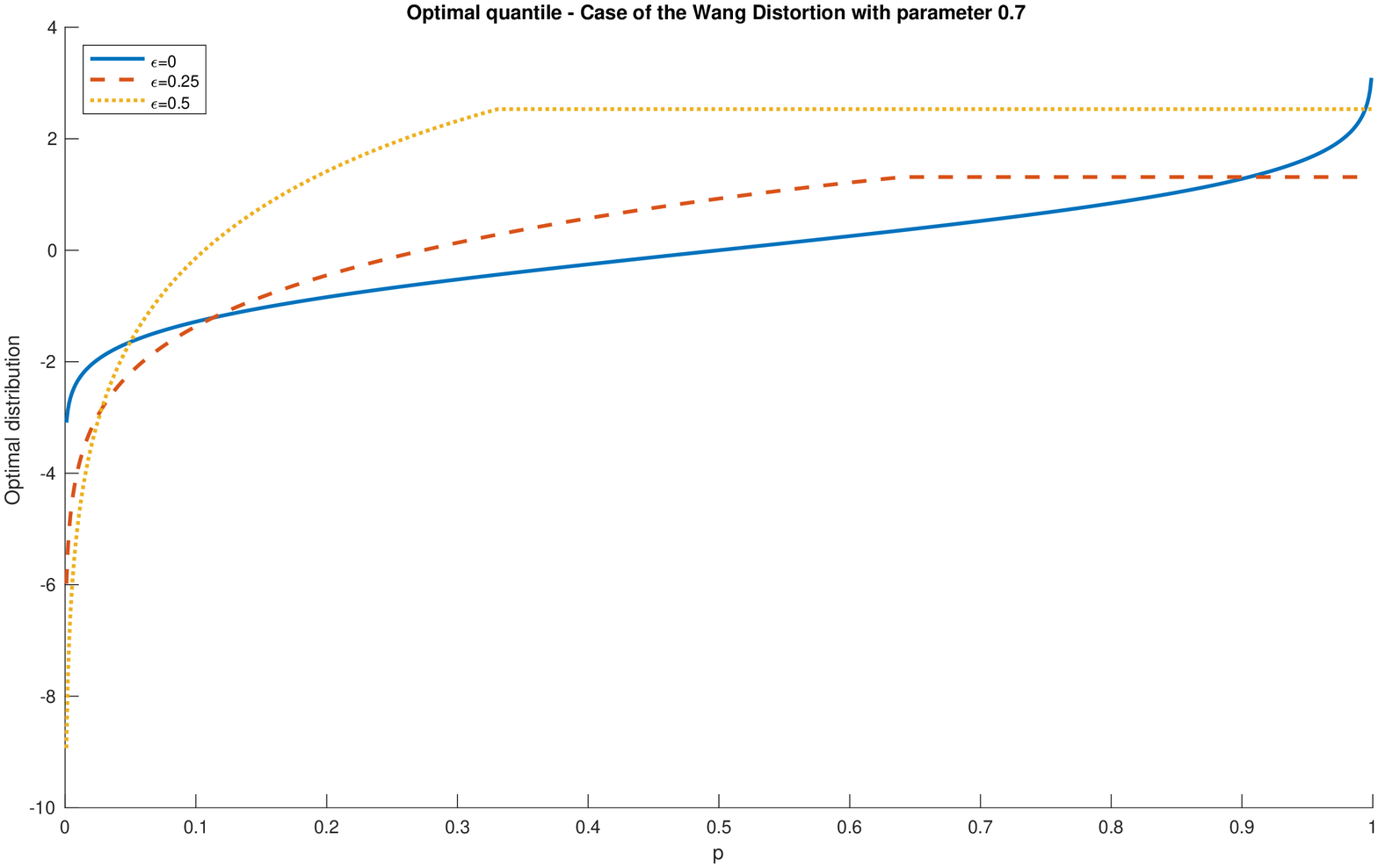} 
&\includegraphics[width=4.5cm, height=5cm]{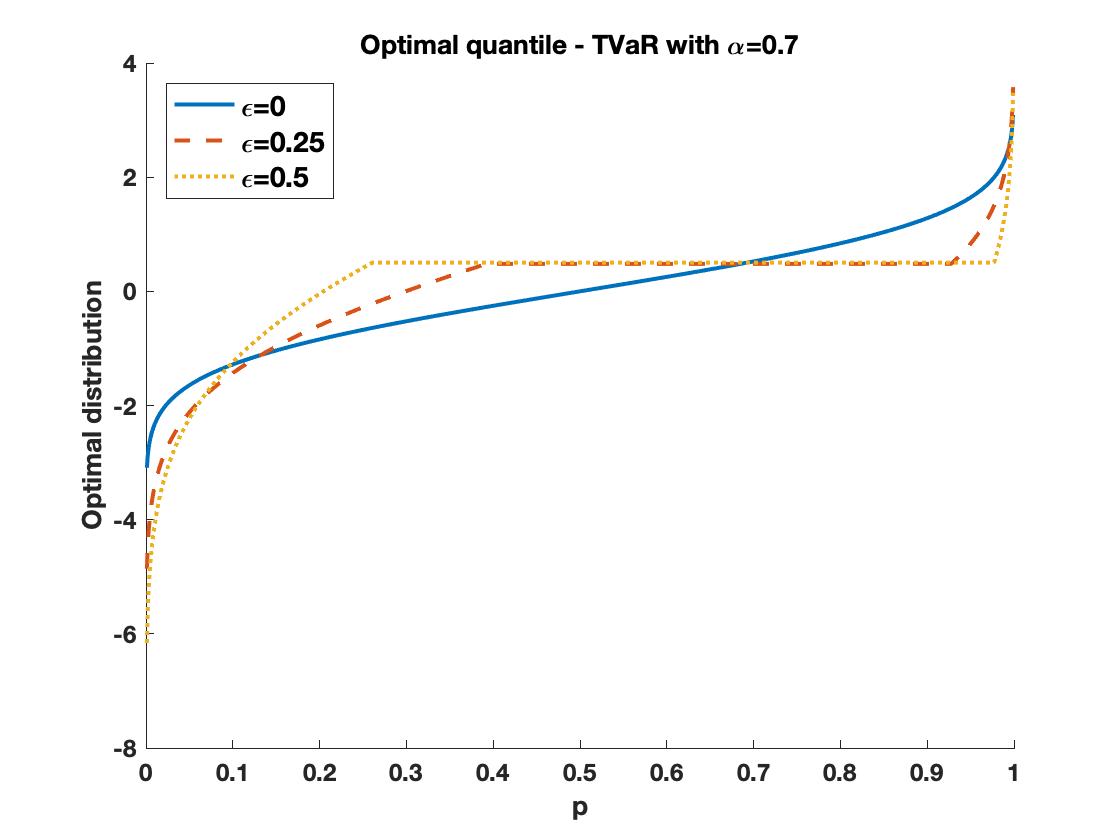} 
\end{tabular}
\centering
\caption{Best- and worst-case values and quantile functions of three concave risk measures; the dual power distortion (left), the Wang transform (middle), and TVaR (right). The top panels display the best- and worst-case values as a function of $\varepsilon$; the bottom panels display the best-case quantile functions for selected values of $\varepsilon$. The vertical turquoise lines in the upper panels correspond to the transition of case 1. to case 2, in the upper bound (Theorem~\ref{thm: general}, solid line) and lower bound (Theorem~\ref{thm: lower bound}, dashed line).
\label{fig: lower-Concave}}
\end{figure}
\end{example}

\section{Extensions}\label{sec: extension}
In this section we consider two extensions. First, we derive explicit bounds on RVaR and (via a direct proof) the bounds on VaR. Second, we extend the results to the  uncertainty sets solely described by the Wasserstein distance. 

\subsection{Bounds on Range Value-at-Risk, Value-at-Risk, and Tail Value-at-Risk}\label{sec: VaR}
In this section we provide bounds on the risk measures RVaR, TVaR, and VaR. Specifically, we first calculate the best- and worst-case values of RVaR and then derive the TVaR bounds as limiting cases. The bounds on VaR are derived via a direct proof as VaR does not satisfy Assumption \ref{asm: Choquet integral}. The $\text{RVaR}_{\alpha, \beta}$ at levels $0 < \alpha < \beta \le  1$ is defined by \citep{cont2010robustness}
\begin{equation*} 
g(x)=\min\left\{\,\max\left\{\,\tfrac{x + \beta - 1}{\beta-\alpha}, ~0\,\right\}, ~1\,\right\}
\quad \text{and} \quad\gamma(u) = \frac{1}{\beta - \alpha} \Id_{(\alpha , \beta]}(u)\,.
\end{equation*}
By letting $\beta \nearrow 1$, we obtain TVaR; that is, for any $G \in \mathcal{M}^2$ it holds that
\begin{align*}
    \text{TVaR}_\alpha (G) &= \lim_{\beta \nearrow 1} \text{RVaR}_{\alpha, \beta}(G)\, .
\end{align*}
Next, recall that $\text{VaR}_\alpha(G) = G^{-1}(\alpha)$ and denote $\text{VaR}_\alpha^+(G) = G^{-1, +}(\alpha)$. Both VaR and VaR\textsuperscript{+} are distortion risk measures ($\text{VaR}_\alpha$ with $g(x) =  \Id_{(1-\alpha , 1]}(x)$ and $\text{VaR}_\alpha^+$ with $g(x) =  \Id_{[1-\alpha , 1]}(x)$, see \cite{Dhaene2012EAJ}); however, they do not belong to the class of distortion risk measures considered in this paper as they do not satisfy Assumption \ref{asm: Choquet integral}. Furthermore, it holds that for any $G \in \mathcal{M}^2$
\begin{equation*}
    \text{VaR}_\alpha(G) = \lim_{\alpha^\prime \nearrow \alpha}\text{RVaR}_{\alpha^\prime, \alpha}(G)\, \quad \text{and} \quad
    \text{VaR}^+_\alpha(G) = \lim_{\beta\searrow \alpha}\text{RVaR}_{\alpha, \beta}(G)\, .
\end{equation*}
The worst-case values of these risk measures under constraints on the first two moments but without a Wasserstein constraint have  been extensively studied; indicatively see 
\cite{kaas1986best}, \cite{lo1987semi},
\cite{grundy1991option}, \cite{hurlimann2005improved}, and \cite{de2010estimate}
for the VaR, \cite{li2018SIAM} for the RVaR, and \cite{Natarajan2010MF} for the TVaR. We collect the worst- and best-case values of these risk measures without a Wasserstein constraint in the subsequent corollaries. The bounds on TVaR and RVaR also follow from Corollary~\ref{cor: momentbound-general} and \ref{cor: momentbound-generallower}.
\begin{corollary}[Worst-Case Values for $\varepsilon = + \infty$]\label{cor: TVaR worst-case}
For $0 < \alpha <  \beta \leq1$, the worst-case values of $\text{VaR}_\alpha,~\text{VaR}^+_\alpha, ~\text{RVaR}_{\alpha, \beta}$, and $\text{TVaR}_\alpha$ coincide; that is, 
\begin{align*}
\sup_{G \in \mathcal{M}(\mu, \sigma)} \text{VaR}_\alpha (G)
	&= \sup_{G \in \mathcal{M}(\mu, \sigma)} \text{VaR}^+_\alpha (G)
	= \sup_{G \in \mathcal{M}(\mu, \sigma)} \text{RVaR}_{\alpha, \beta}(G) \nonumber\\
	&= \sup_{G \in \mathcal{M}(\mu, \sigma)} \text{TVaR}_\alpha(G)
	= \mu  + \sigma\sqrt{\,\frac{\alpha}{1 - \alpha}\,}\,.
\end{align*}
While the worst-case value of $\text{VaR}_\alpha$ cannot be attained, the worst-case distribution of  $\text{VaR}^+_\alpha, ~\text{RVaR}_{\alpha, \beta}$, and $\text{TVaR}_\alpha$  is a two-point distribution with quantile function 
\begin{equation}\label{eq: TVaR worst-case dist}
h_0^\uparrow(u) = 
\begin{cases}
\mu - \sigma\sqrt{\,\tfrac{1 - \alpha}{\alpha}\,} & 0 < u \leq \alpha\,,\\
\mu + \sigma\sqrt{\,\tfrac{\alpha}{1 - \alpha}\,} & \alpha < u < 1\,.
\end{cases}
\end{equation}
\end{corollary}
\begin{corollary}[Best-Case Values for $\varepsilon = + \infty$]\
\label{cor: best-case TVaR}
For $0 < \alpha <  \beta \leq1$, the best-case values of $VaR_\alpha$, $\text{VaR}^+_\alpha$, $RVaR_{\alpha,\beta}$, and $TVaR_{\alpha,\beta}$ are
\begin{align*}
	\inf_{G \in \mathcal{M}(\mu, \sigma)} \text{VaR}_\alpha(G) 
	&= 
	\inf_{G \in \mathcal{M}(\mu, \sigma)} \text{VaR}^+_\alpha(G) 
	= \mu  - \sigma\sqrt{\,\tfrac{1 - \alpha}{\alpha}\,}\,, \\[0.5em]
	\inf_{G \in \mathcal{M}(\mu, \sigma)} \text{RVaR}_{\alpha, \beta}(G) 
	&= \mu  - \sigma\sqrt{\,\tfrac{1 - \beta}{\beta}\,}\,,
	\quad \text{and} \quad
	\inf_{G \in \mathcal{M}(\mu, \sigma)} \text{TVaR}_\alpha(G)
	= \mu\,.
\end{align*}
While the best-case values of $\text{VaR}^+_\alpha$ and $TVaR_{\alpha,\beta}$ cannot be attained, the best-case distribution corresponding to the best-case $\text{VaR}_\alpha$ has a quantile function given by \eqref{eq: TVaR worst-case dist} whereas for the best-case $RVaR_{\alpha, \beta}$ it is a two-point distribution with quantile function 
\begin{equation*}
h_0^\downarrow(u) = 
\begin{cases}
\mu - \sigma\sqrt{\,\tfrac{1 - \beta}{\beta}\,} & 0 < u \leq \beta\,,\\
\mu + \sigma\sqrt{\,\tfrac{\beta}{1 - \beta}\,} & \beta < u < 1\,.
\end{cases}
\end{equation*}
\end{corollary}
Next, we study the lower and upper bound on RVaR when, in addition to the moment constraints, the distributions in the uncertainty set lie within an $\sqrt{\varepsilon}$-Wasserstein ball of the reference distribution $F$.

\begin{proposition}[Worst-Case Quantiles of RVaR]\label{cor: isotonic projection RVaR}
Under the assumptions of Theorem~\ref{thm: general} case 1., a solution to \eqref{eq:problem argsup} with $\text{RVaR}_{\alpha, \beta}$ is unique and is given by $h^\uparrow_\lambda$ defined in \eqref{eq: hup}, where $k^\uparrow_\lambda$ is
\begin{equation}\label{WCformula}
k^\uparrow_\lambda(u)  = 
\begin{cases}
\lambda F^{-1}(u) 	\quad \quad &0 < u \leq \alpha\,,\\
\frac{1}{\beta - \alpha} + \lambda F^{-1}(u) 	& \alpha < u \leq w_0\,,\\
c	& w_0 < u \leq w_1\,,\\
\lambda F^{-1}(u) 		& w_1 < u < 1\,,\\
\end{cases}
\end{equation}
and $w_0$, $w_1$, and $c$, with $\alpha  \leq w_0 \leq \beta \leq w_1$, $c < + \infty$, satisfy
\begin{align*}
\lambda F^{-1}(w_0) &= 
\begin{cases}
c  - \frac{1}{\beta - \alpha}, \quad \quad & \text{if} ~ \frac{1}{\beta - \alpha} \leq c - \lambda  F^{-1}(\alpha)\,,\\
\lambda F^{-1}(\alpha), 			& \text{otherwise}\,,\\
\end{cases}\\[0.5em]
\lambda F^{-1}(w_1) &= 
\begin{cases}
\lambda F^{-1}(1) \quad \quad & \text{if} ~ c \ge \lambda F^{-1}(1)\,,\\
c & \text{otherwise}\,,
\end{cases}
\end{align*}
\begin{align*}
c &= \frac{1}{w_1 - w_0}\frac{\beta - w_0}{\beta - \alpha} + \frac{\lambda}{w_1 - w_0} \int_{w_0}^{w_1} F^{-1}(u) \,\mathrm{d}u\,.
\end{align*}
\end{proposition}

\begin{proposition}[Best-Case Quantiles of RVaR] \label{cor: isotonic projection RVaR lower}
Under the assumptions of Theorem~\ref{thm: lower bound} case 1., a solution to \eqref{eq:problem arginf} with $\text{RVaR}_{\alpha, \beta}$ is unique and is given by $h_\lambda^\downarrow$ defined in Theorem~\ref{thm: lower bound}, where $k_\lambda^\downarrow$ is
\begin{equation}\label{BCformula}
k_\lambda^\downarrow(u)  = 
\begin{cases}
-\lambda F^{-1}(u) 	\quad \quad &0 < u \leq z_0\,,\\
-c	& z_0 < u \leq z_1\,,\\
-\lambda F^{-1}(u) + \frac{1}{\beta - \alpha}, 	& z_1 < u \leq \beta\,,\\
-\lambda F^{-1}(u) 	& \beta < u < 1\,,\\
\end{cases}
\end{equation}
and $z_0$, $z_1$, and $c$, with $z_0 \leq \alpha  \leq z_1 \leq \beta$, $c > -\infty$, solve
\begin{align*}
\lambda F^{-1}(z_1) &= 
\begin{cases}
\lambda F^{-1}(z_0) + \frac{1}{\beta - \alpha} \quad \quad  & \text{if} ~ \frac{1}{\beta - \alpha} < \lambda F^{-1}(\beta) - c\,,
\\
\lambda F^{-1}(\beta) & \text{otherwise}\,,
\end{cases}\\[0.5em]
\lambda F^{-1}(z_0) &= 
\begin{cases}
\lambda F^{-1}(0) \quad \quad & \text{if} ~ c \le \lambda F^{-1}(0)\,,\\
c & \text{otherwise}\,,\nonumber
\end{cases}
\\[0.5em]
c &=  - \frac{z_1 - \alpha}{z_1 - z_0}\frac{1}{\beta - \alpha} + \frac{\lambda}{z_1 - z_0} \int_{z_0}^{z_1} F^{-1}(u) \,\mathrm{d}u\,.\nonumber
\end{align*}
\end{proposition}
We verified with numerical experiments that the closed-form expressions in \eqref{WCformula} and \eqref{BCformula} match  well with numerically obtained isotonic and antitonic projections.\footnote{All numerically obtained isotonic projections are constructed using LSQISOTONIC, a built-in function in Matlab.}

The subsequent corollary and proposition provide the best- and worst-case quantile functions of  $\text{TVaR}$ and $\text{VaR}$, respectively. The results are also summarised in Table \ref{table: var-Tvar-quantile}.

\begin{corollary}[Best- and Worst-Case Quantiles of TVaR] \label{cor: VaR TVaR}
Under the assumptions of case 1. in Theorems \ref{thm: general} and \ref{thm: lower bound}, respectively, it holds that the worst-case quantile function of $\text{TVaR}_\alpha$ is given by Proposition~\ref{cor: isotonic projection RVaR} with $\beta = 1$ and  $w_0 = w_1 = 1$. The corresponding worst-case value of $\text{TVaR}_\alpha$ is given in Corollary~\ref{cor:TVaR-WC-explicit}.
The best-case quantile function of $\text{TVaR}_\alpha$ is given by Proposition~\ref{cor: isotonic projection RVaR lower} with $\beta = 1$.
\end{corollary}

As VaR and $\text{VaR}^+$ do not satisfy Assumption \ref{eq: distortion Choquet integral}, we provide a direct proof of the next proposition.
\begin{proposition}[Best- and Worst-Case Quantile of VaR] \label{prop: VaR TVaR2}
Let $c_0 = \corr\left(F^{-1}(U), \Id_{U \in (\alpha,1)}\right)$ and $(\mu_F - \mu)^2 + (\sigma_F - \sigma)^2 < \varepsilon < (\mu_F - \mu)^2 + (\sigma_F - \sigma)^2 + 2\sigma \sigma_F(1 - c_{0})$. It holds that the worst-case quantile function  of $\text{VaR}_\alpha^+$ is given by Proposition~\ref{cor: isotonic projection RVaR} with $w_0 = \alpha$. The worst-case value of $\text{VaR}_\alpha$ is equal to that of $\text{VaR}_\alpha^+$ but not attained. 
The best-case quantile function of $\text{VaR}_\alpha$ is given by Proposition~\ref{cor: isotonic projection RVaR lower} with $z_1 = \beta$.
The best-case value of $\text{VaR}_\alpha^+$ is equal to that of $\text{VaR}_\alpha$ but is not attained.
\end{proposition}

\begin{table}[t]
\begin{center}
\begin{tabular}{l @{\hspace{2.5em}} l @{\hspace{2.5em}} l }
\toprule \toprule
Risk measure &  Best-case quantile & Worst-case quantile\\
\midrule
$\text{RVaR}_{\alpha, \beta}$ & Proposition~\ref{cor: isotonic projection RVaR lower}; &  Proposition~\ref{cor: isotonic projection RVaR}; \\[0.5em]
 $\text{VaR}_\alpha$ & 
Proposition~\ref{cor: isotonic projection RVaR lower} with $z_1 = \beta$;
& not attained;
 \\[0.5em]
 $\text{VaR}_\alpha^+$ & not attained; & Proposition~\ref{cor: isotonic projection RVaR} with $w_0 = \alpha$;
\\[0.5em]
$\text{TVaR}_\alpha$ & Proposition~\ref{cor: isotonic projection RVaR lower} with $\beta = 1$; & Proposition~\ref{cor: isotonic projection RVaR} with $\beta = 1$ and  $w_0 = w_1 = 1$.\\[0.2em]
\bottomrule\bottomrule
\end{tabular}
\end{center}
\caption{Best- and worst-case quantile functions of $\text{VaR}_\alpha$, $\text{VaR}_\alpha^+$, $\text{RVaR}_{\alpha, \beta}$, and $\text{TVaR}_{\alpha}$
\\
for case 1. of Theorems \ref{thm: general} and \ref{thm: lower bound}, respectively; that is, for $\varepsilon$ sufficiently small.\label{table: var-Tvar-quantile}}
\end{table}

\begin{example}

Figures \ref{fig: worst-case quantile RVaR} and \ref{fig: best-case quantile RVaR} illustrate the isotonic projections for the worst- and best-case quantile function of $\text{RVaR}_{\alpha, \beta}$ with $\alpha = 0.6$ and different values of $\beta \in\{\,0.61, \, 0.85, \, 0.99\}$. Specifically, Figure \ref{fig: worst-case quantile RVaR} displays $\gamma(u)+\lambda F^{-1}(u), \, u \in (0,1)$ and its isotonic projection $k_\lambda^\uparrow$ onto the set of non-decreasing functions (derived in Proposition~\ref{cor: isotonic projection RVaR}).  Note that we chose $\varepsilon$ sufficiently small such that the worst-case quantile functions of $\text{RVaR}_{\alpha, \beta}$ are not two-point distributions, and thus case 1. of Theorem~\ref{thm: general} applies. Figure \ref{fig: best-case quantile RVaR} displays the corresponding graphs for the best-case quantile functions.

\begin{figure}[!h]
\includegraphics[width=1.1\textwidth, height=6cm]{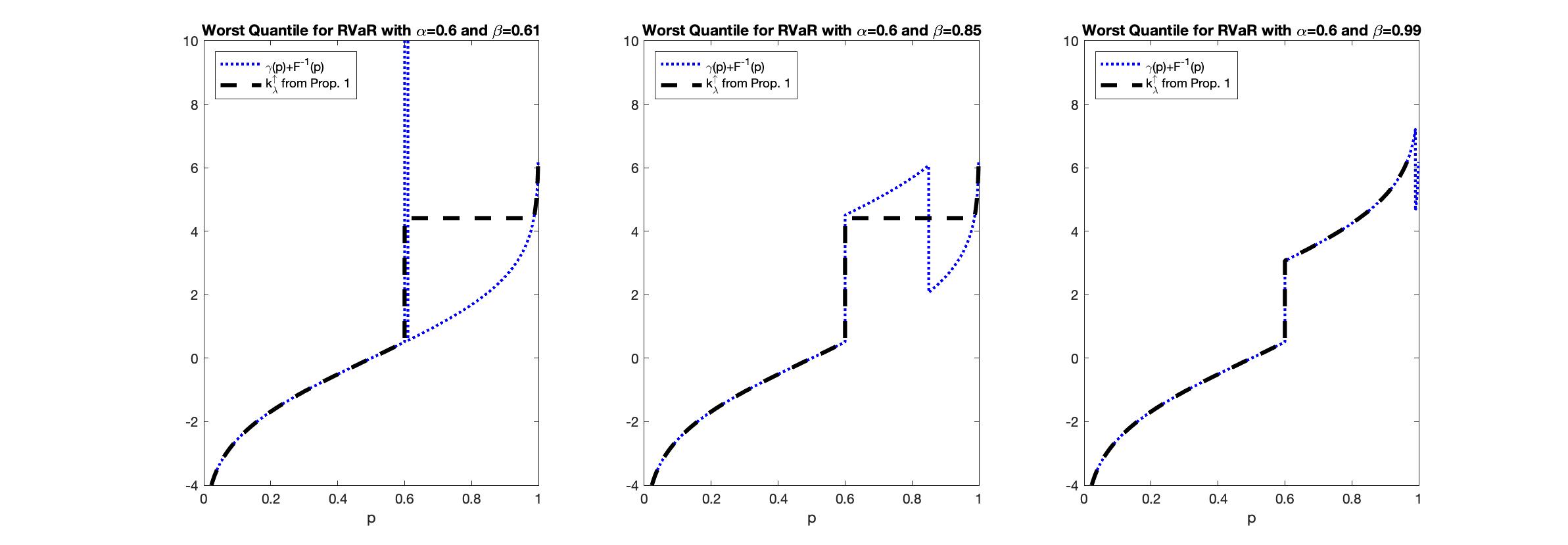}
\caption{For the risk measure $\text{RVaR}_{\alpha, \beta}$, we plot $\gamma(u)+\lambda F^{-1}(u), \, u \in (0,1)$ and its isotonic projection $k_\lambda^\uparrow$ onto the set of non-decreasing functions (derived in Proposition~\ref{cor: isotonic projection RVaR}). The reference distribution $F$ is standard normal with $\mu = \mu_F = 0$ and $\sigma = \sigma_F = 1$, and we set $\varepsilon=0.2$.} 
\label{fig: worst-case quantile RVaR} 
\end{figure}

\begin{figure}[!h]
\includegraphics[width=1.1\textwidth, height=6cm]{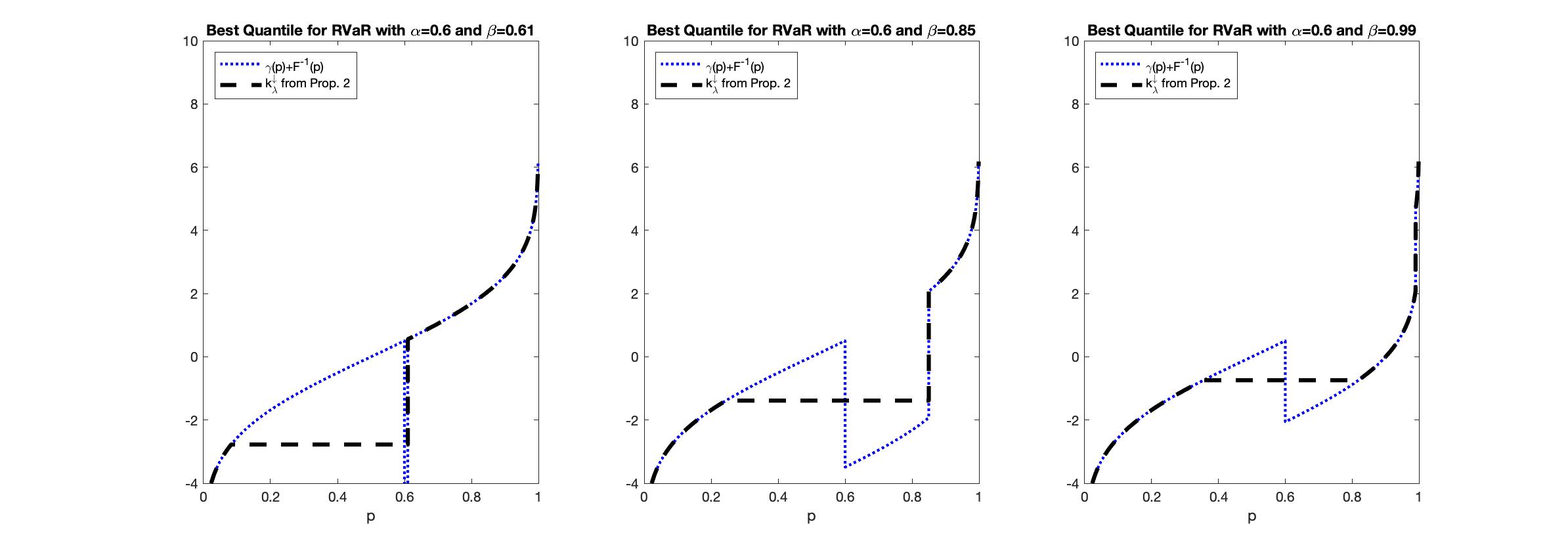}
\caption{For the risk measure $\text{RVaR}_{\alpha, \beta}$, we plot $\gamma(u)-\lambda F^{-1}(u), \, u \in (0,1)$ and its isotonic projection $k_\lambda^\downarrow$ onto the set of non-increasing functions (derived in Proposition~\ref{cor: isotonic projection RVaR lower}). The reference distribution $F$ is standard normal with $\mu = \mu_F = 0$ and $\sigma = \sigma_F = 1$, and we set $\varepsilon=0.2$.} 
\label{fig: best-case quantile RVaR}
\end{figure}

The left plot of Figure \ref{fig: bounds-RVaR}  displays the lower and upper bounds on $\text{VaR}_{\alpha}$, $\text{RVaR}_{\alpha, \beta}$, and $\text{TVaR}_{\alpha}$ as a function of the Wasserstein distance $\varepsilon$. When $\varepsilon$ is sufficiently large (case 2. in Theorems \ref{thm: general} and \ref{thm: lower bound}), the Wasserstein distance  no longer affects the bounds; these bounds coincide with the well-known Cantelli bounds. The normalised lengths of the bounds on $\text{VaR}_{\alpha}$, $\text{RVaR}_{\alpha, \beta}$, and $\text{TVaR}_{\alpha}$ as a function of $\varepsilon$ are displayed in the right plot of Figure \ref{fig: bounds-RVaR}. The normalised length of the bounds are the differences between \eqref{eq:problem sup} and \eqref{eq:problem inf} divided by the risk measure evaluated at the reference distribution and have been introduced to assess model risk by \cite{barrieu2015assessing}. The length of the bounds on $\text{RVaR}_{\alpha, \beta}$, for example, is $(\sup_{G \in \mathcal{M}_\varepsilon(\mu, \sigma)} \text{RVaR}_{\alpha,\beta} (G) - \inf_{G \in \mathcal{M}_\varepsilon(\mu, \sigma)} \text{RVaR}_{\alpha,\beta} (G)) / \text{RVaR}_{\alpha,\beta} (F)$.
\begin{figure}[!tbp]
\begin{tabular}{cc}
\includegraphics*[width=0.5\textwidth, height=6cm]{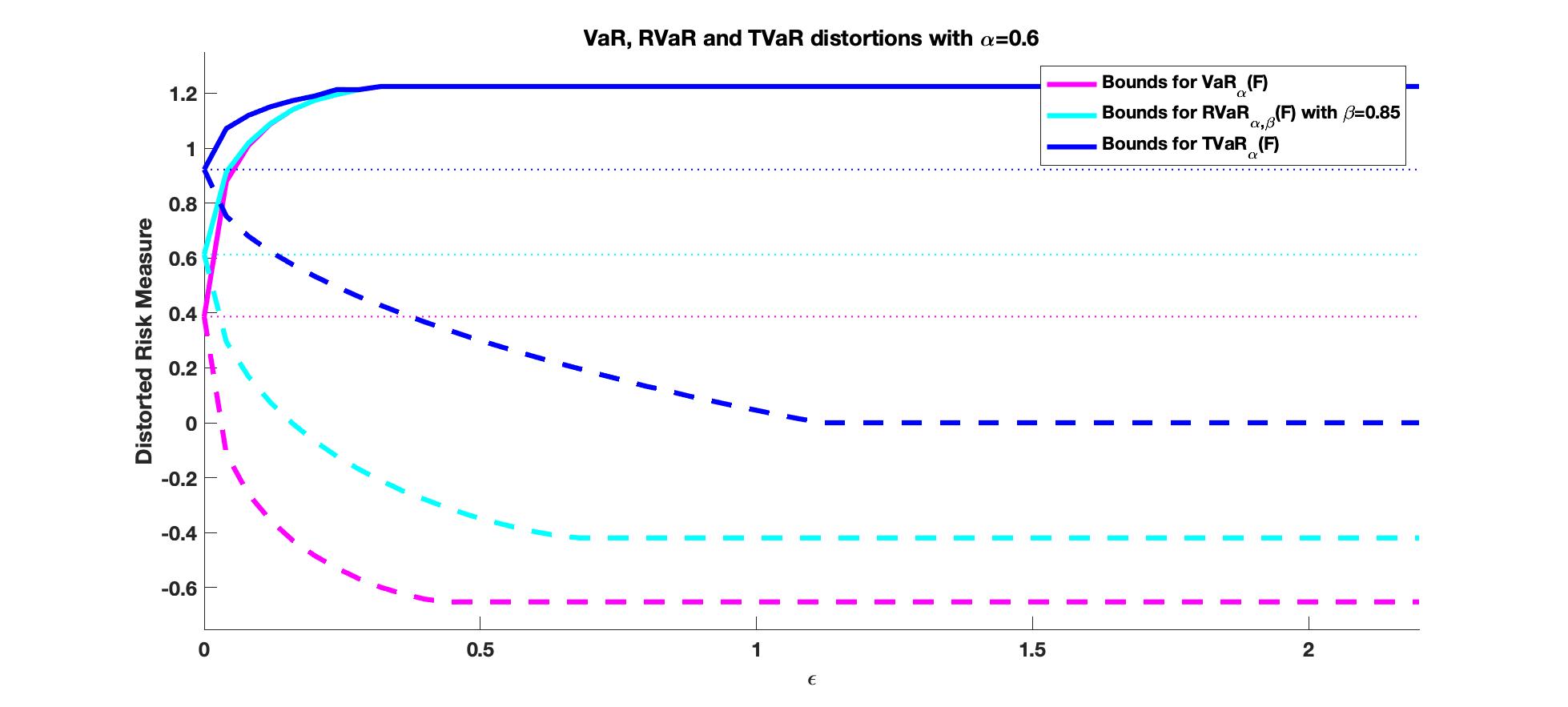} &
\includegraphics*[width=0.5\textwidth, height=6cm]{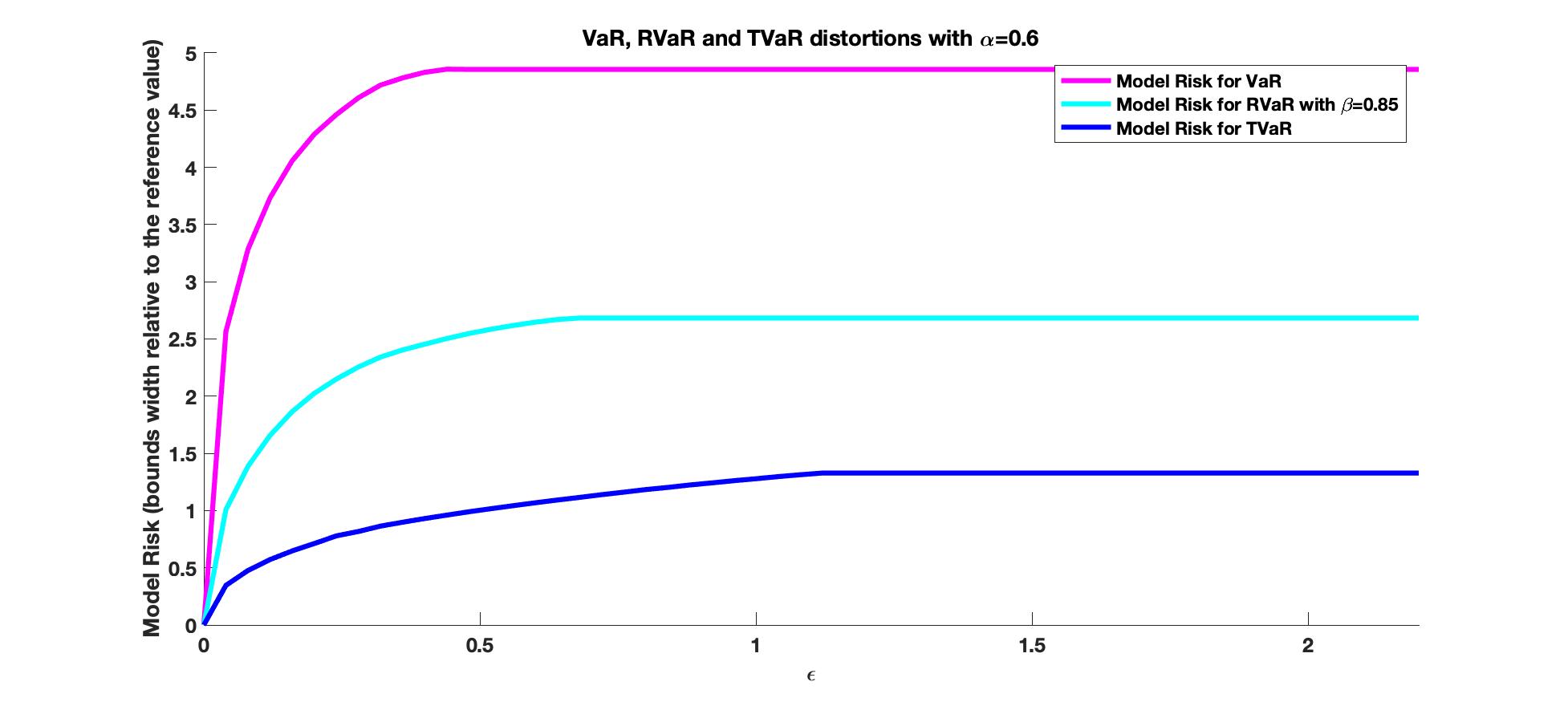} \\
\end{tabular}
\caption{Left: Best- and worst-case values of $\text{VaR}_{\alpha}$, $\text{RVaR}_{\alpha,\beta}$, and $\text{TVaR}_{\alpha}$ as a function of the tolerance level $\varepsilon$. VaR and TVaR are at level $\alpha=0.6$ and RVaR is at levels $\alpha=0.6$ and $\beta=0.85$. Solid lines represent the worst-case values, dashed lines represent the best-case values.
Right: Length of the normalised bounds of VaR, RVaR, and TVaR.}
\label{fig: bounds-RVaR}
\end{figure}
We observe that the normalised length of the bounds on all three risk measures increases with respect to $\varepsilon$. Moreover, the normalised length of the VaR bounds is significantly larger than in the case of TVaR, a fact that is well-known for the case when $\varepsilon = +\infty$ \citep{Embrechts2015FS}.  
\end{example}

\subsection{Robustness to Moment Uncertainty}\label{sec: uncertainty sets}

In practical applications, the mean and variance of loss distributions may themselves be prone to uncertainty, e.g., when estimated jointly, see \cite{Delage2010OR}. When uncertainty in the mean is prevalent, \eqref{eq:problem inf} and \eqref{eq:problem sup} should be considered with uncertainty set $\mathcal{M}_\varepsilon(K) = \{ G  \in \mathcal{M}_\varepsilon (\mu, \sigma) ~| ~ (\mu, \sigma) \in K\}$, where $K \subset \mathbb{R}\times \mathbb{R}_{+}$ specifies the range of ambiguity on $\mu$ and $\sigma$. Proposition \ref{prop: uncertainty} provides the worst-case values when $\mu$ and $\sigma$ are prone to marginal and elliptical uncertainty. Extensions to best-case values follow in a similar fashion and are omitted.

Consider the following cases 
 \begin{enumerate}
	\item (\textit{marginal})  $K = \{\,(\mu, \sigma)~|~ \underline{\mu}\leq \mu \leq \overline{\mu}, ~|\underline{\mu}|\leq |\overline{\mu}|, ~ \underline{\sigma}\leq \sigma \leq \overline{\sigma}\,\}$. 

	\item (\textit{circlic})  $K = \left\{(\,\mu, \sigma)~|~ \sigma >0, ~(\mu_F - \mu)^2 + (\sigma_F - \sigma)^2 \leq r^2 \,\right\}, ~ r>0$. 
 
	\item (\textit{elliptical})  $K = \left\{\,(\mu, \sigma)~\left|~\sigma >0, ~ \frac{(\mu_F - \mu)^2}{c^2} + \frac{(\sigma_F - \sigma)^2}{d^2} \leq r^2 \,\right\}\right.$, with $c, d, r>0$. 
\end{enumerate}

\begin{proposition}[Upper Bounds with Moment Uncertainty]\label{prop: uncertainty}
Let $g$ be a distortion function and $K$ given by case 1. 2. or 3 defined above. If for all $(\mu, \sigma) \in K$, it holds that $(\mu_F - \mu)^2 + (\sigma_F - \sigma)^2 < \varepsilon < (\mu_F - \mu)^2 + (\sigma_F - \sigma)^2 + 2\sigma \sigma_F(1 - c_{0})$, then
\begin{equation*}
\sup_{G \in \mathcal{M}_\varepsilon (K)} H_g(G)~
	= \sup_{G \in \mathcal{M}_\varepsilon (\mu_{\max}^K, \sigma_{\max}^K)} H_g(G)
\end{equation*}
and its solution is given in Theorem \ref{thm: general}, case 1., with  $\mu_{\max}^K$ and $\sigma_{\max}^K$ given by

\begin{enumerate}
	\item (\textit{marginal}) If $\mu_F  < -\frac{1}{\lambda}$, then
	\begin{equation*}
	    (\mu_{\max}^K, \sigma_{\max}^K) = (\underline{\mu},  \overline{\sigma})\,,
	    \quad \text{otherwise } \quad
	    (\mu_{\max}^K,\, \sigma_{\max}^K) = (\overline{\mu},\,\overline{\sigma})\,.
	\end{equation*}

	\item (\textit{circlic}) We have
	\begin{equation}
	 \left(\mu_{\max}^K, \;\sigma_{\max}^K\right) = \left(\mu_F \pm \frac{r }{ \sqrt{\,1 + (\text{cv}_\lambda)^2}\,}, \; \sigma_F +  \frac{r\, |\text{cv}_\lambda|}{\sqrt{\, 1 + (\text{cv}_\lambda)^2\,}}\right),
 	\end{equation}
	where $\text{cv}_\lambda = \sqrt{\,\var (Z_\lambda)\,} \corr\left(Z_\lambda, k_\lambda^\uparrow(U)\right)/ ( 1 + \lambda \,\mu_F) $ with $Z_\lambda = \gamma(U) + \lambda F^{-1}(U)$, and $\mu_{\max}^K$ takes on the smaller value whenever $\mu_F < -\frac{1}{\lambda}$.
	
	\item (\textit{elliptical}) We have
 	\begin{equation}
 	    \left(\mu_{\max}^K, \;\sigma_{\max}^K\right) = \left(\mu_F \pm \frac{r c}{ \sqrt{\,1 + \left(\frac{d}{c} \,\text{cv}_\lambda \right)^2\,} },\; \sigma_F + \frac{r d^2\,|\text{cv}_\lambda|}{ c\,\sqrt{\,1 + \left(\frac{d}{c} \,\text{cv}_\lambda \right)^2\,}}\right),
 	\end{equation}
	where $\mu_{\max}^K$ takes on the smaller value whenever $\mu_F < -\frac{1}{\lambda}$.
\end{enumerate}

If for all $(\mu, \sigma) \in K$, it holds that $ (\mu_F - \mu)^2 + (\sigma_F - \sigma)^2 + 2\sigma \sigma_F(1 - c_{0}) \le \varepsilon $, then the solution to $\sup_{G \in \mathcal{M}_\varepsilon (K)} H_g(G)$ is given in Theorem \ref{thm: general}, case 2.
\end{proposition}

\subsection{Wasserstein Uncertainty Set}\label{sec:Wasserstein-only}
In this section we consider worst-case concave distortion risk measures under complete moment uncertainty. Specifically, we consider an uncertainty set that contains all distribution functions that lie within an $\sqrt{\varepsilon}$-Wasserstein ball around the reference distribution, i.e., without a first and second moment constraint. Thus, the optimisation problem becomes
\begin{equation}\label{opt:wasser-only}
    \argmax_{G \in \mathcal{M}_{\varepsilon}} H_g(G)\,,
\end{equation}
where $\mathcal{M}_{\varepsilon} = \left\{G \in \mathcal{M}^2 ~|~ d_W(F,G) \le \sqrt{\varepsilon}  \right\}$ is the set of all distribution functions that lie within a $\sqrt{\varepsilon}$-Wasserstein ball around the reference distribution $F$.

\begin{theorem}[Wasserstein Uncertainty]\label{thm:wasser-only}
Let Assumptions \ref{asm: Choquet integral} and \ref{asm: gamma not 1} be fulfilled. Further, let $g$ be a concave distortion function; then, the solution to problem \eqref{opt:wasser-only} exists, is unique, and has quantile function given by 
	\begin{equation*}
	h_\lambda(u) = 
	\mu^* + \sigma^* \left(\frac{\gamma(u) + \lambda F^{-1}(u)-a_\lambda}{b_\lambda}\right), \quad  0 < u < 1\,,
	\end{equation*}
	with
	\begin{equation*}
    (\mu^*,\sigma^*)=\left(\ \mu_F+\sqrt{\frac{\varepsilon}{1+V}}\ \ ,\ \  \sqrt{\sigma_F^2+\frac{2C_{\gamma,F}\sqrt{\varepsilon}}{\sqrt{1+V}}+\frac{\varepsilon V}{1+V}}\ \right)\,,
    \end{equation*}
where $a_\lambda, b_\lambda$, $V$, and $C_{\gamma, F}$ are given in Theorem~\ref{thm: concave}
and $\lambda>0$ denotes the unique positive solution to $ d_W(F^{-1}, h_\lambda)= \sqrt{\varepsilon} $, which is explicitly given in Theorem~\ref{thm: concave}.
The corresponding worst-case value is
\begin{equation*}\label{eq:wasser-only-max}
	H_g(h_\lambda) 
	= \mu^* + \sigma^* \std(\gamma(U)) \corr\left(\gamma(U),\, \lambda F^{-1}(U)\right)\,.
\end{equation*}
\end{theorem}
Note that if $\varepsilon >0$, then the mean and standard deviation of the worst-case distribution fulfil $\mu^* > \mu_F$ and $\sigma^* > \sigma_F$ (recall that $C_{\gamma, F} \ge 0$). Thus, the worst-case distribution when the uncertainty set is only characterised by a Wasserstein ball has a larger mean and standard deviation than the benchmark. Thus, the uncertainty set $\mathcal{M}_{\varepsilon}(\mu, \sigma)$ with fixed $\mu$ and $\sigma$ results in a different worst-case distribution compared to the uncertainty set characterised solely by the Wasserstein distance.

\section{Applications}\label{sec: application}

The tolerance distance $\varepsilon,$ i.e., the degree of the uncertainty, should be adequately chosen. \cite{pesenti2021Wasser} study optimal portfolio choice and assume that the investor is prepared to accept terminal wealth distributions that stay within an $\varepsilon$-Wasserstein ball around some benchmark distribution. In this case, the Wasserstein tolerance distance $\varepsilon$ reflects the investor's tolerance of deviating from the benchmark portfolio strategy and its value is thus driven by the investor's risk preferences. In most applications, however, the uncertainty set is constructed to contain all ``plausible'' distributions, i.e., the uncertainty set should be large enough to contain with high probability the true data-generating distribution and at the same time small enough to exclude pathological distributions, which would incentivise overly conservative decisions (\cite{Esfahani2018dMP}). In these settings, the tolerance distance should be driven by data considerations rather than being exogenously specified.
In this regard, in various (data) contexts and under various assumptions \cite{Wozabal2014OR} chooses $\varepsilon$ via cross-validation whereas \cite{Blanchet2021MS} choose $\varepsilon$ such that the true distribution function lies with a given probability within a Wasserstein ball around the empirical reference distribution. Additionally, including expert opinion in constructing the uncertainty set may provide valuable information, particularly in the case of limited available data (\cite{clemen1999combining}). In the applications hereafter we do not pursue the estimation of $\varepsilon$ in great length, but rather provide some guidelines for choosing it.

\subsection{Portfolio Optimisation}
We consider a portfolio optimisation problem in which an investor aims to construct a robust portfolio $\x = (x_1, \ldots, x_n) \in \mathcal{P}$ among returns $\Rr = (R_1, \ldots, R_n)$, $n \in \N$, in which $\mathcal{P}$ is a suitable polyhedral set. We denote the multivariate distribution function of $\Rr$ by $G_\Rr$ and assume that it is subject to uncertainty in that only its mean vector $\boldsymbol \mu$ and covariance matrix $\Sigma$ are known. For each portfolio $\x$, we denote by $G_{\x}$ the (unknown) distribution function of the aggregate portfolio loss $-\x^\top \Rr$ having mean $-\mu_\x=-\boldsymbol{\x}^\top \boldsymbol{\mu}$ and variance $\sigma^2_\x=\x^\top \Sigma \x$. There is a benchmark model under which the investor assumes\footnote{This assumption is fulfilled, for instance, when an elliptical multivariate distribution function is taken as reference model for $\Rr$.} that the aggregate portfolio loss has distribution function $F_\x$ belonging to a location-scale family, i.e., there is a distribution function $F$ such that $$F^{-1}_\x(u)=-\mu_\x+\sigma_\x F^{-1}(u),\quad  0<u<1.$$  
Then, for fixed $\x \in \mathcal{P}$, the Wasserstein distance between $F_\x$ and $G_\x$ is given by
\begin{equation}
 \label{WSD}
     d_W(F_\x, G_\x)^2 = 2 \sigma_\x^2 (1 - corr(F_\x^{-1}(U), G_\x^{-1}(U))\,.
\end{equation}
The investor now aims to find the portfolio $\x$ that has the smallest risk -- measured via a concave distortion risk measure -- among all possible portfolios whose aggregate loss lie within a Wasserstein distance of the benchmark portfolio. Specifically,
we consider the optimisation problem
\begin{equation} \label{opt: portfolio}
    \min_{\x \in \mathcal{P}}\; 
    \max_{\stackrel{G_\Rr \in \M(\boldsymbol{\mu}, \Sigma)}{d_W(F_{\x},G_{\x} ) \le \sqrt{\varepsilon_\x}}} 
    H_g \left(G_{\x}\right)\,,
\end{equation}
where $\M(\boldsymbol{\mu}, \Sigma)$ is the set of all $n$-dimensional distribution functions with mean vector $\boldsymbol{\mu}$ and covariance matrix $\Sigma$, and where we allow the tolerance distance to depend on $\x$, i.e., $\varepsilon_\x \geq 0$, $\x \in \mathcal{P} $. 

\begin{proposition}\label{Propopt} 
Let Assumptions \ref{asm: Choquet integral}, \ref{asm: gamma not 1}, and \ref{asm: epsilon big enough} be fulfilled.
Further, let $g$ be a concave distortion function. Then, 
Problem \eqref{opt: portfolio} is equivalent to
\begin{equation}\label{eq:opt-portfolio-min}
    \min_{\x \in \mathcal{P}}\;
    \left\{\;-\mu_\x + \left(\sigma_\x - \frac{\underline{\varepsilon_\x}}{2 \sigma_\x}\right){\sqrt{V}c_0} +\
    \sqrt{ \underline{\varepsilon_\x} -\frac{{\underline{\varepsilon_\x}}^2}{4 \sigma_\x^2} } \;\sqrt{V(1 - c^2_0)}\; \right\}
    \,,
\end{equation}
where $\underline{\varepsilon_\x}=\min\{\varepsilon_\x,2\sigma_\x^2(1-c_0)\}$, $V = \var(\gamma(U))$, and $c_0=\corr(F^{-1}(U),\gamma(U)).$
\end{proposition}

Note that problem formulation  \eqref{opt: portfolio} is general in that the tolerances $\varepsilon_\x$ may depend on the portfolio $\x \in \mathcal{P}$ at hand. As a consequence, it is not always possible to obtain the minimum in \eqref{eq:opt-portfolio-min} in explicit form, and numerical procedures may be required. 

From \eqref{WSD}, however, we observe that the Wasserstein distance scales linearly with the standard deviation of the portfolio loss. Hence, it appears natural to take the tolerance $\varepsilon_\x$  of the form\footnote{If historical asset returns are available the value of $A$ can be determined as follows: using the historical returns we estimate for a given portfolio $\x = (x_1, \ldots, x_n) \in \mathcal{P}$, the empirical distribution $\hat G_\x$ of the portfolio loss, its standard deviation $\hat \sigma_\x$, and the Wasserstein distance $\hat d_\x$ between $\hat G_\x$ and the (discretised version of the) reference distribution $F_\x$. By doing so for a set of portfolios  $\x^{(i)}\in \mathcal{P}, i\in \{1,\ldots, N\}$, one can estimate $A$ by $\hat{A}=\max\{\frac{\hat d_\x}{2\hat \sigma_\x^2} ~|~ \x^{(i)} \in \mathcal{P},\, i=1,\ldots,N$\}. This procedure in particular guarantees that for each portfolio  $\x^{(i)} ,\, i\in\{1,\ldots,N\}$ its corresponding empirical distribution $\hat G_{\x^{(i)}}$ lies in the uncertainty set.} 
$\varepsilon_\x = 2\sigma^2_\x A$ for $0 \leq A \leq 1$. The next result considers this case. Its proof follows from Proposition~\ref{Propopt} via direct calculation and is omitted.
\begin{corollary}\label{corropt} 
Let Assumptions \ref{asm: Choquet integral}, \ref{asm: gamma not 1}, and \ref{asm: epsilon big enough} be fulfilled. Further, let $g$ be a concave distortion function and assume  that $\varepsilon_\x=2\sigma^2_\x A$ for $A \in[0,1]$. Let $V = \var(\gamma(U))$ and $c_0=\corr(F^{-1}(U),\gamma(U)).$ Then, the following holds:
\begin{enumerate}

\item If $A=0$, then Problem \eqref{opt: portfolio} is equivalent to
\begin{equation*}
    \min_{\x \in \mathcal{P}}\;
    \left\{\;-\mu_\x + \sigma_\x \, \sqrt{V}c_0  
    \; \right\}
    \,.
\end{equation*}

\item If $0<A<\,1-c_0$, then Problem \eqref{opt: portfolio} is equivalent to
\begin{equation*}
    \min_{\x \in \mathcal{P}}\;
   \left\{\; -\mu_\x + \sigma_\x \, \sqrt{V}  \left (c_0(1-A) +
    \sqrt{A(2-A)}\sqrt{1-c^2_0}\right)   \; \right\}
    \,.
\end{equation*}
   
    \item If $\,1-c_0 \leq A \leq 1$, then Problem \eqref{opt: portfolio} is equivalent to
\begin{equation*}
    \min_{\x \in \mathcal{P}}\;
    \left\{\;-\mu_\x + \sigma_\x\sqrt{V}   \; \right\}
    \,.
\end{equation*}
\end{enumerate}
\end{corollary}
In all cases of Corollary~\ref{corropt}, the robust portfolio optimisation problem reduces to solving a second order cone program. Specifically, for $A=0$, there is no ambiguity and the investor aims to obtain the optimal portfolio under the assumption that for each portfolio $\x$ the portfolio (loss) has known quantile function $F^{-1}_\x(u)=-\mu_\x + \sigma_\x \, F^{-1}_Z(u).$ The optimal choice $\x^*$ is the portfolio\footnote{If an elliptical multivariate distribution function is taken as the reference model for $\Rr$, then $F_{\x^*}$ is also an elliptical distribution.} for which $H_g(F_{\x^*})$ attains 
\begin{equation*}
\min_{\x \in \mathcal{P}}\;
    \left\{\;-\mu_\x + \sigma_\x \, H_g(F)
    \; \right\}\,,
\end{equation*}
where we recall that $\sqrt{V}c_0=H_g(F)$. 
When $ 1-c_0 \leq A \leq \,1$, the Wasserstein distance becomes irrelevant and the optimal portfolio problem reduces to the one considered in \cite{Li2018OR}, i.e., to
\begin{equation*}
\min_{\x \in \mathcal{P}}\;
    \left\{\;-\mu_\x + \sigma_\x \, \sqrt{V} 
    \; \right\}\,.
\end{equation*} 
In this regard, $\sqrt{V}c_0=H_g(F) \leq \sqrt{V}$ implies that in the case of no ambiguity ($A=0$), more emphasis is placed on the expected return component than in the case in which the Wasserstein distance constraint is irrelevant ($1-c_0 \leq A \leq 1\, $.) Moreover, in this instance, the optimal terminal wealth has a quantile function that is linear in the (non-negative) weight function $\gamma$ (see Corollary \ref{cor: properties}), and thus is bounded from below, a feature that might be considered as not very realistic. The cases $0<A<\,(1-c_0)$ deal with situations that are between these extreme scenarios. Observe that $c_0(1-A) +
\sqrt{A(2-A)}\sqrt{1-c^2_0}$ is increasing in $A$ on the interval $[0,1-c_0]$ taking value $c_0$ when $A=0$, and value  $1$ when $A= 1-c_0.$ As $A$ and thus the degree of ambiguity increases, less emphasis is placed on the expected return leading to optimal portfolios that become more conservative. Note that in all intermediate cases the optimal quantile function inherits structure of the reference quantile function $F_{x^*}^{-1}$ (see Theorem \ref{thm: concave}).

\subsection{Insurance Portfolio of Risks}\label{sec: applications-insurance}

This section illustrates the best- and worst-case values of $\text{VaR}$ on a simulated portfolio of dependent risks, that could e.g., arise from insurance activities. We consider the Pareto-Clayton model that offers a flexible way of modelling portfolios with dependent risks, see \cite{Oakes1989JASA} and \cite{Albrecher2011IME} for applications in insurance and \cite{Dacorogna2016SCOR} and \cite{Bernard2018JRI} for applications in finance. In the Pareto-Clayton model, the portfolio components $X_1, \ldots, X_d$ are, given $\Theta = \theta>0$, independent and Exponentially distributed with parameter $\theta$, that is drawn from a Gamma distribution, i.e., $\Theta \sim Gamma(a, b)$. The aggregate portfolio risk $S = \sum_{i=1}^{d}X_{i}$ thus follows, conditionally on $\Theta =\theta $, a Gamma$(d,\theta )$ distribution. By \cite{Dubey1970Metrika} Equations 1.1 and 1.3, the scaled aggregate portfolio risk $S/b$ is a Beta distribution of the second kind. Thus, the quantile function of the aggregate portfolio risk $S$ and the first five moments are explicitly given by
\begin{equation*}
F_{S}^{-1}(u) =b ~ \frac{G_B^{-1}(u)}{1-G_B^{-1}(u)}
\quad \text{and} \quad
\E\left(S^{k}\right) =\frac{b^k\,B(a-k,\,d+k)}{B(d,\,a)}\, , \quad \text{for} \quad k = 1, \ldots, 5\,.
\end{equation*}
where $B(\cdot,\cdot )$ denotes the Beta function and $G_B^{-1}$ the quantile function of the Beta distribution with parameters $d$ and $a$. 

\begin{figure}[!htbp]
\centering
\includegraphics[width=0.8\textwidth, height=6cm]{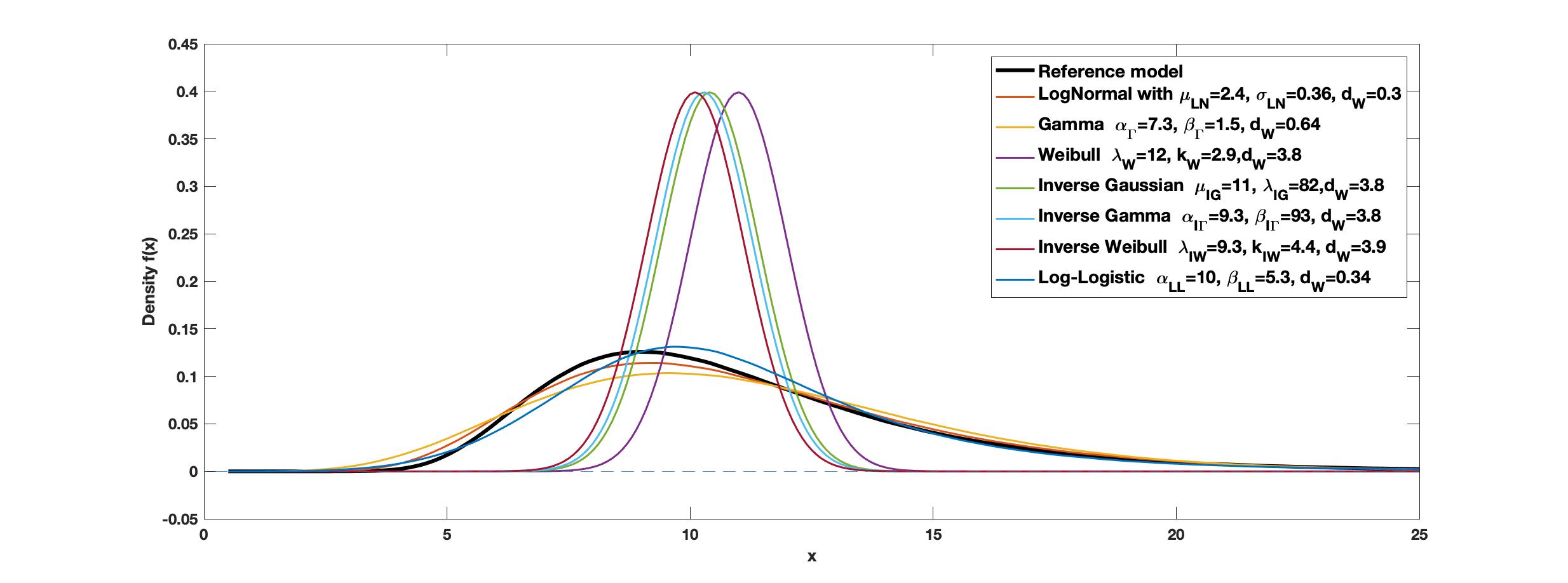} 
\caption{Densities of the reference model (solid black), of the Pareto-Clayton model, and of the alternative models.}
\label{fig: Pareto-Clayton quantile density}
\end{figure}

We consider the situation where modellers or experts may not fully agree on the reference Pareto-Clayton model, as available data or estimation may induce uncertainties or the distribution of $\Theta$ might not seem appropriate. As alternative models for the distribution of the aggregate portfolio risk $S$, we consider two-parameter distributions with support in $(0, +\infty)$ that are commonly utilised for modelling insurance portfolios. Specifically, we consider a Lognormal model (LN), a Gamma model ($\Gamma$), a Weibull model (W), an Inverse Gaussian model (IG), an Inverse Gamma model (I$\Gamma$), an Inverse-Weibull model (IW), and a Log-Logistic model (LL). The different models are described in Table \ref{table: alternative models} and additional information is provided in Appendix \ref{appendix Pareto-Clayton}. For the numerical simulations of the Pareto-Clayton model we use $a =10$, $b =1$, and $d=100$, as in \cite{Bernard2018JRI}. The parameters of the alternative models, reported in Table \ref{table: alternative models}, are obtained by matching the first two moments to those of the reference Pareto-Clayton model, i.e., $\mathbb{E}(S) = 11.11$ and $\mathbb{E}(S^2) = 140.28$. 
\begin{table}[t]
\begin{center}
\begin{tabular}{l @{\hspace{2.5em}} l @{\hspace{2em}} c c @{\hspace{2em}} l }
\toprule \toprule
Model & Abb.  &  \multicolumn{2}{l}{Parameters} & $d_W$\\
\midrule
Lognormal model & LN & $\mu_{LN} = 2.4$ & $\sigma_{LN} = 0.36 $ & 0.298 \\
Gamma & $\Gamma$ & $\alpha_\Gamma = 7.3$ & $\beta_\Gamma = 1.5$ & 0.637\\
Weibull & W & $\lambda_W = 12$ & $k_W = 2.9$ & 3.787\\
Inverse Gaussian & IG & $\mu_{IG} = 11$ &  $\lambda_{IG} = 82$ & 3.802\\
Inverse Gamma & I$\Gamma$ & $\alpha_{I\Gamma} = 9.3$ & $ \beta_{I\Gamma} = 93$ & 3.792\\
Inverse-Weibull & IW & $\lambda_{IW} = 9.3$ & $k_{IW} = 4.4$ & 3.868\\
Log-Logistic & LL & $\alpha_{LL} = 10$ & $\beta_{LL} = 5.3$ & 0.345\\
\bottomrule\bottomrule
\end{tabular}
\end{center}
\caption{\label{table: alternative models} Alternative models for the aggregate portfolio risk. The last column reports the Wasserstein distance between the Pareto-Clayton and the alternative model.}  
\end{table}
We illustrate the reference Pareto-Clayton model (solid black) and the alternative models in Figure \ref{fig: Pareto-Clayton quantile density} via their densities. It is apparent that the considered alternative models can be grouped into light-tailed models (W, IG, I$\Gamma$, IW) and heavy-tailed models (LN, $\Gamma$, LL). 

Next, we discuss the choice of Wasserstein distance by means of model uncertainty. Specifically, we let $\varepsilon$ be the maximum of all Wasserstein distances between the reference and a selection of alternative models. The constructed Wasserstein ball then includes all distribution functions whose Wasserstein  distance is smaller or equal to the maximum of all Wasserstein distances between the reference and the considered alternative models and, in particular, all considered alternative models. We report the Wasserstein distances between the Pareto-Clayton model and all alternative models in Table \ref{table: alternative models} and observe that the Wasserstein distances between the reference distribution and the alternatives from the heavy-tailed models are significantly smaller compared to the alternatives from the light-tailed models. This illustrates that model uncertainty is considered with respect to the reference distribution, which in our case is heavy-tailed. Thus, in a model uncertainty context, changing from the heavy-tailed reference model, i.e., Pareto-Clayton, to a light-tailed model, e.g., Weibull, entails a higher level of uncertainty compared to other models from the heavy-tailed family, e.g., Lognormal.

\begin{table}[!htbp]
\begin{center}
\begin{tabular}{c@{\hspace{2.5em}} c@{\hspace{1.5em}}  c@{\hspace{1.5em}} c}
\toprule \toprule
Uncertainty Set & $\alpha=0.9$  &  $\alpha=0.95$  &  $\alpha=0.99$  \\
\midrule
$\mathcal{M}_\varepsilon(\mu, \sigma)$ with $\varepsilon=0.637$& ( 12.8 ; 18.8 ) & ( 14.6 ; 22.8 ) & ( 19.0 ; 35.1 )\\
$\mathcal{M}_\varepsilon(\mu, \sigma)$ with $\varepsilon=3.868$& ( 10.7 ; 21.6 ) & ( 12.1 ; 26.1 ) & ( 15.0 ; 41.5 )\\
\# of moments $k=1$ &  ( 0.0 ; 111.1 ) & ( 0.0 ; 222.2 ) & ( 1.1 ; 1000 ) \\ 
\# of moments $k=2$ &  ( 9.7 ; 23.4 ) & ( 10.2 ; 29.0 ) & ( 10.7 ; 51.9 ) \\ 
\# of moments $k=3$ &  ( 10.1 ; 22.5 ) & ( 10.8 ; 26.2 ) & ( 11.7 ; 38.7 ) \\ 
\# of moments $k=4$ &  ( 10.1 ; 22.5 ) & ( 11.5 ; 25.6 ) & ( 15.1 ; 32.9 ) \\ 
\# of moments $k=5$ &  ( 10.1 ; 22.4 ) & ( 11.5 ; 25.5 ) & ( 15.1 ; 32.9 )\\
$\E=\mu$, $\std\leq \sigma$, unimodal &  ( 9.9 ; 18.7 ) & ( 10.3 ; 22.6 ) & ( 10.7 ; 38.1 )\\
\bottomrule \bottomrule\\[0.25em]
\end{tabular}%
\end{center}
\caption{Best- and worst-case values of $\text{VaR}_\alpha$, for $\alpha = 0.9, \, 0.95, \, 0.99$, for various uncertainty sets. The VaR bounds are reported as ( ; ) in which the first value is the lower and the second is the upper bound.}
\label{table: Pareto-Clayton-bounds}
\end{table}

The best- and worst-case values of $\text{VaR}_\alpha$, for $\alpha = 0.9,\, 0.95, \,0.99$, for various uncertainty sets are report in Table \ref{table: Pareto-Clayton-bounds}. The values of $\text{VaR}_\alpha$ of the Pareto-Clayton model are respectively ${\text{VaR}}_{0.90}=16.29$, ${\text{VaR}}_{0.95}=18.75$, and ${\text{VaR}}_{0.99}=24.79$. The uncertainty sets we consider are  $\mathcal{M}_\varepsilon(\mu, \sigma)$ with $\mu = \mu_F =11.11$, $\sigma_F^2 = \sigma^2=16.82$, $\varepsilon=0.637$ and $\varepsilon=3.868$. The tolerance $\varepsilon=0.637$ corresponds to model uncertainty with respect to heavy-tailed alternative models (LN, $\Gamma$, LL), whereas $\varepsilon=3.868$ allows for alternative models that are light-tailed (W, IG, I$\Gamma$, IW). To compare these bounds with results in the literature, we calculate the lower and upper bounds on VaR when only the first $k$, $k = 1, \ldots, 5$, moments are known \citep{Bernard2018JRI}. Note that the forth row in Table \ref{table: Pareto-Clayton-bounds} ($k = 2$) corresponds to $\mathcal{M}(\mu, \sigma)$, i.e., with Wasserstein tolerance $\varepsilon = + \infty$. It is apparent that the bounds with a tolerance distance that contain the heavy-tailed alternative models ($\varepsilon = 0.637$) are significantly smaller than the ones corresponding to both heavy-tailed and light-tailed alternative models ($\varepsilon = 3.868)$. Furthermore, we report in the last row of Table \ref{table: Pareto-Clayton-bounds} the bounds on VaR for an uncertainty set with fixed mean, bounded standard deviation, and where the uncertainty set further contains only unimodal distributions \citep{bernard2019risk}. We observe in Table \ref{table: Pareto-Clayton-bounds}, that the Wasserstein constraint significantly reduces the bounds and that information on higher moments only affects the bounds in a minor way. Moreover, when only heavy-tailed models are considered as alternatives ($\varepsilon = 0.637)$, the length of the bounds is $\frac{18.8-12.8}{21.6 - 10.7} = 55\%$ of the length  when both light- and heavy-tailed are valid alternative models ($\varepsilon = 0.3868)$.

\section{Conclusion}
We derive quasi explicit best- and worst-case values of a large class of distortion risk measures when the underlying loss distribution has a given mean and variance and lies within a $\sqrt{\, \varepsilon\,}$-Wasserstein ball around a given reference distribution. We find that for small Wasserstein tolerance distances the worst-case distribution function (the distribution function attaining the worst-case value) is probabilistically close to the reference distribution, whereas for large tolerance distances, the worst-case distribution function is no longer affected by the reference distribution. We also derive the worst-case distribution function for VaR and RVaR and show that for small Wasserstein tolerance distances, the worst-case distribution functions of VaR and TVaR are no longer two-point distributions, thus making the bounds attractive for risk management applications. 

Our results generalise findings in \cite{Li2018OR} and \cite{Zhu2018SSRN}, which correspond to a Wasserstein tolerance of $\varepsilon = + \infty$. Furthermore, we illustrate the risk bounds on two applications, to portfolio optimisation and to model risk assessment of an insurance portfolio.
\vspace{1cm}








%
%
%
\begin{APPENDICES}
\section{Isotonic Projection.}\label{sec: isotone projection}
Here, we collect properties of the metric projection defined in \eqref{eq: isotonic non-decreasing}. For this, we denote by $\mathcal{N} = \{ \ell ~|~ \ell \colon (0,1) \to \mathbb{R}\, , \, \int_0^1 \ell(u)^2 \mathrm{d}u < + \infty\}$ the space of square-integrable functions on $(0,1)$ and consider the following metric projection from $\mathcal{N}$ to $\K$ (set defined by \eqref{defK}):
\begin{equation*}
	P_\K (\ell) = \arginf_{k \in \K} \, || \ell - k ||^2, \qquad  \ell \in \mathcal{N}.
\end{equation*}	 
A metric projection is called \emph{isotone} if the projection preserves the order induced by $\K$ \citep{Nemeth2003isothonic}. The partial order $\preceq_\K$ on $\mathcal{N}$ induced by $\K$ is given, for any $\ell_1, \ell_2 \in \mathcal{N}$, by 
\begin{equation*}
\ell_1 \, \preceq_\K \, \ell_2 \qquad \text{if} \quad \ell_2 - \ell_1 \in \K\,.
\end{equation*}

\begin{proposition}[Theorem 3. \cite{Nemeth2003isothonic}]\label{prop:iso-property-appendix}
The metric projection $P_\K$ is isotone and subadditive; that is, $P_\K$ fulfils for all $\ell_1, \ell_2 \in \mathcal{N}$:
\begin{enumerate}
	\item \textbf{Isotone:} If $\ell_1 \preceq_\K \ell_2$, then $P_\K (\ell_1) \preceq_\K  P_\K (\ell_2)$.

	\item \textbf{Subadditive:} $P_\K (\ell_1 + \ell_2) \, \preceq_\K \,  P_\K (\ell_1)  + P_\K (\ell_2)$.
\end{enumerate}
\end{proposition}

In this paper, we compare only projections of functions of the type $\gamma + \lambda F^{-1}$, for fixed $F$ and $\gamma$. Thus, we view the projection of $\gamma + \lambda F^{-1}$ as a function of $\lambda$ and define $k_\lambda = P_\K(\gamma  + \lambda F^{-1})$, $\lambda \geq 0$.  

\begin{proposition}\label{prop: prop isotonic projection}
For fixed $\gamma$ and $F$, the metric projection $k_\lambda$ is an isotonic projection with the following properties:
\begin{enumerate}
	\item \textbf{Isotone:} For $\lambda_1 \le \lambda_2$, it holds that $k_{\lambda_1} \,\preceq_\K \, k_{\lambda_2}$.
	
	\item \textbf{Continuous:} If $\lambda_n \to \lambda$, then $\lim\limits_{n \nearrow +\infty} ||k_{\lambda_n} - k_\lambda||^2 = 0.$
\end{enumerate}
\end{proposition}

\proof{Proof of Proposition~\ref{prop: prop isotonic projection}.}
Property 1. (isotonicity of $k_\lambda$) follows by noting that $\lambda_1 \le \lambda_2$ implies $\gamma + \lambda_2 F^{-1} - (\gamma + \lambda_1 F^{-1})  = (\lambda_2 - \lambda_1) F^{-1}\in \K$. Thus, by isotonicity of $P_\K$ we obtain
\begin{equation*}
k_{\lambda_1} = P_\K (\gamma + \lambda_1 F^{-1}) \, \preceq_\K \,  P_\K (\gamma + \lambda_2 F^{-1}) = k_{\lambda_2}.
\end{equation*} 
To prove continuity of $k_\lambda$ (Property 2.), denote $\lambda_n = \lambda + e_n$, for any sequence $e_n$ with $e_n \to 0$, as $n \to +\infty$. Then by Theorem 2.3 in \cite{Brunk1965AMS}, it holds that
\begin{align*}
\lVert k_{\lambda_n} - k_\lambda\rVert^2 
& \le
\Vert \gamma + (\lambda + e_n) F^{-1} - \gamma - \lambda F^{-1}  \rVert^2
=
|e_n|^2\, \lVert F^{-1} \rVert^2 \to 0, \quad \text{for} \quad n \to +\infty.
\end{align*}
\hfill \Halmos
\endproof
The next proposition collects a few properties of the isotonic projection $P_\K$ and thus of $k_\lambda$. These properties follow from \cite[Theorem 2.2. and Corollary 2.3]{Brunk1965AMS} and we provide short proofs adapted to our setting. 
For this we denote by $\langle \cdot, \cdot\rangle$ the inner product on $\mathcal{N}$ -- that is, $\langle \ell_1, \ell_2 \rangle = \int_0^1 \ell_1(u) \ell_2(u) \, \mathrm{d}u $, for $\ell_1,\, \ell_2 \in \mathcal{N}$. 
\begin{proposition}\label{prop isotonic prop}
The isotonic projection $P_\K$ satisfies the following properties for all $\ell \in \mathcal{N}$ and $k \in \K$:
\begin{enumerate}
	\item $\langle \ell - P_\K(\ell)\, ,\, P_\K(\ell) - k\rangle \ge 0$;
	
	\item $\langle \ell \,, \,P_\K(\ell) \rangle  = \langle P_\K(\ell), P_\K(\ell) \rangle$;

	\item $\langle \ell\, , \,k \rangle  \le \langle P_\K(\ell), k \rangle$.
\end{enumerate}
\end{proposition}

\proof{Proof of Proposition~\ref{prop isotonic prop}.}
We follow the arguments of the proof of Theorem 1.3.2 by \cite{barlow1972statistical}. 

\textit{Property 1.} For $\ell \in \mathcal{N}$, $f\in \K$, and $w\in [0,1]$ define the function $(1 - w) P_\K(\ell) + w f$. By definition of the isotonic projection it holds that $(1 - w) P_\K(\ell) + w f \in \K$. Next, we consider the function
	\begin{equation*}
	w \mapsto \left\langle \ell - (1 - w) P_\K(\ell) - w f \, , \, \ell - (1 - w) P_\K(\ell) - w f \right\rangle\, , 
	\end{equation*}
	which obtains its minimum in $w = 0$, by definition of the isotonic projection of $P_\K(\ell)$. The first order condition at $w = 0$, implies that 
	\begin{equation*}
	2\, \big\langle \ell - (1 - w) P_\K(\ell) - w f \, , \,P_\K(\ell) - f\big\rangle\big|_{w = 0}
	=  2\, \big\langle \ell - P_\K(\ell) \, , \,P_\K(\ell) - f\big\rangle  \ge 0\, .
	\end{equation*}
	
\textit{Property 2.} By the definition of an isotonic projection, it holds that 	$\langle  P_\K(\ell) - \ell\, , \, P_\K(\ell) - \ell \rangle
	 \le 	\langle  k -  \ell \, , \, k -  \ell \rangle\, $ for all  $\ell, \, k\in \mathcal{N}$. 
Choosing $k = c\, P_\K(\ell)$ with $c>0$, we obtain
\begin{equation*}
2\, (c-1)\,\langle P_\K(\ell)\, , \, \ell \rangle 
\le 
(c^2-1)\,\langle P_\K(\ell)\, , \, P_\K(\ell) \rangle\, .
\end{equation*}
Thus, for $0<c_1<1$ and $c_2>1$ the above inequality becomes
\begin{equation*}
\tfrac{1 - c_1^2}{2(1 - c_1)}\, \langle P_\K(\ell)\, , \, P_\K(\ell) \rangle
\le 
\langle P_\K(\ell)\, , \, \ell \rangle 
\le 
\tfrac{c_2^2-1}{2(c_2-1)}\, \langle P_\K(\ell)\, , \,  P_\K(\ell) \rangle\, .
\end{equation*}
Taking limits with $c_1 \nearrow 1$ and $c_2\searrow 1$ concludes the statement of the property.

\textit{Property 3.} For $\ell \in \mathcal{N}$ and $k \in \K$, we obtain, by first applying property 2. and then property 1.
	\begin{align*}
	\langle \ell - P_\K(\ell)\, ,\, P_\K(\ell) - k\rangle 
	&= 
	\langle \ell \, ,\, P_\K(\ell) \rangle 
		-\langle \ell \, ,\, k \rangle 		
		- \,\langle P_\K(\ell)\, ,\, P_\K(\ell) \rangle 
		+ 	\langle k \, ,\, P_\K(\ell) \rangle 
		= -\langle \ell \, ,\, k \rangle 	
		+\langle k \, ,\, P_\K(\ell) \rangle 
		\ge 0\, .
	\end{align*}
\hfill \Halmos
\endproof

\section{Proofs.} \label{sec: proofs}
We denote by $\mathcal{M}^{-1}_\varepsilon(\mu, \sigma) = \{G^{-1} ~|~G \in \mathcal{M}_\varepsilon(\mu, \sigma)\}$ the corresponding set of quantile functions in $\mathcal{M}_\varepsilon(\mu, \sigma)$. Further, we denote $\mathcal{M}^{-1}(\mu, \sigma) = \{G^{-1} ~|~G \in \mathcal{M}(\mu, \sigma)\}$.
\begin{lemma}\label{lemma-asm-1}
   If $\int_0^1 \gamma(u)\, du = +\infty$, then 
   $ \inf_{G \in \mathcal{M}_\varepsilon(\mu, \sigma)} H_g(G)
       =
       \sup_{G \in \mathcal{M}_\varepsilon(\mu, \sigma)} H_g(G)
       = 
       +\,\infty\,.$
\end{lemma}
\proof{Proof of Lemma~\ref{lemma-asm-1}.}
For $v\in(0,1)$ define 
\begin{equation*}
    h_{v, \lambda}(u) = \mu + \sigma \left(\frac{\min\{k^\uparrow_{\lambda}(u), v\} - a_{v, \lambda}}{b_{v, \lambda}}\right)\,,
    \quad 0<u<1\,,
\end{equation*}
where $k_\lambda^\uparrow$ is defined in \eqref{eq: isotonic non-decreasing}, $a_{v, \lambda} = \E(\min\{k^\uparrow_{\lambda}(U), v\})$, and $b_{v, \lambda} = \std(\min\{k^\uparrow_{\lambda}(U), v\})$. By construction $h_{v, \lambda}\in\mathcal{M}^{-1}(\mu, \sigma)$ for all $v \in(0,1)$. Moreover, 
\begin{align*}
    \int_0^1 h_{v, \lambda}(u) \gamma(u)\, du
    &= 
    \mu + \frac{\sigma}{b_{v, \lambda}}
    a_{v, \lambda}\left(\int_0^1\frac{\min\left\{ k_\lambda^\uparrow(u), v\right\}}{a_{v, \lambda}} \gamma(u) du- 1\right)
    \\
    &\ge 
    \mu + \frac{\sigma}{b_{v, \lambda}}
    a_{v, \lambda}\left(\int_0^1\frac{(\min\left\{ k_\lambda^\uparrow(u), v\right\})^2}{a_{v, \lambda}}  du- 1\right)
    =
    \mu + \sigma \,\frac{b_{v, \lambda}}{a_{v, \lambda}}\,,
\end{align*}
which diverges to $+\infty$, for $v \nearrow 1$.
 \hfill \Halmos
\endproof

\subsection{Proof of Theorem~\ref{thm: concave} and Corollary~\ref{cor: properties}.}
For the proof of Theorem~\ref{thm: concave} we need the following lemmas.
\begin{lemma}\label{lemma: concave}
Let $\lambda \geq 0$, and $\gamma$ be non-decreasing; then 
\begin{equation*}
\sup_{h \in \mathcal{M}^{-1}(\mu, \sigma)} \langle \gamma + \lambda F^{-1}, \,h \rangle
\end{equation*}
has a unique solution given by
\begin{equation} \label{proof:eqn:h-lambda}
h_\lambda(u) = 
\mu + \sigma \left(\frac{\gamma(u) + \lambda F^{-1}(u)-a_\lambda}{b_\lambda}\right), \quad  0 < u < 1,
\end{equation}
where $a_\lambda = \E(\gamma(U) + \lambda F^{-1}(U)) $ and $b_\lambda ={\std(\gamma(U) + \lambda F^{-1}(U))}$. Moreover, $\corr(F^{-1}(U), h_\lambda(U) )$ is continuous in $\lambda$, for $\lambda \ge 0$.
\end{lemma}

\proof{Proof of Lemma~\ref{lemma: concave}.}
The function $h_\lambda$, defined in \eqref{proof:eqn:h-lambda}, belongs to $ \mathcal{M}^{-1}(\mu, \sigma)$. Let $h \in \mathcal{M}^{-1}(\mu, \sigma)$; then, application of the Cauchy-Schwarz inequality implies that $\langle \gamma + \lambda F^{-1}, h \rangle \leq \langle \gamma + \lambda F^{-1}, h_\lambda \rangle$, with equality if and only if $h$ is linear in $\gamma + \lambda F^{-1}$. Continuity of $\corr(F^{-1}(U), h_\lambda(U) )$ is immediate. \hfill \Halmos
\endproof

\begin{lemma} \label{lemma:explicitLambda} The solution $\lambda$ to  $d_W(F^{-1}, h_\lambda)^2=\E((F^{-1}(U) - h_\lambda(U))^2)= \varepsilon$, where $h_\lambda(\cdot)$ is defined in Lemma~\ref{lemma: concave} is explicit and is given by 
\begin{equation}\label{lamSOL}
\lambda=\frac{\sqrt{\Delta}-C_{\gamma, F}}{\sigma_F^2}\,,
\quad \text{where} \quad
\Delta=\frac{K^2(C_{\gamma, F}^2-V\sigma_F^2)}{K^2-\sigma^2\sigma_F^2}\,,
\end{equation}
with $V=\var(\gamma(U))$, $C_{\gamma, F}=\cov\left(	F^{-1}(U),  \gamma(u)\right)$, and 
\begin{equation}\label{defa}K=\tfrac{1}{2}\left(\mu_F^2 +\sigma_F^2 + \mu^2 + \sigma^2- 2 \mu \mu_F-\varepsilon\right) \ge0.
\end{equation}
\end{lemma}
\proof{Proof of Lemma~\ref{lemma:explicitLambda}.} 
Let $h_\lambda(u) \in \mathcal{M}^{-1}(\mu, \sigma)$ defined in \eqref{proof:eqn:h-lambda}. Solving for $\lambda$ such that $d_W(F^{-1}, h_\lambda)^2= \varepsilon$ amounts to solving
\begin{align}\label{interm0}
\varepsilon
	& = \mu_F^2 + \sigma_F^2 + \mu^2 + \sigma^2
		- 2 ~ \cov\left(	F^{-1}(U), h_\lambda(U)\right) - 2 \mu \mu_F.
\end{align}
Using the expression of $K$ in \eqref{defa}, Equation \eqref{interm0} is simply
\begin{equation}\label{comcov}\cov\left(	F^{-1}(U), h_\lambda(U)\right)=K,
\end{equation}
which ensures that $K\geq0$. Using the expression of $h_\lambda$,
\begin{align*}
\cov\left(	F^{-1}(U), h_\lambda(U)\right)
&=\frac{\sigma}{b_\lambda}\cov\left(	F^{-1}(U),  \gamma(U) + \lambda F^{-1}(U)\right)
=\frac{\sigma}{b_\lambda}\Big(\cov\left(	F^{-1}(U),  \gamma(U)\right) + \lambda \sigma_F^2\Big)\,.
\end{align*}
Rearranging, we obtain
\begin{equation}
K b_\lambda=\sigma\cov\left(	F^{-1}(U),  \gamma(U)\right) + \lambda \sigma\sigma_F^2.\label{temp2}
\end{equation}
Next, denote by $V=\var(\gamma(U))$ and $C_{\gamma, F}=\cov\left(	F^{-1}(U),  \gamma(U)\right)$, then the expression of $b_\lambda ={\std(\gamma(U) + \lambda F^{-1}(U))}$ becomes
$b_\lambda^2=V+2\lambda C_{\gamma, F}+\lambda^2\sigma_F^2$
and the square of Equation \eqref{temp2} becomes
$$K^2(V+2\lambda C_{\gamma, F}+\lambda^2\sigma_F^2)=(\sigma C_{\gamma, F} + \lambda \sigma\sigma_F^2)^2.$$
This is a second degree equation in $\lambda$, which can also be written as
\begin{equation}\label{LambdaEQ}
\sigma_F^2\lambda^2+2C_{\gamma, F}\lambda  + \frac{K^2V-\sigma^2 C_{\gamma, F}^2}{K^2-\sigma^2\sigma_F^2}
=
\sigma_F^2\lambda^2+2C_{\gamma, F}\lambda  - \frac{1}{\sigma_F^2}\left(\Delta - C_{\gamma, F}^2\right)
=
0\,, 
\quad \text{as}
\end{equation}
\begin{equation}\label{Deltaproof}
\sigma_F^2\frac{\sigma^2 C_{\gamma, F}^2-K^2V}{K^2-\sigma^2\sigma_F^2}+C_{\gamma, F}^2=\Delta\,.
\end{equation}
Observe that if $\gamma(\cdot)\neq F^{-1}(\cdot)$, then  $C_{\gamma, F}>0$ as the covariance between two comonotonic variables.  $K^2-\sigma^2\sigma_F^2< 0$ by definition of $K$ as a covariance between two variables with respective variances $\sigma^2$ and $\sigma_F^2$ and that are not perfectly correlated, and  $C_{\gamma, F}^2-V\sigma_F^2\leq0$, by definition of $C_{\gamma, F}$ as a covariance between two variables with respective variances $V$ and $\sigma_F^2$. Thus, from the expression \eqref{Deltaproof} we have that $\Delta>0$. As a consequence, \eqref{LambdaEQ} then has two roots, of which only one, $\lambda=\frac{\sqrt{\Delta}-C_{\gamma, F}}{\sigma_F^2}$, is positive.\hfill \Halmos
\endproof

\proof{Proof of Theorem~\ref{thm: concave}.}
For concave distortion risk measures, \eqref{eq:problem sup} is equivalent to 
\begin{equation}\label{eq: problem concave}
\sup_{h \in \mathcal{M}^{-1}_\varepsilon(\mu, \sigma)}\langle \gamma, h \rangle,
\end{equation}
where $\gamma$ is non-decreasing. For $\lambda \geq 0$, let $h_\lambda\in \mathcal{M}^{-1}(\mu, \sigma)$ denote the quantile function defined in \eqref{proof:eqn:h-lambda}. The Wasserstein distance between $F^{-1}$ and $h_\lambda$ is given by
\begin{align*}
d_W(F^{-1}, h_\lambda)^2
	& 
	= (\mu_F - \mu)^2 + (\sigma_F - \sigma)^2 + 2 \sigma \sigma_F(1 - c_\lambda),
\end{align*}
where $c_\lambda = \corr(F^{-1}(U), h_\lambda (U))$. Note that by Lemma~\ref{lemma: concave} $c_\lambda \colon [0, + \infty) \to [-1, 1]$ is continuous with $c_0 = \corr(F^{-1}(U), h_0(U))= \corr(F^{-1}(U), \gamma(U))$ and $\lim_{\lambda \nearrow +\infty}c_\lambda = 1$. 

\textit{Case 1.} Let $(\mu_F - \mu)^2 + (\sigma_F - \sigma)^2 < \varepsilon < (\mu_F - \mu)^2 + (\sigma_F - \sigma)^2 + 2\sigma \sigma_F(1 - c_0)$.
We split this part into three steps. First, we show that the optimal quantile function has to be of the form $h_\lambda$ for some $\lambda \ge 0$. Second, we show that the solution lies at the boundary of the Wasserstein ball. Third, we show uniqueness. 
	
For the first  step, let $h_1 \in \mathcal{M}_{\varepsilon}(\mu, \sigma)$ be a solution to problem \eqref{eq: problem concave} with $d_W(F^{-1}, h_1) = \sqrt{ \varepsilon_1}$ and $(\mu_F - \mu)^2 + (\sigma_F - \sigma)^2 < \varepsilon_1 \le \varepsilon$. Then, by continuity of $c_\lambda$, there exists  $\lambda_1>0$ and a corresponding $h_{\lambda_1}\in \mathcal{M}_{\varepsilon}(\mu, \sigma)$ such that $d_W(F^{-1}, h_{\lambda_1}) = \sqrt{\varepsilon_1}$. Moreover, it holds that
\begin{equation*}
d_W(F^{-1}, h_1)  = d_W(F^{-1}, h_{\lambda_1}) \quad \Leftrightarrow \quad \langle F^{-1}, h_1 \rangle  = \langle F^{-1}, h_{\lambda_1} \rangle \,.
\end{equation*}
Applying Lemma~\ref{lemma: concave}, we obtain $\langle \gamma, h_1 \rangle  \le \langle \gamma, h_{\lambda_1} \rangle $; thus, $h_{\lambda_1}$ improves upon $h_1$ and we conclude the first step. 
	
For the second step, let $\lambda\ge0$ be such that $h_\lambda$ satisfies $d_W(F^{-1}, h_\lambda) = \sqrt{\varepsilon}$. By construction, it holds that
\begin{equation}
\label{pf: concave-F-lambda}
d_W(F^{-1}, h_{\lambda_1}) \le d_W(F^{-1}, h_\lambda) \quad
\Leftrightarrow \quad \langle F^{-1}, h_{\lambda_1} \rangle  \ge \langle F^{-1}\,, h_\lambda \rangle\,. 
\end{equation}
Furthermore, by Lemma~\ref{lemma: concave} we have that $\langle \gamma + \lambda F^{-1}, h_\lambda \rangle \ge \langle \gamma + \lambda F^{-1}, h_{\lambda_1} \rangle $, which, together with \eqref{pf: concave-F-lambda}, implies 
\begin{equation*}
\langle \gamma , h_\lambda - h_{\lambda_1} \rangle 
\ge 
\lambda \,\langle F^{-1}, h_{\lambda_1} - h_\lambda  \rangle 
\ge 0\,.
\end{equation*}
Therefore, we obtain that $\langle \gamma , h_\lambda \rangle  \ge \langle \gamma , h_{\lambda_1} \rangle,$ and hence that the solution lies at the boundary of the Wasserstein ball. Uniqueness follows from the uniqueness of $h_\lambda$ as the solution of Lemma~\ref{lemma: concave}. The worst-case value $H_g(h_\lambda)$ is
\begin{align*}
    H_g(h_\lambda)
    &= \int_0^1 h_\lambda(u) \gamma(u) \mathrm{d}u
    = \mu + \frac{\sigma}{b_\lambda} \left( \int_0^1 \gamma(u) \left(\gamma(u)  + \lambda F^{-1}(u)\right)\mathrm{d}u - a_\lambda\right)\\
    &= \mu + \sigma \std\left(\gamma(U)\right)\corr\left( \gamma(U),\gamma(U)  + \lambda F^{-1}(U) \, \right)\,.
\end{align*}
The explicit expression for $\lambda$ then follows from Lemma~\ref{lemma:explicitLambda} and is given in \eqref{lamSOL}.

\textit{Case 2.} Let $(\mu_F - \mu)^2 + (\sigma_F - \sigma)^2 + 2\sigma \sigma_F (1 - c_0) \leq \varepsilon $; then $h_0 \in \mathcal{M}^{-1}_\varepsilon(\mu, \sigma)$ and thus is admissible. Moreover, it is straightforward that for all $\lambda >0$, $\langle \gamma , h_0 \rangle  \ge \langle \gamma , h_{\lambda} \rangle $. Hence, $h_0$ solves problem \eqref{eq: problem concave}. 
\hfill \Halmos
\endproof

\proof{Proof of Corollary~\ref{cor: properties}.} This result follows directly from Theorem~\ref{thm: concave}, case 2.
\hfill \Halmos
\endproof

\subsection{Proof of Theorem~\ref{thm: general} and Corollary~\ref{cor: momentbound-general}.}
The following lemma is needed to prove Theorem~\ref{thm: general}.
\begin{lemma}\label{proof:lemma: general}
Let $\lambda > 0$ and $g$ be a distortion function with weight function $\gamma$. Then 
\begin{equation*}
\sup_{h \in \mathcal{M}^{-1}(\mu, \sigma)} \langle \gamma + \lambda F^{-1}, h \rangle
\end{equation*}
has a unique solution given by
\begin{equation}\label{proof:eqn:h_lambda_uparrow}
h^{\uparrow}_\lambda(u) = \mu + \sigma  \left(\frac{k_\lambda^\uparrow(u)-a_\lambda}{b_\lambda}\right) \,,\quad 0  < u < 1\,,
\end{equation} 
where $a_\lambda = \E(k_\lambda^\uparrow(U))$ and $b_\lambda =\std(k_\lambda^\uparrow(U))$, and $k_\lambda^\uparrow$ is given in \eqref{eq: isotonic non-decreasing}. Moreover, $\corr(F^{-1}(U), h^{\uparrow}_\lambda(U) )$ is continuous in $\lambda$, for $\lambda > 0$.
\end{lemma}

\proof{Proof of Lemma~\ref{proof:lemma: general}.}
For $\lambda > 0 $, note that the isotonic projection $k_\lambda^\uparrow$ is non-constant, which implies that $h^{\uparrow}_\lambda$, as defined in \eqref{proof:eqn:h_lambda_uparrow}, fulfils $h^{\uparrow}_\lambda \in \mathcal{M}^{-1}(\mu, \sigma)$. Let $h \in \mathcal{M}^{-1}(\mu, \sigma)$; then $k(u) = b_\lambda(h(u) - \mu) / \sigma + a_\lambda$ is a non-decreasing function. Moreover, it holds that $||k_\lambda^\uparrow||^2 = || k||^2$. Thus, the following inequalities are equivalent: 
\begin{align*}
    ||  \gamma + \lambda F^{-1} - k_\lambda^\uparrow||^2  \leq || \gamma + \lambda F^{-1} - k||^2 
 	\quad & \Leftrightarrow  \quad \langle \gamma + \lambda F^{-1}, ~ k_\lambda^\uparrow \rangle 
	 \geq \langle \gamma + \lambda F^{-1}, ~k \rangle
	 \\
	 &\Leftrightarrow  \quad \langle \gamma + \lambda F^{-1}, ~h_\lambda^\uparrow \rangle \geq  \langle \gamma + \lambda F^{-1}, ~h \rangle\,.
\end{align*}
Note that unless $k_\lambda^\uparrow=k$ the inequalities are strict, which implies uniqueness of the solution. Finally, continuity of $\corr(F^{-1}(U), h^{\uparrow}_\lambda(U) ) = \corr(F^{-1}(U), k^{\uparrow}_\lambda(U) )$ follows from continuity of the isotonic projection $k_\lambda$, property 2. in Proposition~\ref{prop: prop isotonic projection}. 
\hfill \Halmos
\endproof

\proof{Proof of Theorem~\ref{thm: general}.}
For $\lambda > 0$, define $h^{\uparrow}_\lambda$ as in \eqref{proof:eqn:h_lambda_uparrow} and note that $h^{\uparrow}_\lambda \in \mathcal{M}^{-1}(\mu, \sigma)$. The Wasserstein distance between $F^{-1}$ and $h^\uparrow_\lambda$ is given by (see also Theorem~\ref{thm: concave})
$d_W(F^{-1}, h_\lambda^\uparrow)^2
	 = (\mu_F - \mu)^2 + (\sigma_F - \sigma)^2 + 2 \sigma \sigma_F(1 - c_\lambda)$,
where $c_\lambda  = \corr(F^{-1}(U), h^\uparrow_\lambda(U))$. By Lemma~\ref{proof:lemma: general}, $c_\lambda \colon (0, + \infty) \to [-1, 1]$ is continuous in $\lambda > 0$, with $\lim_{\lambda \nearrow +\infty}c_\lambda = 1$. 

\textit{Case 1.} Let $(\mu_F - \mu)^2 + (\sigma_F - \sigma)^2 < \varepsilon < (\mu_F - \mu)^2 + (\sigma_F - \sigma)^2 + 2\sigma \sigma_F(1 - c_0)$.	Similar to the proof of Theorem~\ref{thm: concave}, we split the proof into three steps: first, the worst-case quantile function is of the form $h_\lambda$, second, the solution lies at the boundary of the Wasserstein ball, and third we show that it is unique. The first two steps follow along the lines of the proof of Theorem~\ref{thm: concave} case 1., replacing $h_\lambda$ with $h^\uparrow_\lambda$ and using in the second step the properties of the isotonic projection $h_\lambda^\uparrow$. Uniqueness follows from the uniqueness of $h^\uparrow_\lambda$ as the solution to \eqref{proof:lemma: general} in Lemma~\ref{proof:lemma: general}.

\textit{Case 2.} If $ (\mu_F - \mu)^2 + (\sigma_F - \sigma)^2 + 2\sigma \sigma_F(1 - c_{0}) \leq \varepsilon$ and $\gamma^\uparrow$ is not constant, then Lemma~\ref{proof:lemma: general} applies for all $\lambda \ge 0$. Let $h^{\uparrow}_{0}$ be as defined in \eqref{proof:eqn:h_lambda_uparrow} with $\lambda = 0$. Then $h^{\uparrow}_{0}\in \mathcal{M}^{-1}_\varepsilon(\mu, \sigma)$. Moreover, for all $h \in \mathcal{M}^{-1}(\mu, \sigma)$ it holds, applying Lemma~\ref{proof:lemma: general} with $\lambda=0$, that
	$\langle \gamma, h \rangle \leq \langle \gamma, h^{\uparrow}_0 \rangle$.
Hence, $h^{\uparrow}_0$	cannot be improved and thus yields the unique optimal solution. 

If $\gamma^\uparrow$ is constant, then we find that for all $h \in \mathcal{M}^{-1}(\mu, \sigma)$, 
	$\langle \gamma, h \rangle \leq \langle \gamma^\uparrow, h \rangle=\mu\,$,
where the first inequality follows from the properties of the isotonic projection (Proposition~\ref{prop isotonic prop}, property 3.) and the equality follows from the fact that $\gamma^\uparrow =1$ as $\langle \gamma, 1 \rangle =  \langle \gamma^\uparrow, 1 \rangle$ (applying Proposition~\ref{prop isotonic prop}, property 3. with $k  \equiv 1$ and $k  \equiv -1$). Finally, we show that $\mu$ is indeed the supremum. Consider the case in which $\mu \geq 0$ and define $t_\alpha=\alpha\,\mu\, \gamma^\uparrow+(1-\alpha)z_\alpha$, for $0<\alpha<1$, and where $z_\alpha$ is a non-negative non-decreasing function such that  $\E(z_\alpha(U))=\mu$ and $\std(z_\alpha(U))=\sigma/(1-\alpha)$. Hence, $t_\alpha \in \mathcal{M}^{-1}(\mu,\sigma)$, for $0<\alpha<1$. Moreover, for every $\varepsilon >0$ that are sufficiently small, there exists $0<\alpha<1$ such that 
\begin{equation*}
	\langle \gamma, t_\alpha \rangle =\langle \gamma, \,\alpha\,\mu\, \gamma^\uparrow  \rangle + \langle \gamma, \,(1-\alpha)\,z_\alpha  \rangle \geq \alpha \,\mu \geq \mu - \varepsilon\,.
\end{equation*}
For $\mu <0$ the proof is similar and is omitted.	
\hfill \Halmos
\endproof

\proof{Proof of Corollary~\ref{cor: momentbound-general}.}
Corollary~\ref{cor: momentbound-general} follows from Theorem~\ref{thm: general}.
We prove only that equation \eqref{eq:problem sup bpv} (when $\gamma^\uparrow$ is not constant) is equivalent to the worst-case value in Theorem~\ref{thm: general} (case 2.). From the characterisation of the isotonic projection - that is, applying  Property 3. of Proposition~\ref{prop isotonic prop} with $k = \pm 1$ - we obtain that $\langle  \gamma^{\uparrow},\, 1 \rangle=\langle  \gamma,\, 1 \rangle=1$; thus, $\E(\gamma^{\uparrow}(U))=\E(\gamma(U))=1$. Further, by Property 2. of Proposition~\ref{prop isotonic prop}, it holds that $\E\left(\gamma^{\uparrow}(U)^2\right)=\E\left(\gamma^{\uparrow}(U)\gamma(U)\right)$. Hence, 
\begin{equation*}
\corr\left(\gamma(U), \gamma^{\uparrow}(U)\right)
    = \frac{\cov\left(\gamma^{\uparrow}(U),\gamma(U)\right)}{\std\left(\gamma^{\uparrow}(U)\right) \std\left(\gamma(U)\right)}
    = \frac{\var\left(\gamma^{\uparrow}(U)\right)}{\std\left(\gamma^{\uparrow}(U)\right) \std\left(\gamma(U)\right)}
    = \frac{\std\left(\gamma^{\uparrow}(U)\right)}{ \std\left(\gamma(U)\right)}\,.
\end{equation*}
\hfill \Halmos
\endproof

\subsection{Proof of Theorem~\ref{thm: lower bound}.}
The following lemma is needed for the proof of Theorem~\ref{thm: lower bound}.
\begin{lemma}\label{lemma: lower bound}
Let $\lambda > 0$ and $g$ be a distortion function. Then 
\begin{equation*}
\inf_{h \in \mathcal{M}^{-1}(\mu, \sigma)} \langle \gamma - \lambda F^{-1}, h \rangle
\end{equation*}
has a unique solution given by 
\begin{equation}\label{proof:eqn:h_lambda_downarrow}
h^{\downarrow}_\lambda(u) = \mu + \sigma  \left(\frac{a_\lambda-k_\lambda^\downarrow(u)}{b_\lambda}\right) ,\quad 0  < u < 1\,,
\end{equation} 
where $a_\lambda = \E\left(k_\lambda^\downarrow(U)\right)$ and $b_\lambda =\std\left(k_\lambda^\downarrow(U)\right)$,
and $k_\lambda^\downarrow$ is given in \eqref{eq: isotonic non-increasing}. Moreover, $\corr(F^{-1}(U), h^{\downarrow}_\lambda(U) )$ is continuous in $\lambda$, for $\lambda > 0$.
\end{lemma}

\proof{Proof of Lemma~\ref{lemma: lower bound}.}
For $\lambda > 0 $, note that $k_\lambda^\downarrow$ is not constant and that $h^{\downarrow}_\lambda$, as defined in \eqref{proof:eqn:h_lambda_downarrow}, fulfils $h^{\downarrow}_\lambda \in \mathcal{M}^{-1}(\mu, \sigma)$. Let $h \in \mathcal{M}^{-1}(\mu, \sigma)$; then $k(u) = b_\lambda\left(\mu - h(u) \right)/ \sigma + a_\lambda$ is a non-increasing function. Moreover, it holds that $||k_\lambda^\downarrow||^2 = || k||^2$. Thus, the following inequalities are equivalent: 
\begin{align*}
|| \gamma - \lambda F^{-1} - k_\lambda^\downarrow||^2  \leq || \gamma - \lambda F^{-1} - k||^2 
\quad &\Leftrightarrow  \quad 
\langle \gamma - \lambda F^{-1}, ~ k_\lambda^\downarrow \rangle \geq \langle \gamma - \lambda F^{-1}, ~k \rangle
\\
&\Leftrightarrow  \quad \langle \gamma - \lambda F^{-1}, ~h_\lambda^\downarrow \rangle \leq  \langle \gamma - \lambda F^{-1}, ~h \rangle\,.
\end{align*}
Note that unless $k_\lambda^\downarrow = k$ the inequalities are strict, which implies that the solution is unique. Finally, continuity of $\corr(F^{-1}(U), h^{\downarrow}_\lambda(U) ) = \corr(F^{-1}(U), k^{\downarrow}_\lambda(U) )$ follows from continuity of the isotonic projection $k_\lambda^\downarrow$; that is, Proposition~\ref{prop: prop isotonic projection}, property 2. 
\hfill \Halmos
\endproof

\proof{Proof of Theorem~\ref{thm: lower bound}.}
For $\lambda > 0 $, define $h^{\downarrow}_\lambda$ as in \eqref{proof:eqn:h_lambda_downarrow} and observe that  $h^{\downarrow}_\lambda \in \mathcal{M}^{-1}(\mu, \sigma)$. Then,
$d_W(F^{-1}, h^{\downarrow}_\lambda)^2
	 = (\mu_F - \mu)^2 + (\sigma_F - \sigma)^2 + 2 \sigma \sigma_F(1 - c_\lambda)$,
where $c_\lambda =  \corr(F^{-1}(U), h^{\downarrow}_\lambda(U))$. By Lemma~\ref{lemma: lower bound}, $c_\lambda \colon (0, + \infty) \to [-1, 1]$ is continuous in $\lambda, ~\lambda > 0$, with $\lim_{\lambda \nearrow +\infty}c_\lambda = 1$. The remainder of the proof follows along the arguments of the proof of Theorem~\ref{thm: general}.
\hfill \Halmos
\endproof

\subsection{Proofs of Corollaries \ref{cor: TVaR worst-case} \& \ref{cor: best-case TVaR} and Propositions \ref{cor: isotonic projection RVaR} \& \ref{cor: isotonic projection RVaR lower}.}

\proof{Proof of Corollary~\ref{cor: TVaR worst-case}.}
These results are well-known and we provide here a proof using the statements derived in this paper. First, we prove the worst-case value of the TVaR. Recall that its weight function $\gamma(u) = \frac{1}{1 - \alpha}\Id_{(\alpha, 1)}(u)$ is non-decreasing and, hence, $\gamma^\uparrow(u) = \gamma(u)$ for all $u\in(0,1)$. Applying Corollary~\ref{cor: momentbound-general} we obtain
    $\int_0^1 \left(\gamma^\uparrow(u) - 1\right)^2 \, \mathrm{d}u  = \frac{\alpha}{1 - \alpha}$
and therefore
\begin{equation*}
\sup_{G \in \mathcal{M}(\mu, \sigma)} \text{TVaR}_\alpha(G)
	= \mu  + \sigma\sqrt{\,\tfrac{\alpha}{1 - \alpha}\,}\,.
\end{equation*}
Next, we calculate the worst-case value of $\text{RVaR}_{\alpha , \beta}$ with weight function $\gamma(u) = \frac{1}{\beta - \alpha} \Id_{(\alpha , \beta]}(u)$. To apply Corollary~\ref{cor: momentbound-general}, we first derive the isotonic projection of $\gamma$. Using similar arguments as in the proof of Proposition~\ref{cor: isotonic projection RVaR} below, we find that the isotonic projection of $\gamma(u)$ is equal to $\gamma(u)$ on $u \in (0, \alpha]$ and constant on $(\alpha, 1)$, on which it takes the value $c>0$, which must be such that $\langle \gamma, 1 \rangle =  \langle \gamma^\uparrow, 1 \rangle=1$. Hence, $c=\frac{1}{1 - \alpha}$, and the worst-case $\text{RVaR}_{\alpha, \beta}$ equals that of $\text{TVaR}_\alpha$. Consider the right-continuous quantile function $H^{-1}$ taking the value $\mu-\sigma\sqrt{\frac{1-\alpha}{ \alpha}}$ on $(0,\alpha)$, and the value $\mu+\sigma\sqrt{\frac{\alpha}{ 1-\alpha}}$ on $[\alpha,1)$, and observe that it has mean $\mu$ and standard deviation $\sigma$. As $\text{VaR}_{\alpha}^+(H^{-1})= \mu  + \sigma\sqrt{\,\frac{\alpha}{1 - \alpha}\,},$ it follows that   
\begin{equation*}
\sup_{G \in \mathcal{M}(\mu, \sigma)} \text{VaR}_\alpha^+(G)
	= \mu  + \sigma\sqrt{\,\tfrac{\alpha}{1 - \alpha}\,}\,.
\end{equation*}
Finally, assume that $ \sup_{G \in \mathcal{M}(\mu, \sigma)} \text{VaR}_\alpha(G)
	= \mu  + \sigma\sqrt{\,\frac{\alpha}{1 - \alpha}\,}-\delta$, for some $\delta >0$.
One can easily construct a left-continuous quantile function $H^{-1}$ having mean $\mu$ and standard deviation $\sigma$, such that $\text{VaR}_\alpha(H) >  \mu  + \sigma\sqrt{\,\frac{\alpha}{1 - \alpha}\,}-\delta$. Hence, \begin{equation*}
\sup_{G \in \mathcal{M}(\mu, \sigma)} \text{VaR}_\alpha(G)
	= \mu  + \sigma\sqrt{\,\tfrac{\alpha}{1 - \alpha}\,}\,.
\end{equation*}
\hfill \Halmos
\endproof

\proof{Proof of Corollary~\ref{cor: best-case TVaR}.}
The results for the case of $\text{RVaR}_{\alpha,\beta}$, $\text{VaR}_\alpha,$ and $\text{VaR}_\alpha^+,$  $0<\alpha<\beta<1$  follow from Remark~\ref{remark1} and Corollary~\ref{cor: momentbound-general} in a straightforward manner. 

For the lower bound on $\text{TVaR}_\alpha, 0<\alpha<1$, observe that the isotonic projection $\gamma^\downarrow(u)$ of $\gamma(u) = \frac{1}{1 - \alpha}\Id_{(\alpha, 1)}(u)$ is constant, and thus that $\gamma^\downarrow =1$, as $\langle \gamma, 1 \rangle =  \langle \gamma^\downarrow, 1 \rangle$ (applying Proposition~\ref{prop isotonic prop}, property 3. with $k  \equiv 1$ and $k  \equiv -1$). Hence, 
\begin{equation*}
\inf_{G \in \mathcal{M}(\mu, \sigma)} \text{TVaR}_{\alpha}(G) = \mu\,.    
\end{equation*}
\endproof

\proof{Proof of Proposition~\ref{cor: isotonic projection RVaR}.}
By Theorem~\ref{thm: general} case 1., we have to calculate the isotonic projection of $\frac{1}{\beta - \alpha}\Id_{(\alpha, \beta]} +\lambda F^{-1}$ onto the space of non-decreasing functions. According to Lemma 5.1 in \cite{Brighi1994SIAM}, the isotonic projection has the form
\begin{equation}\label{pf: RVAR isotonic}
k^{\uparrow}_\lambda(u) = \left(\tfrac{1}{\beta - \alpha}\Id_{(\alpha, \beta]}(u) + \lambda F^{-1}(u)\right) \Id_{(\bigcup_{j\in \mathcal{J}} I_j)^\complement} (u)+ \sum_{j \in \mathcal{J}} c_j \Id_{I_j}(u)\,, \quad 0<u<1\,,
\end{equation}
where $\mathcal{J}$ is a countable index set, and $I_j \subset [0,1]$, $j \in \mathcal{J}$, are disjoint intervals, and for any $j \in \mathcal{J}$, $I_j^\complement= [0,1] \backslash I_j$ denotes the complement of the set $I_j$.

We first show that $\mathcal{J}$ contains only one element. For this, let $k^*$ be an optimal solution with representation given by \eqref{pf: RVAR isotonic}. By definition $k^*$ minimises $\lVert \frac{1}{\beta - \alpha}\Id_{(\alpha, \beta]} + \lambda F^{-1} - k \rVert_2^2$ over all non-decreasing functions $k$. Moreover, 
\begin{align}\label{pf: RVaR isotonic distance}
\Big\lVert \tfrac{1}{\beta - \alpha}\Id_{(\alpha, \beta]} + \lambda F^{-1} - k^* \Big\rVert^2_2
	= \sum_{j \in \mathcal{J}} \;\int_{I_j} \left(\tfrac{1}{\beta - \alpha}\Id_{(\alpha, \beta]}(u) + \lambda F^{-1}(u) - c_j\right)^2\mathrm{d}u\,.
\end{align}

\textit{Case 1}: 
	If $I_{j} \cap  (\alpha, \beta]= \emptyset$, then the $j^{\text{th}}$ summand in \eqref{pf: RVaR isotonic distance} can be improved by choosing $k^*(u) =\min\{\,  \lambda F^{-1}(u)\, , \, k^*(u_j^+)\,\}$, for all $u \in I_j$, and where $u_j^+$ denotes the right endpoint of $I_j$.
	
\textit{Case 2}:
	If $I_j \subset  [\alpha, \beta)$, then the $j^{\text{th}}$ summand in \eqref{pf: RVaR isotonic distance} can be improved by setting $k^*(u) = \min\{\,\frac{1}{\beta - \alpha} +  \lambda F^{-1}(u)\, , \, k^*(u_j^+)\, \}$ on $I_j$, and where $u_j^+$ denotes the right endpoint of $I_j$.
Therefore, $|\mathcal{J}| = 1$ and the isotonic projection $k^*$ is constant on one interval $I = (w_0, w_1]$, with $\alpha \leq w_0 \leq \beta \leq w_1$. Define $c = k^*(u)$, $u \in I$. Then, $c$ crosses $\lambda F^{-1}(u)$ at $u=w_1$ and, moreover, $c$ crosses $1 / (\beta - \alpha) + \lambda F^{-1}(u)$ when $u=w_0$, provided the jump $1 / (\beta - \alpha)$ is not too large. More precisely, for a given $c$, $w_0$ and $w_1$ satisfy
\begin{align*}
\lambda F^{-1}(w_0) &= 
\begin{cases}
c  - \frac{1}{\beta - \alpha}, \quad \quad & \text{if} ~ \frac{1}{\beta - \alpha} \leq c - \lambda  F^{-1}(\alpha)\,,\\[0.5em]
\lambda F^{-1}(\alpha), 			
& \text{otherwise}\,,
\end{cases}
\quad \text{and}\quad
\lambda F^{-1}(w_1) = 
\begin{cases}
\lambda F^{-1}(1) \quad \quad & \text{if} ~ c > F^{-1}(1)\,,\\
c & \text{otherwise}\,,
\end{cases}
\end{align*}

The optimal $c$, which minimises \eqref{pf: RVaR isotonic distance}, fulfils
\begin{align*}
& \argmin_{w_0, w_1, c }~\int_{w_0}^{w1}\left(\tfrac{1}{\beta - \alpha}\Id_{(\alpha , \beta]}(u) + \lambda F^{-1}(u) - c\right)^2\mathrm{d}u
\\
	&\quad= \argmin_{w_0, w_1, c } ~ - 2c \left(\frac{\beta - w_0}{\beta - \alpha} + \lambda \int_{w_0}^\beta F^{-1}(u) \,\mathrm{d}u\right) + c^2 (w_1 - w_0)\,.
\end{align*}
Thus, the optimal $c, w_0$, and $w_1$ fulfil
$c = \frac{1}{w_1 - w_0}\frac{\beta - w_0}{\beta - \alpha} + \frac{\lambda}{w_1 - w_0} \int_{w_0}^{w_1} F^{-1}(u)\, \mathrm{d}u$.
\hfill \Halmos
\endproof

\proof{Proof of Proposition~\ref{cor: isotonic projection RVaR lower}.}
The proof follows analogously to the proof of Proposition~\ref{cor: isotonic projection RVaR} and is omitted. 
\hfill \Halmos
\endproof

\proof{Proof of Corollary~\ref{cor: VaR TVaR}.}
For the worst-case value, let $G^{-1,*}$ denote the quantile function given in Proposition~\ref{cor: isotonic projection RVaR} with $\beta = 1$, $w_0 = 1$, and $w_1 = 1$. Direct calculations show that $G^{-1,*}$ indeed attains the supremum given in Corollary~\ref{cor:TVaR-WC-explicit}. For the best-case value, recall that $\text{TVaR}_\alpha = \lim_{\beta \nearrow 1}\text{RVaR}_{\alpha, \beta}$ and that $\text{RVaR}_{\alpha, \beta}$ is an increasing function in $\beta$. Thus, by Theorem~\ref{thm: lower bound}
\begin{equation}\label{eq:sup-TVaR-lim-sup-RVaR}
\inf_{G\in\mathcal{M}_\varepsilon(\mu, \sigma)} \text{TVaR}_\alpha 
\ge
\lim_{\beta \nearrow 1} \inf_{G\in\mathcal{M}_\varepsilon(\mu, \sigma)}\text{RVaR}_{\alpha, \beta}
=
  \lim_{\beta \nearrow 1}
  \mu  - \frac{\sigma}{b_\lambda} \left(\E\left(\frac{1}{\beta - \alpha}\Id_{\{U \in (\alpha, \beta]\}} k_\lambda^\downarrow(U)\right) - a_\lambda\right)\,,
\end{equation}
where $a_\lambda = \E(k_\lambda^\downarrow(U))$, $b_\lambda = std(k_\lambda^\downarrow(U))$, and $k_\lambda^\downarrow$ is given in Proposition~\ref{cor: isotonic projection RVaR lower}. Tedious calculations show that the limit in \eqref{eq:sup-TVaR-lim-sup-RVaR} exists and is equal to $\text{TVaR}_\alpha(G^*)$, where $G^*$ has a quantile function given in Proposition  \ref{cor: isotonic projection RVaR lower} with $\beta= 1$.
\hfill \Halmos
\endproof

\proof{Proof of Proposition~\ref{prop: VaR TVaR2}.}
We first consider the worst-case of $\text{VaR}_\alpha^+$ and second that of $\text{VaR}_\alpha$.\\
To show the worst-case value of $\text{VaR}_\alpha^+$ we split the proof into three steps.\\
\textit{Step 1:} We show that for any $\eta,\delta >0 $ and $m \in \mathbb{R}$ the unique quantile function attaining 
\begin{equation}\label{eq:pf-var+-wasser-only}
    \sup_{d_W( G^{-1},\eta F^{-1} - m) \le  \sqrt{\delta}} \; \text{VaR}_\alpha^+(G)\,.
\end{equation}
is given by
\begin{equation}\label{eq:pf-var+-wasser-only-H*}
    H_{\eta,\delta,m}^{*, -1}(u) = \left(\eta F^{-1}(u) - m \right)
    + \left( c - \eta F^{-1}(u)+m\right)\Id_{\{u \in (\alpha, w_1]\}}\,,
\end{equation}
where $w_1 \in (0,1]$ and $c \in \mathbb{R}$ are the unique solutions to 
\begin{equation*}
    \delta
    =
    \int_\alpha^{w_1} \left(c - \eta F^{-1}(u)+m\right)^2\, du  \,.
\end{equation*} 
First, we show that the solution to \eqref{eq:pf-var+-wasser-only}, denoted here by $G^*$, lies at the boundary of the Wasserstein ball, i.e., $d_W(G^{*, -1}, F^{-1}) = \sqrt{\delta}$. Assume by contradiction that $G^*$ lies in the interior of the Wasserstein ball; then there exists a $\delta_0>0$ such that $d_W(G^{*,-1} + \delta_0, F^{-1}) \le \delta$ and $VaR_\alpha^+(G^* + \delta) = VaR_\alpha^+(G^*) + \delta > VaR_\alpha^+(G^*) $, which is a contradiction to the optimality of the solution. Next, we show that the optimal solution to \eqref{eq:pf-var+-wasser-only} has to be of the form \eqref{eq:pf-var+-wasser-only-H*}. For this, assume that $G^*$ is a solution with $VaR_\alpha^+(G^*) > VaR_\alpha^+(H^*_{\eta, \delta, m}) = c$. By non-decreasingness of $G^*$ it holds that $G^{*,-1}(u) > c \ge \eta F^{-1}(u) - m$, for all $u \in [\alpha, w_1]$. Thus, the Wasserstein distance  has a lower bound given by
\begin{align*}
    d_W(G^{*,-1}, \eta F^{-1} - m)^2
    &\ge
    \int_\alpha^{w_1} \left(G^{*,-1}(u) - \eta F^{-1}(u) + m\right)^2du
     > 
    \int_\alpha^{w_1} \left(c - \eta F^{-1}(u) + m\right)^2du
    \\
    &= 
    d_W(H^*_{\eta, \delta, m}, F)^2 = \delta\,,
\end{align*}
which is a contradiction to the fact that the optimal solution lies at the boundary of the Wasserstein ball. Uniqueness of the solution follows from the uniqueness of $\delta$ and $w_1$.

\textit{Step 2:}
Denote by $\eta^*$, $\delta^*$, and $m^*$  the unique parameters such that $H_{\eta^*,\delta^*,m^*}^{*, -1}$ 
has mean $\mu$, variance $\sigma^2$, and Wasserstein distance $d_W(  H_{\eta^*,\delta^*,m^*}^{*},F)=\sqrt\epsilon$. These indeed exist as the three equations for the optimal $\eta^*$, $\delta^*$, and $m^*$ are not linearly dependent. 
Next, we claim that $H_{\eta^*,\delta^*,m^*}^{*}$  is the solution to our problem:
\begin{equation}\label{eqtoto-a}
    \sup_{\stackrel{G \in \mathcal{M}(\mu, \sigma)}{d_W(G^{-1}, F^{-1}) \le \sqrt{\epsilon}}} \; \text{VaR}_\alpha^+(G)\,.
\end{equation}
To show this, note that problem \eqref{eqtoto-a} is equivalent to
\begin{equation}\label{eq:pf-var+-alternative-a}
    \sup_{\stackrel{G \in \mathcal{M}(\mu, \sigma)}{d_W(G^{-1},\eta^* F^{-1} - m^*) \le \sqrt{\delta^*}}} \; \text{VaR}_\alpha^+(G)\,.
\end{equation}
Indeed, a quantile function $H^{-1}$ is admissible to \eqref{eqtoto-a} if and only if it is admissible to \eqref{eq:pf-var+-alternative-a}. This is because $H^{-1}$ and $H_{\eta^*,\delta^*,m^*}^{*, -1}$ have the same mean and variance, which implies that $d_W(  H^{-1}, F^{-1}) \leq d_W(  H_{\eta^*,\delta^*,m^*}^{*, -1}, F^{-1})=\sqrt\epsilon$ if and only if $\corr(H_{\eta^*,\delta^*,m^*}^{*, -1}(U),F^{-1}(U)) \leq \corr(H^{-1}(U),F^{-1}(U))$ if and only if $\corr(H_{\eta^*,\delta^*,m^*}^{*, -1}(U),\eta^* F^{-1}(U) - m^*) \leq \corr(H^{-1}(U),\eta^* F^{-1}(U) - m^*)$ if and only if $d_W(H^{-1},\eta^* F^{-1} - m^*) \leq d_W(H_{\eta^*,\delta^*,m^*}^{*, -1},\eta^* F^{-1} - m^*) = \sqrt{\delta^*}$. However, $H_{\eta^*,\delta^*,m^*}^{*, -1}$ is the unique solution to \eqref{eq:pf-var+-alternative-a} (because it solves \eqref{eq:pf-var+-wasser-only} with $\eta=\eta^*$, $m=m^*$, and $\delta=\delta^*$). Hence, $H^{-1}$ cannot be a solution to \eqref{eqtoto-a} unless it coincides with $H_{\eta^*,\delta^*,m^*}^{*, -1}$.  

\textit{Step 3:} Finally, we can verify that $H_{\eta^*,\delta^*,m^*}^{*, -1}$ coincides with the solution as stated in the statements of the corollary. For this, note first that both $H_{\eta^*,\delta^*,m^*}^{*, -1}$ and $k_\lambda^\uparrow$ are affine functions of $F^{-1}$, and we only need to show that for every $c_0 \leq \rho<1$ there exist $\lambda>0$ such that  $\corr(F^{-1}(U), k_\lambda^\uparrow (U))=\rho.$ Let us first consider the case in which $\rho$ is close to one. First, assume that $F^{-1}$ is constant on $(\alpha,w_1^*]$, where $w_1^*=\sup\{ \,\beta \in (\alpha,1) | F^{-1}(\beta)=F^{-1}(\alpha)\,\}$, then we take $w_1=w_1^*$. By increasing $\lambda$ we can bring $k_\lambda^\uparrow$ arbitrarily close to $\lambda F^{-1}$ on $(\alpha,w_1^*],$ and $k_\lambda^\uparrow$ will coincide with $\lambda F^{-1}$ on $(w_1^*,1]$ as well as on $(0,\alpha].$ Second, assume that $F^{-1}$ is not constant on the right of $\alpha$, then for any $w_1> \alpha$ close to $\alpha$ we can find $\lambda>0$ such that $k_\lambda^\uparrow$ will coincide with $\lambda F^{-1}$ on $(w_1,1]$ as well as on $(0,\alpha].$ Hence, for any value of $\rho$ close to 1, it is possible to specify $k_\lambda^\uparrow$ such that $\corr(F^{-1}(U), k_\lambda^\uparrow (U))=\rho.$ When $\rho=c_0$, we can take $\lambda=0$, i.e., $k_0^\uparrow(u)=c \Id_{\{u \in (\alpha,1]\}}$ and $\corr(F^{-1}(U), k_0^\uparrow (U))=c_0.$ By continuity of  $\corr(F^{-1}(U), k_\lambda^\uparrow (U))$ it thus follows that for all $c_0 \leq \rho<1$ there exist $\lambda>0$ such that  $\corr(F^{-1}(U), k_\lambda^\uparrow (U))=\rho.$

Next,we consider the worst-case $\text{VaR}_\alpha$.
Clearly, the worst-case values of $\text{VaR}_\alpha$ and $\text{VaR}_\alpha^+$ are equal. The worst-case value of $\text{VaR}_\alpha$, however, is not attained. To see this, assume by contradiction that the worst-case $\text{VaR}_\alpha$ is attained by $G^*$. Then, there exists $\delta_0>0$ such that $G^{*,-1}(u) > \eta F^{-1}(u) - m$, for all $u \in (\alpha - \delta, \alpha]$. This, however, implies that $d_W(G^*, F) > d_W(\hat{G}, F) = \sqrt{\delta}$, where $\hat{G}$ attains the worst-case $\text{VaR}_\alpha^+$.

The proof of the best-case values of $\text{VaR}_\alpha$ and $\text{VaR}_\alpha^+$ follows similar steps to the worst-case values. Note that the best-case $\text{VaR}_\alpha$ is attained while the best-case $\text{VaR}_\alpha^+$ is not.
\Halmos
\endproof

\subsection{Proof of Proposition \ref{prop: uncertainty}.}
\begin{lemma}\label{lemma: uncertainty properties}
Let $\lambda > 0$, and $K$ be one of the cases in Proposition \ref{prop: uncertainty}. Then 
\begin{equation*}
\sup_{h \in \mathcal{M}^{-1}(K)} \langle \gamma + \lambda F^{-1}, h \rangle
\end{equation*}
has a unique solution with quantile function
\begin{equation*}
h_\lambda(u) = a_{\max}^K + b_{\max}^K k_{\lambda}^{\uparrow}(u)\,, \quad 0 < u < 1\,,
\end{equation*}
where $a^K_{\max}\in \mathbb{R}$ and $b^K_{\max}>0$ are such that $\int_0^1 h_\lambda (u) \mathrm{d}u = \mu_{\max}^K$ and $\int_0^1 h_\lambda (u)^2 \mathrm{d}u = (\mu_{\max}^K)^2 + (\sigma_{\max}^K)^2$, and $Z_\lambda$, $\text{cv}_\lambda$ are defined in Proposition \ref{prop: uncertainty}. Then,  $(\mu_{\max}^K,~ \sigma_{\max}^K)$ is given by the respective case in Proposition \ref{prop: uncertainty}.
\end{lemma}

\proof{Proof of Lemma \ref{lemma: uncertainty properties}.}
We only prove the case $(\mu_F - \mu)^2 + (\sigma_F - \sigma)^2 < \varepsilon < (\mu_F - \mu)^2 + (\sigma_F - \sigma)^2 + 2\sigma \sigma_F(1 - c_{0})$. By Theorem \ref{thm: general} it holds that
\begin{equation}\label{eq: proof uncertainty}
\sup_{h^{-1} \in \mathcal{M}(K)} \langle \gamma + \lambda F^{-1}, h \rangle 	
 	=  \sup_{(\mu, \sigma) \in K} ~ \mu \E(Z_\lambda) + \sigma \std( Z_\lambda) \corr\big(Z_\lambda, k_\lambda^\uparrow(U)\big)\,.
\end{equation}
As the isotonic projection $k^\uparrow_\lambda$ does not depend on $\mu$ nor $\sigma$, the supremum in \eqref{eq: proof uncertainty} can be taken over $(\mu, \sigma) \in K$. For the three different cases of $K$ as defined in Proposition \ref{prop: uncertainty}, and since $ \corr\big(Z_\lambda, k_\lambda^\uparrow(U)\big) \ge 0$, we have:
\begin{enumerate}
	\item If $\E(Z_\lambda) = 1 + \lambda \mu_F<0$, then the maximum is attained at $(\mu_{\max}^K,\, \sigma_{\max}^K) =(\underline{\mu},\, \overline{\sigma})$ otherwise at $(\mu_{\max}^K,\, \sigma_{\max}^K) =(\overline{\mu},\, \overline{\sigma})$.
	\item This case follows from case 3 with $c = d = 1$.
	\item If $\mu_F < - 1 / \lambda$, $\mu_{\max}^K$ is the negative root of the ellipse. Thus, optimisation \eqref{eq: proof uncertainty} is equivalent to, replacing $\mu$ from the formula of the ellipse,
\begin{align*}
\sup_{0\leq |\sigma_F -\sigma| \leq r}\left( -\,\sqrt{\,c^2r^2 - \frac{c^2}{d^2}\,(\sigma_F - \sigma)^2\,} + \mu_F \right) \E(Z_\lambda) + \sigma \std(Z_\lambda)\corr\big(Z_\lambda, k_\lambda^\uparrow(U)\big)\,.
\end{align*}
Taking the derivative with respect to $\sigma$, the maximum fulfils 
\begin{align*}
(\sigma_F - \sigma )^2
	= 	\text{cv}_\lambda^2 ~ \frac{d^4}{c^4}\left(c^2r^2 - \frac{c^2}{d^2}(\sigma_F - \sigma)^2\right).
\end{align*}
Thus, $\sigma_{\max}^K = \sigma_F + r d\frac{\frac{d}{c}|\text{cv}_\lambda|}{ \sqrt{1 + (\frac{d}{c} \text{cv}_\lambda )^2}}$ and $\mu_{\max}^K = \mu_F - \frac{r c}{ \sqrt{1 + (\frac{d}{c} \text{cv}_\lambda )^2} }$.

If $\mu_F> -1/\lambda$, $\mu_{\max}^K$ is the positive root of the ellipse and the proof follows similar steps.
\end{enumerate}
\hfill \Halmos
\endproof

\proof{Proof of Proposition \ref{prop: uncertainty}.}
Let $K$ be one of the cases in Proposition \ref{prop: uncertainty}. For $\lambda > 0$, define the quantile function $h_\lambda(u) = a_{\max}^K + b_{\max}^K k_\lambda^\uparrow(u)$, $0 < u < 1$, where $a_{\max}^K\in\mathbb{R}$ and $b_{\max}^K>0$ are unique and such that $h_\lambda \in \mathcal{M}^{-1}(\mu_{\max}^K, \sigma_{\max}^K)$, and $\mu_{\max}^K$, $\sigma_{\max}^K$ are given in Proposition \ref{prop: uncertainty}. The Wasserstein distance between $F$ and $h_\lambda^{-1}$ is $ d_W(F^{-1}, h_\lambda)^2 = (\mu_F - \mu_{\max}^K)^2 + (\sigma_F - \sigma_{\max}^K)^2 + 2 \sigma_{\max}^K \sigma_F(1 - c_\lambda)$, where $c_\lambda = \corr(F^{-1}(U), h_\lambda (U))$. The remainder of the poof follows along the arguments of the proof of Theorem \ref{thm: general}.
\hfill \Halmos
\endproof

\subsection{Proof of Theorem~\ref{thm:wasser-only}.}
\begin{lemma}\label{lemma:wasser-only}
Optimisation problem \eqref{opt:wasser-only} can  be equivalently written as
\begin{equation*}
    \sup_{G \in \mathcal{M}_{\varepsilon}} H_g(G)\quad
    = 
    \sup_{\stackrel{(\mu_F - \mu)^2 + (\sigma_F - \sigma)^2 \le \varepsilon}{\sigma >0}} \mu + \sigma std\left(\gamma(U)\right) \, corr\left(\gamma(U), \gamma(U) + \lambda F^{-1}(U) \right)\,,
\end{equation*}
where $\lambda$ is given in Theorem~\ref{thm: concave}.
\end{lemma}
\proof{Proof of Lemma~\ref{lemma:wasser-only}.}
First, we define the set
\begin{equation*}
    \mathcal{N}_{\varepsilon}
    = 
    \left\{ G \in \mathcal{M}_{\varepsilon}(\mu, \sigma) ~|~ (\mu_F - \mu)^2 + (\sigma_F - \sigma)^2 \le \varepsilon, \,\sigma >0 \right\}\,
\end{equation*}
and show that $\mathcal{N}_\varepsilon = \mathcal{M}_\varepsilon$. Clearly, if $G\in\mathcal{N}_\varepsilon $, then $G\in\mathcal{M}_\varepsilon $. Conversely, let $G \in \mathcal{M}_\varepsilon$ with mean $\mu_G$ and standard deviation $\sigma_G>0$; then the Wasserstein distance between $F$ and $G$ fulfils
$(\mu_F - \mu_G)^2 + (\sigma_F - \sigma_G)^2 \le d_W(F,G)^2 \le \varepsilon$.    
Thus, $G\in\mathcal{N}_\varepsilon $ and we indeed obtain that $\mathcal{M}_{\varepsilon} = \mathcal{N}_\varepsilon$. 
Applying Theorem~\ref{thm: concave} case 1.\, concludes the proof.
\Halmos
\endproof

\proof{Proof of Theorem~\ref{thm:wasser-only}.}
By Lemma~\ref{lemma:wasser-only} we only have to solve 
\begin{equation*}
\sup_{\stackrel{(\mu_F-\mu)^2+(\sigma_F-\sigma)^2\le\varepsilon }{\sigma >0}}
\Theta(\mu,\sigma)\,,
\quad \text{where} \quad
\Theta(\mu,\sigma)=\mu+\sigma\std(\gamma(U))\corr(\gamma(U),\gamma(U)+\lambda F^{-1}(U))\,.
\end{equation*}
We can rewrite $\Theta$ as follows, using the expression of $\lambda$ in \eqref{lamSOL}:
\begin{equation*}
\Theta(\mu,\sigma)=\mu+\frac{\sigma}{b_\lambda}\left(V+\frac{\sqrt{\Delta}-C_{\gamma, F}}{\sigma_F^2} C_{\gamma, F}\right),    
\end{equation*}
where  $b_\lambda$ can be simplified as $b_\lambda=\frac{\sigma\sqrt{\Delta}}{K}$ using \eqref{lamSOL} and \eqref{temp2}. Thus,
$$\Theta(\mu,\sigma)
=\mu+\frac{K\left(V\sigma_F^2+C_{\gamma, F}\sqrt{\Delta}-C_{\gamma, F}^2\right)}{\sqrt{\Delta}\sigma_F^2}.$$
Furthermore, recall that $\Delta=\frac{K^2(C_{\gamma, F}^2-V\sigma_F^2)}{K^2-\sigma^2\sigma_F^2}$ and   that $K\geq0$ since $K$ is the covariance between two comonotonic variables in \eqref{comcov}. We therefore obtain that
\begin{align*}
\Theta(\mu,\sigma)
&=
\mu+\frac{C_{\gamma, F}K+\sqrt{{V\sigma_F^2-C_{\gamma, F}^2}}\sqrt{\sigma^2\sigma_F^2-K^2}}{\sigma_F^2}
=
\mu+\frac{C_{\gamma, F}K}{\sigma_F^2}+\sqrt{{V-\frac{C_{\gamma, F}^2}{\sigma_F^2}}}\sqrt{\sigma^2-\frac{K^2}{\sigma_F^2}}.
\end{align*}
Observe that from  the expression of $K$ in \eqref{defa}, $K=\frac{\mu_F^2 +\sigma_F^2 + \mu^2 + \sigma^2- 2 \mu \mu_F-\varepsilon}{2}$, we have
\begin{equation*}
\frac{\partial K}{\partial \mu}=\mu-\mu_F\,,
\quad
\frac{\partial K}{\partial \sigma^2}=\frac{1}{2}\,,
\quad \text{and} \quad    
\frac{\partial \Theta(\mu,\sigma)}{\partial K}=\frac{C_{\gamma, F}}{\sigma_F^2}-\frac{K \sqrt{V-\frac{C_{\gamma, F}^2}{\sigma_F^2}}}{\sigma_F^2\sqrt{\sigma^2-\frac{K^2}{\sigma_F^2}}}.
\end{equation*}
Thus, when differentiating $\Theta$ with respect to $\mu$ interpreted as a function of $(\mu,K(\mu))$, we obtain
\begin{equation}\label{diffmu}\frac{\partial \Theta(\mu,\sigma)}{\partial \mu}=1+\left(\frac{C_{\gamma, F}}{\sigma_F^2}-\frac{K \sqrt{V-\frac{C_{\gamma, F}^2}{\sigma_F^2}}}{\sigma_F^2\sqrt{\sigma^2-\frac{K^2}{\sigma_F^2}}}\right)(\mu-\mu_F),\end{equation}
and when differentiating $\Theta$ with respect to $\sigma^2$, we obtain
\begin{equation}\label{diffsig}\frac{\partial \Theta(\mu,\sigma)}{\partial \sigma^2}=\left(\frac{1}{2}-\frac{K }{2\sigma_F^2}\right)\frac{ \sqrt{V-\frac{C_{\gamma, F}^2}{\sigma_F^2}}}{\sqrt{\sigma^2-\frac{K^2}{\sigma_F^2}}}+\frac{C_{\gamma, F}}{2\sigma_F^2}\,.
\end{equation}
To solve $\frac{\partial \Theta(\mu,\sigma)}{\partial \sigma^2}=0$, we first use that  $\sigma^2=2K+\varepsilon-(\mu-\mu_F)^2-\sigma_F^2$ and then square the equation we aim to solve. There are two roots for $K$ characterised by $K=\sigma_F^2\pm\frac{C_{\gamma, F}}{\sqrt{V}}\sqrt{\varepsilon-(\mu-\mu_F)^2}.$ 
But only one solves the original equation \eqref{diffsig}, which is given by
\begin{equation}\label{Kstar}K=\sigma_F^2+\frac{C_{\gamma, F}}{\sqrt{V}}\sqrt{\varepsilon-(\mu-\mu_F)^2},\end{equation}
which then corresponds to
$\sigma^2=\varepsilon-(\mu-\mu_F)^2+\sigma_F^2+ \frac{2C_{\gamma, F}}{\sqrt{V}}\sqrt{\varepsilon-(\mu-\mu_F)^2}$.
From the fact that $\frac{\partial \Theta(\mu,\sigma)}{\partial \sigma^2}=0$ in \eqref{diffsig}, we have that
$$\frac{ \sqrt{V-\frac{C_{\gamma, F}^2}{\sigma_F^2}}}{\sqrt{\sigma^2-\frac{K^2}{\sigma_F^2}}}=\frac{C_{\gamma, F}}{K-\sigma_F^2}\;.$$
Replacing this expression in \eqref{diffmu}, we find that $\frac{\partial \Theta(\mu,\sigma)}{\partial \mu}=0$ is equivalent to 
$C_{\gamma, F}(\mu_F-\mu)+K-\sigma_F^2=0$.
Using the expression of $K$ in \eqref{Kstar}, this gives
$\sqrt{\varepsilon-(\mu-\mu_F)^2}=(\mu-\mu_F)\sqrt{V}$
which can be solved explicitly and has a unique root:
$$\mu=\mu_F+\sqrt{\frac{\varepsilon}{1+V}}.$$
Recall the equation
$\sigma^2=\varepsilon-(\mu-\mu_F)^2+\sigma_F^2+ \frac{2C_{\gamma, F}}{\sqrt{V}}\sqrt{\varepsilon-(\mu-\mu_F)^2},$
 which can be simplified to the unique point $$(\mu^*,\sigma^*)=\left(\mu_F+\sqrt{\frac{\varepsilon}{1+V}}\ ,\ \sqrt{\sigma_F^2+\frac{2C_{\gamma, F}\sqrt{\varepsilon}}{\sqrt{1+V}}+\frac{\varepsilon V}{1+V}}\right),$$ 
in which $\frac{\partial \Theta(\mu^*,\sigma^*)}{\partial \sigma}=\frac{\partial \Theta(\mu^*,\sigma^*)}{\partial \mu}=0$. Next, we verify that the  Hessian matrix is negative-definite at this point  (negative diagonal coefficients and positive determinant), which ensures that it is a global maximum.

In fact, the Hessian matrix at any point $(\mu,\sigma)$ can be written as
$$\left(\begin{array}{cc}
\frac{\partial^2 \Theta(\mu,\sigma)}{\partial \mu^2} & \frac{\partial^2 \Theta(\mu,\sigma)}{\partial \sigma^2\partial\mu}\\
\frac{\partial^2 \Theta(\mu,\sigma)}{\partial \mu\partial \sigma^2}&\frac{\partial^2 \Theta(\mu,\sigma)}{(\partial \sigma^2)^2} \\
\end{array}\right)=\left(\begin{array}{cc}
\frac{C_{\gamma, F}}{\sigma_F^2}-M\left(\frac{K}{\sigma_F^2}+\frac{(\mu-\mu_F)^2\sigma^2}{\sigma_F^2\sigma^2-K^2}\right) & \quad\quad\frac{M}{\sigma_F^2\sigma^2-K^2}\frac{(K-\sigma^2)(\mu-\mu_F)}{2}\\
\frac{M}{\sigma_F^2\sigma^2-K^2}\frac{(K-\sigma^2)(\mu-\mu_F)}{2} & \quad\quad \frac{M}{\sigma_F^2\sigma^2-K^2}\frac{(\mu-\mu_F)^2-\varepsilon}{4}\\
\end{array}\right),$$
where $M= \sqrt{V-\frac{C_{\gamma, F}^2}{\sigma_F^2}}/\sqrt{\sigma^2-\frac{K^2}{\sigma_F^2}}$.
Evaluating this Hessian matrix at  $(\mu^*,(\sigma^*)^2)$, it can be simplified to
$$\left(\begin{array}{cc}
\frac{\partial^2 \Theta(\mu,\sigma)}{\partial \mu^2} & \frac{\partial^2 \Theta(\mu,\sigma)}{\partial \sigma^2\partial\mu}\\
\frac{\partial^2 \Theta(\mu,\sigma)}{\partial \mu\partial \sigma^2}&\frac{\partial^2 \Theta(\mu,\sigma)}{(\partial \sigma^2)^2} \\
\end{array}\right)=\left(\begin{array}{cc}
-\frac{\sqrt{1+V}}{\sqrt{\varepsilon}}-\frac{\sqrt{\varepsilon}}{\sqrt{1+V}}\frac{(\sigma^*)^2}{\sigma_F^2(\sigma^*)^2-K^2} & \quad\quad\frac{1}{\sigma_F^2(\sigma^*)^2-K^2}\frac{K-(\sigma^*)^2}{2}\\
\frac{1}{\sigma_F^2(\sigma^*)^2-K^2}\frac{K-(\sigma^*)^2}{2} & \quad\quad \frac{-1}{\sigma_F^2(\sigma^*)^2-K^2}\frac{V\sqrt{\varepsilon}}{4\sqrt{1+V}}\\
\end{array}\right).$$
The diagonal coefficients are clearly negative and the determinant $$\frac{V}{4(\sigma_F^2(\sigma^*)^2-K^2)}+\frac{\varepsilon(V\sigma_F^2-C_{\gamma, F}^2)}{4(\sigma_F^2(\sigma^*)^2-K^2)^2(1+V)}$$ is positive, which ensures that $\Theta(\mu,\sigma)$ is locally concave at $(\mu^*,(\sigma^*)^2)$, which completes the proof.
\Halmos
\endproof

\proof{Proof of Proposition~\ref{Propopt}.} 

The general projection property (Theorem 1) in \cite{Popescu2007OR} shows that for every portfolio $\x \in \mathcal{P}$ and every distribution function $G_{\x}$ having mean $-\mu_\x=-\boldsymbol{\x}^\top \boldsymbol{\mu}$ and variance $\sigma^2_\x=\x^\top \Sigma \x$ there exists a multivariate distribution function ${G_\Rr \in \M(\boldsymbol{\mu}, \Sigma)}$ such that the portfolio loss $-\x^\top \Rr$  is distributed with $G_{\x}.$  As a consequence, optimisation \eqref{opt: portfolio} is equivalent to 
\begin{equation} \label{opt: portfolio-univ}
    \min_{\x \in \mathcal{P}}\; 
    \max_{\stackrel{G_\x \in \M(-\mu_\x, \sigma_\x)}{d_W(F_\x,G_\x ) \le \sqrt{\varepsilon_\x}}} 
    H_\gamma \left(G_\x\right)\,
\end{equation}
where $\M(-\mu_\x, \sigma_\x)$ denotes the set of univariate distributions with mean $-\mu_\x$ and standard deviation $\sigma_\x$. Applying Theorem~\ref{thm: concave} yields that \eqref{opt: portfolio-univ} is equivalent to the minimisation problem 
\begin{equation} \label{opt: portfolio-univ2}
    \min_{\x \in \mathcal{P}} \;\left\{\;-\mu_\x + \frac{\sigma_\x}{b_{\lambda_\x}} \left(V + \lambda_\x C_\x \right)\; \right\},
\end{equation}
in which\footnote{Note that when ${\varepsilon_\x} \geq 2\sigma_\x^2(1-c_0)$, i.e., when $\underline{\varepsilon_\x}=2\sigma_\x^2(1-c_0)$, it follows that $K_\x=\sigma_\x^2c_0$ and thus that $\lambda_\x=0$.} $\lambda_\x=\frac{K_\x}{\sigma_\x^2}\sqrt{\frac{{V\sigma_\x^2-{C_\x^2}}}{{\sigma_\x^4-{K_\x}^2}}}-\frac{C_\x}{\sigma_\x^2}$
with 
$C_\x= \cov(\gamma(U),F_{\x}^{-1}(U))$, $K_\x=\sigma_\x^2-\frac{1}{2}{\underline{\varepsilon_\x}}$ 
, and $b_{\lambda_\x}=\sigma_\x\sqrt{\frac{{V\sigma_\x^2-{C_\x}^2}}{{\sigma_\x^4-{K_\x}^2}}}$, where we used equation \eqref{temp2} to obtain the expression for $b_{\lambda_\x}$. Note that $C_\x =\sigma_\x \widetilde{C}$, where $\tilde{C} = \cov(\gamma(U),F^{-1}(U))$. Then, minimisation \eqref{opt: portfolio-univ2} becomes
\begin{equation*}
\min_{\x \in \mathcal{P}} \;\left\{\;-\mu_\x +\frac{V{\sqrt{\sigma_\x^4-{K_\x}^2}}}{\sigma_\x{\sqrt{V-\widetilde{C}^2}}}+ \frac{{\sqrt{\sigma_\x^4-{K_\x}^2}}}{\sigma_\x{\sqrt{V-\widetilde{C}^2}}}
\left(\frac{K_\x}{\sigma_\x}\frac{\sqrt{V-\widetilde{C}^2}}{\sqrt{\sigma_\x^4-{K_\x}^2}}-\frac{\widetilde{C}}{\sigma_\x}\right)\sigma_\x \widetilde{C}
\; \right\}, 
\end{equation*}
which reduces to 
\begin{equation*}
\min_{\x \in \mathcal{P}} \;\left\{\;-\mu_\x +\frac{V{\sqrt{\sigma_\x^4-{K_\x}^2}}}{\sigma_\x{\sqrt{V-\widetilde{C}^2}}}
+ \frac{K_\x \widetilde{C}}{\sigma_\x}  
- \frac{{\sqrt{\sigma_\x^4-{K_\x}^2}}}{\sigma_\x{\sqrt{V-\widetilde{C}^2}}} \widetilde{C}^2 \; \right\}.   
\end{equation*}
Substituting $K_\x = \sigma_\x^2 - \frac{\varepsilon_\x}{2}$ and noting that $\tilde{C}=\sqrt{V}c_0$ concludes the proof.
\Halmos\endproof

\section{Application to a Portfolio of Risks}
\label{appendix Pareto-Clayton}
Here we collect the models used in Section~\ref{sec: applications-insurance}. We provide the first two moments, which are used to calculate the models' parameters by matching the first two moments to those of the reference Pareto-Clayton model. 
\begin{enumerate}
	\item \textbf{Lognormal Model (LN)}: We write $S$ as $S=e^Z$, where $Z$ is a normally distributed random variable with mean $\mu$ and variance $\sigma^2$. The first two moments of $S$ are $\E(S) =e^{\mu+\frac{1}{2}\sigma^2}$ and $\E(S^{2}) =e^{\mu+2\sigma^2}$. 	

	\item \textbf{Gamma model ($\Gamma$)}: The portfolio loss $S$ is Gamma distributed with parameters $\alpha, \beta>0 $, $\E(S) =\alpha \beta$, $\E(S^2)= (1 + \alpha)\, \alpha\, \beta^2$, and density
	 $f_S(s)=\frac{\beta^{\alpha}s^{\alpha-1}e^{-\beta s}}{\Gamma{(\alpha)}}$, $s>0$, 
     where $\Gamma(\cdot)$ denotes the Gamma function. 

	\item \textbf{Weibull model (W)}: The portfolio loss $S$ follows a Weibull distribution with parameters $\lambda, k\in (0, + \infty)$, distribution function 
	    $F_S(s)=1-e^{-(s/\lambda)^k}$, $s>0$,
     $\E(S) =\lambda \Gamma(1+1/k)$, and $\E(S^2)=\lambda^2 \Gamma(1+2/k)$.

    \item \textbf{Inverse Gaussian model (IG)}: The distribution of $S$ is Inverse Gaussian with parameters $\mu, \lambda>0$, distribution function
    $F_S(s)=\Phi \left({\sqrt{\,\tfrac{\lambda}{s}\left(\tfrac{\lambda}{\mu}-1\right)\;}}\right)+e^{\frac{2\lambda}{\mu}}\,\Phi\left(-\,{\sqrt{\,\tfrac{\lambda}{s}\left(\tfrac{s}{\mu}+1\right)}\;}\right)$, $s>0$,
    $\E(S) =\mu$, and $\E(S^2)= \frac{\mu^3}{\lambda}+\mu^2$.

	\item \textbf{Inverse Gamma model (I$\Gamma$)}: The portfolio loss $S$ follows a Inverse Gamma distribution with parameters $\alpha, \beta>0$, density function $f_S(s)=\frac{\beta^{\alpha}{(1/s)}^{\alpha+1}e^{-\beta / s}}{\Gamma{(\alpha})}$, $s>0$, 
    $\E(S) = \frac{\beta}{\alpha-1}$, and $\E(S^2)= \frac{\beta^2}{(\alpha-1)(\alpha-2)}$. 

	\item \textbf{Inverse Weibull model (IW)}: The distribution function of $S$ is Inverse Weibull with parameters $\lambda,\, k >0,$ and density 	    $f_S(s)=e^{-(\lambda/s)^k}$, $s>0$,
    $\E(S) =\lambda \Gamma(1-1/k)$, and $\E(S^2)=\lambda^2 \Gamma(1-2/k)$. 

	\item \textbf{Log-Logistic Model (LL)}: The portfolio loss $S$ follows a Log-Logistic distribution with parameters $\alpha, \beta >0$, quantile function 
	$F^{-1}_S(s)=\alpha \left (\frac{p}{1-p}\right )^{1/\beta}$, $s>0$,
    $\E(S) =\frac{\alpha \pi/\beta}{\sin(\pi/\beta)}$, and $\E(S^2)=\frac{\alpha^2 2\pi/\beta}{\sin(2\pi/\beta)}$. 
\end{enumerate}

\end{APPENDICES}
\OneAndAHalfSpacedXI
\ACKNOWLEDGMENT{S. Pesenti acknowledges the support of the Natural Sciences and Engineering Research Council of Canada (NSERC) with funding reference numbers DGECR-2020-00333 and RGPIN-2020-04289. C. Bernard and S. Vanduffel acknowledge funding from FWO (FWOODYSS11).}


\OneAndAHalfSpacedXI
\bibliography{bibliography.bib}
\bibliographystyle{informs2014} 



\end{document}